\documentclass[superscriptaddress,aps,prx,twocolumn,floatfix]{revtex4-2}

\usepackage{graphicx}
\usepackage[dvipsnames]{xcolor}

\definecolor{bdiv2-0}{HTML}{7ed9e1}
\definecolor{bdiv2-1}{HTML}{58aebf}
\definecolor{bdiv2-2}{HTML}{3f8597}
\definecolor{bdiv2-3}{HTML}{345a68}
\definecolor{bdiv2-4}{HTML}{2b343e}
\definecolor{bdiv2-5}{HTML}{4f2330}
\definecolor{bdiv2-6}{HTML}{91263c}
\definecolor{bdiv2-7}{HTML}{bd494c}
\definecolor{bdiv2-8}{HTML}{de725a}
\definecolor{bdiv2-9}{HTML}{ff9b68}

\usepackage[
colorlinks=true,
linkcolor=bdiv2-6,
citecolor=bdiv2-3,
urlcolor=bdiv2-2,
]{hyperref}
\usepackage{changes}
\usepackage{makerobust}
\usepackage{amsmath}
\usepackage{cancel}
\usepackage{amssymb}
\usepackage{comment}
\usepackage{mathtools}
\usepackage{bm}
\usepackage{braket}
\usepackage[capitalize]{cleveref}
\usepackage[export]{adjustbox}
\usepackage{dsfont}
\usepackage{upgreek}
\usepackage{multirow, colortbl}

\usepackage{xcolor}
\usepackage{comment}
\newcommand{\dosExamplesImplement}{%
\begin{tikzpicture}[x=10pt, y=10pt]
  \begin{scope}[shift={(0,0)}]
    \coordinate (f) at (0,7);
    \coordinate (c) at (0,2);

    \draw[bdiv2-6, fill=bdiv2-6!50!white]
      ($(f)+(0,2)$) to[out=-20, in=90] ($(f)+(4,0)$) to[out=-90, in=20]
      ($(f)-(0,2)$) -- ($(f)+(0,2)$);
    \node[bdiv2-6] at ($(f)-(0.6,0)$) {$f$};

    \draw[bdiv2-3, fill=bdiv2-3!50!white]
      ($(c)+(0,2)$) to[out=-20, in=90] ($(c)+(4,0)$) to[out=-90, in=20]
      ($(c)-(0,2)$) -- ($(c)+(0,2)$);
    \node[bdiv2-3] at ($(c)-(0.6,0)$) {$c$};

    \draw[thick, ->] (0,0) -- (0,14) node[anchor=south west] (E) {$E$};
    \draw[thick, ->] (0,0) -- (5,0) node[anchor=south west] {$\mathrm{DOS}$};
    \draw[thick] ($(f)+(4,0)$) -- ($(f)+(4.5,0)$) node[anchor=west] {$E_F$};
    \node[anchor=east, xshift=-1em] at (E.west) {(a)};

    \draw[gray] ($(c)-(1,0)$) -- ++(-0.5,0)
        --node[midway, above, rotate=90] {$\Delta$} ++(0,5) -- ++(0.5,0);
  \end{scope}
  \begin{scope}[shift={(13,0)}]
    \coordinate (f) at (0,7);
    \coordinate (c) at (0,2);
    \coordinate (c2) at (0,12);

    \draw[bdiv2-6, fill=bdiv2-6!50!white]
      ($(f)+(0,2)$) to[out=-20, in=90] ($(f)+(4,0)$) to[out=-90, in=20]
      ($(f)-(0,2)$) -- ($(f)+(0,2)$);
    \node[bdiv2-6] at ($(f)-(0.6,0)$) {$f$};

    \draw[bdiv2-3, fill=bdiv2-3!50!white]
      ($(c)+(0,2)$) to[out=-20, in=90] ($(c)+(4,0)$) to[out=-90, in=20]
      ($(c)-(0,2)$) -- ($(c)+(0,2)$);
    \node[bdiv2-3] at ($(c)-(0.6,0)$) {$c$};

    \draw[bdiv2-3, fill=bdiv2-3!50!white]
      ($(c2)+(0,1.5)$) to[out=-20, in=90] ($(c2)+(4,0)$) to[out=-90, in=20]
      ($(c2)-(0,1.5)$) -- ($(c2)+(0,1.5)$);
    \node[bdiv2-3] at ($(c2)-(0.6,0)$) {$c$};

    \draw[thick, ->] (0,0) -- (0,14) node[anchor=south west] (E) {$E$};
    \draw[thick, ->] (0,0) -- (5,0) node[anchor=south west] {$\mathrm{DOS}$};
    \draw[thick] ($(f)+(4,0)$) -- ($(f)+(4.5,0)$) node[anchor=west] {$E_F$};
    \node[anchor=east, xshift=-1em] at (E.west) {(b)};

    \draw[gray] ($(c)-(1,0)$) -- ++(-0.5,0)
        --node[midway, above, rotate=90] {$\Delta_-$} ++(0,5) -- ++(0.5,0)
        -- ++(-0.5,0)
        --node[midway, above, rotate=90] {$\Delta_+$} ++(0,5) -- ++(0.5,0);
        
  \end{scope}
\end{tikzpicture}%
}

\definecolor{A22color}{HTML}{3C798A}
\definecolor{B22color}{HTML}{9B7FC5}
\definecolor{A31color}{HTML}{FFB18A}
\definecolor{HoleColor}{HTML}{282128}
\definecolor{PartColor}{HTML}{EEEEEE}
\definecolor{gcolor}{HTML}{BCE2E6}
\definecolor{fspacebg}{HTML}{E5E5E5}
\colorlet{notconn}{bdiv2-6}
\colorlet{G2c}{B22color}

\definecolor{Cud}{HTML}{91263C}
\definecolor{Oem}{HTML}{7ED9E1}
\definecolor{Onem}{HTML}{345A68}

\usepackage{tikz}
\usetikzlibrary{arrows, positioning, calc, shapes, shapes.geometric,
decorations.markings, decorations.pathreplacing, math, backgrounds}

\tikzset{Cvert/.pic={
  \node[inner sep=0, outer sep=0] (-c) at (0,0) {};
  \node[inner sep=0, outer sep=0] (-1) at (-0.25,-0.25) {};
  \node[inner sep=0, outer sep=0] (-2) at (-0.25, 0.25) {};
  \node[inner sep=0, outer sep=0] (-3) at ( 0.25,-0.25) {};
  \node[inner sep=0, outer sep=0] (-4) at ( 0.25, 0.25) {};
  \draw[HoleColor,fill=gcolor] (-1) rectangle (-4);
  \draw[draw=HoleColor, fill=PartColor] (-1) circle [radius=2pt];
  \draw[draw=HoleColor, fill=HoleColor] (-4) circle [radius=2pt];
  \draw[draw=HoleColor, fill=HoleColor] (-2) circle [radius=2pt];
  \draw[draw=HoleColor, fill=PartColor] (-3) circle [radius=2pt];
}}

\tikzset{invisi/.pic={
  \node[inner sep=0, outer sep=0] (-c) at (0,0) {};
  \node[inner sep=0, outer sep=0] (-1) at (-0.25,-0.25) {};
  \node[inner sep=0, outer sep=0] (-2) at (-0.25, 0.25) {};
  \node[inner sep=0, outer sep=0] (-3) at ( 0.25,-0.25) {};
  \node[inner sep=0, outer sep=0] (-4) at ( 0.25, 0.25) {};
  \draw[white,fill=white] (-1) rectangle (-4);
  \draw[draw=white, fill=white] (-1) circle [radius=2pt];
  \draw[draw=white, fill=white] (-4) circle [radius=2pt];
  \draw[draw=white, fill=white] (-2) circle [radius=2pt];
  \draw[draw=white, fill=white] (-3) circle [radius=2pt];
}}

\tikzset{A22/.pic={
  \node[inner sep=0, outer sep=0] (-c) at (0,0) {};
  \node[inner sep=0, outer sep=0] (-1) at (-0.25,-0.25) {};
  \node[inner sep=0, outer sep=0] (-2) at (-0.25, 0.25) {};
  \node[inner sep=0, outer sep=0] (-3) at ( 0.25,-0.25) {};
  \node[inner sep=0, outer sep=0] (-4) at ( 0.25, 0.25) {};
  \fill[A22color] (-1) rectangle (-4);
  \draw[draw=HoleColor, fill=PartColor] (-1) circle [radius=2pt];
  \draw[draw=HoleColor, fill=HoleColor] (-4) circle [radius=2pt];
  \draw[draw=HoleColor, fill=HoleColor] (-2) circle [radius=2pt];
  \draw[draw=HoleColor, fill=PartColor] (-3) circle [radius=2pt];
}}
\tikzset{A222/.pic={
  \node[inner sep=0, outer sep=0] (-c) at (0,0) {};
  \node[inner sep=0, outer sep=0] (-1) at (-0.25,-0.25) {};
  \node[inner sep=0, outer sep=0] (-2) at (-0.25, 0.25) {};
  \node[inner sep=0, outer sep=0] (-3) at ( 0.25,-0.25) {};
  \node[inner sep=0, outer sep=0] (-4) at ( 0.25, 0.25) {};
  \fill[A22color] (-1) rectangle (-4);
  \draw[draw=HoleColor, fill=HoleColor] (-1) circle [radius=2pt];
  \draw[draw=HoleColor, fill=PartColor] (-4) circle [radius=2pt];
  \draw[draw=HoleColor, fill=HoleColor] (-2) circle [radius=2pt];
  \draw[draw=HoleColor, fill=PartColor] (-3) circle [radius=2pt];
}}
\tikzset{B22/.pic={
  \node[inner sep=0, outer sep=0] (-c) at (0,0) {};
  \node[inner sep=0, outer sep=0] (-1) at (-0.25,-0.25) {};
  \node[inner sep=0, outer sep=0] (-2) at (-0.25, 0.25) {};
  \node[inner sep=0, outer sep=0] (-3) at ( 0.25,-0.25) {};
  \node[inner sep=0, outer sep=0] (-4) at ( 0.25, 0.25) {};
  \fill[B22color] (-1) rectangle (-4);
  \draw[draw=HoleColor, fill=PartColor] (-1) circle [radius=2pt];
  \draw[draw=HoleColor, fill=HoleColor] (-4) circle [radius=2pt];
  \draw[draw=HoleColor, fill=HoleColor] (-2) circle [radius=2pt];
  \draw[draw=HoleColor, fill=PartColor] (-3) circle [radius=2pt];
}}
\tikzset{A31/.pic={
  \node[inner sep=0, outer sep=0] (-c) at (0,0) {};
  \node[inner sep=0, outer sep=0] (-1) at (-0.25,-0.25) {};
  \node[inner sep=0, outer sep=0] (-2) at (-0.25, 0.25) {};
  \node[inner sep=0, outer sep=0] (-3) at ( 0.25,-0.25) {};
  \node[inner sep=0, outer sep=0] (-4) at ( 0.25, 0.25) {};
  \fill[A31color] (-1) rectangle (-4);
  \draw[draw=HoleColor, fill=PartColor] (-1) circle [radius=2pt];
  \draw[draw=HoleColor, fill=HoleColor] (-4) circle [radius=2pt];
  \draw[draw=HoleColor, fill=HoleColor] (-2) circle [radius=2pt];
  \draw[draw=HoleColor, fill=PartColor] (-3) circle [radius=2pt];
}}
\tikzset{Vff/.pic={
  \node[inner sep=0, outer sep=0] (-c) at (0,0) {};
  \node[inner sep=0, outer sep=0] (-1) at (-0.25,-0.25) {};
  \node[inner sep=0, outer sep=0] (-2) at (-0.25, 0.25) {};
  \node[inner sep=0, outer sep=0] (-3) at ( 0.25,-0.25) {};
  \node[inner sep=0, outer sep=0] (-4) at ( 0.25, 0.25) {};
  \fill[bdiv2-6!50!white, draw=bdiv2-6] (-1) rectangle (-4);
  \draw[draw=HoleColor, fill=PartColor] (-1) circle [radius=2pt];
  \draw[draw=HoleColor, fill=HoleColor] (-4) circle [radius=2pt];
  \draw[draw=HoleColor, fill=HoleColor] (-2) circle [radius=2pt];
  \draw[draw=HoleColor, fill=PartColor] (-3) circle [radius=2pt];
}}
\tikzset{Sff/.pic={
  \node[inner sep=0, outer sep=0] (-c) at (0,0) {};
  \node[inner sep=0, outer sep=0] (-1) at (-0.25,0) {};
  \node[inner sep=0, outer sep=0] (-2) at ( 0.25,0) {};
  \draw[double, double=bdiv2-6!50!white, bdiv2-6, line width=0.3pt,
  double distance=3.8pt] (-1) -- (-2);
  \draw[draw=HoleColor, fill=PartColor] (-1) circle [radius=2pt];
  \draw[draw=HoleColor, fill=HoleColor] (-2) circle [radius=2pt];
}}
\tikzset{A11/.pic={
  \node[inner sep=0, outer sep=0] (-c) at (0,0) {};
  \node[inner sep=0, outer sep=0] (-1) at (-0.25,0) {};
  \node[inner sep=0, outer sep=0] (-2) at ( 0.25,0) {};
  \draw[A22color, line width=4.4pt] (-1) -- (-2);
  \draw[draw=HoleColor, fill=PartColor] (-1) circle [radius=2pt];
  \draw[draw=HoleColor, fill=HoleColor] (-2) circle [radius=2pt];
}}
\tikzset{A11i/.pic={
  \node[inner sep=0, outer sep=0] (-c) at (0,0) {};
  \node[inner sep=0, outer sep=0] (-1) at (-0.25,0) {};
  \node[inner sep=0, outer sep=0] (-2) at ( 0.25,0) {};
  \draw[A22color, line width=4.4pt] (-1) -- (-2);
  \draw[draw=HoleColor, fill=PartColor] (-2) circle [radius=2pt];
  \draw[draw=HoleColor, fill=HoleColor] (-1) circle [radius=2pt];
}}
\tikzset{gprop/.style={
  double,
  line width=0.5pt,
  double distance=1pt,
  double=gcolor,
  HoleColor,
}}
\tikzset{fprop/.style={
  line width=1.5pt,
  dashed,
  bdiv2-6,
}}
\tikzset{gfprop/.style={
  line width=1.5pt,
  B22color,
}}

\newcommand{\bg}[1]{\begin{pgfonlayer}{bg}#1\end{pgfonlayer}}

\newcommand{\newDoubleDiagramsImplement}{
\begin{tikzpicture}
  \pgfdeclarelayer{bg}
  \pgfsetlayers{bg,main}

  \node[anchor=west] at (-0.5,0.8) {(a)\strut};
  \pic (Sff) at (0,0) {Sff};
  \node[anchor=south, yshift=2pt, bdiv2-6] at (Sff-c) {$\Sigma^f$};
  \pic (tfc) at (1.2,0) {A11};
  \node[anchor=center] at ($0.5*(Sff-2)+0.5*(tfc-1)$) {$=$};
  \pic (tcf) at (2.4,0) {A11};
  \pic (Vcf) at (3.8,0) {A222};
  \bg{ \node[ellipse, minimum height=0.8cm, minimum width=1.0cm, fill=fspacebg] at
          ($0.5*(tcf-2)+0.5*(tfc-1)$) {};
       \node[ellipse, minimum height=0.7cm, minimum width=1.1cm, fill=fspacebg] at
          ($0.5*(Vcf-2) + 0.5*(Vcf-4) + (0,0.1)$) {};
       \draw[gprop] (Vcf-2) to[out=135, in=45, looseness=2] (Vcf-4);
       \draw[gprop] (tfc-2) -- node[midway,above,gcolor!80!black,yshift=-1.5pt] {\scriptsize$g^{(1)}$} (tcf-1); }
  \node[anchor=center] at (3.1,0) {$+$};
  \node[anchor=south, yshift=2pt, A22color] at (tfc-c) {\scriptsize$A^{1:1}$};
  \node[anchor=south, xshift=4pt, yshift=2pt, A22color] at (tcf-c) {\scriptsize$A^{1:1}$};
  \node[anchor=center, A22color!10!white] at (Vcf-c) {\scriptsize$A^{2:2}$};

  \begin{scope}[shift={(5.1,0)}]
    \node[anchor=west] at (-0.5,0.8) {(b)\strut};
    \pic (Gff) at (0,0) {Vff};
    \node[anchor=center, bdiv2-6] at (Gff-c) {$\Gamma^f$};
    \begin{scope}[shift={(-0.25,0)}]
    \pic[rotate=-45] (tfc) at (1.2,0.5) {A11};
    \pic[rotate= 45] (tcf) at (2.3,0.5) {A11};
    \pic[rotate=-90] (Vcf) at (1.75,-0.25) {A22};
    \pic (V1) at (3.5,0) {A22};
    \pic[rotate=180] (V2) at (5,0) {A22};
    \node[anchor=west, xshift=2pt] at ($0.5*(Gff-4)+0.5*(Gff-3)$) {$=$};
    \node[anchor=center] at (2.7,0) {$\displaystyle{}+\frac12$};
    \bg{ \node[ellipse, fill=fspacebg, minimum width=1.1cm,
               minimum height=0.8cm, yshift=-4pt]
            at ($0.5*(tfc-2)+0.5*(tcf-1)$) {};
         \node[ellipse, fill=fspacebg, minimum width=1.5cm, minimum height=1.2cm]
            at ($0.5*(V1-c)+0.5*(V2-c)$) {};
         \draw[gprop] (tfc-2) -- (Vcf-1);
         \draw[gprop] (Vcf-2) -- (tcf-1);
         \draw[gprop] (V1-3) to[in=-135, out=-45] (V2-4);
         \draw[gprop] (V1-4) to[in=135, out=45] (V2-3); }
    \end{scope}
  \end{scope}

  \begin{scope}[shift={(11,0)}]
    \node[anchor=west] at (-0.5,0.8) {(c)\strut};
    \node[regular polygon, regular polygon sides=6,
      inner sep=-2pt, outer sep=0pt, draw=bdiv2-6, fill=bdiv2-6!50!white,
      text=bdiv2-6] (G6) at (0.2,0) {\scriptsize$\Gamma^{f;(6)}$};
    \draw[draw=HoleColor, fill=PartColor] (G6.corner 1) circle [radius=2pt];
    \draw[draw=HoleColor, fill=HoleColor] (G6.corner 2) circle [radius=2pt];
    \draw[draw=HoleColor, fill=PartColor] (G6.corner 3) circle [radius=2pt];
    \draw[draw=HoleColor, fill=HoleColor] (G6.corner 4) circle [radius=2pt];
    \draw[draw=HoleColor, fill=PartColor] (G6.corner 5) circle [radius=2pt];
    \draw[draw=HoleColor, fill=HoleColor] (G6.corner 6) circle [radius=2pt];

    \pic[rotate=-60] (A1) at (1.6,0.5) {A11};
    \pic[rotate= 60] (A2) at (3.6,0.5) {A11};
    \pic[rotate= 240] (V1) at (2.0,-0.4) {A22};
    \pic[rotate=-60] (V2) at (3.2,-0.4) {A22};
    \bg{ \node[ellipse, fill=fspacebg, minimum width=2.3cm, minimum
               height=1.4cm, yshift=2pt] at
                  ($0.25*(A1-c)+0.25*(A2-c)+0.25*(V1-c)+0.25*(V2-c)$) {};}
    \bg{ \draw[gprop] (A1-2) -- (V1-1); }
    \bg{ \draw[gprop] (V1-2) -- (V2-1); }
    \bg{ \draw[gprop] (V2-2) -- (A2-1); }

    \begin{scope}[shift={(4.9,0)}]
      \pic (X1) at (0,0) {A22};
      \pic[rotate=120] (X2) at (1,-0.57) {A22};
      \pic[rotate=240] (X3) at (1, 0.57) {A22};
      \bg{ \node[circle, fill=fspacebg, minimum width=1.3cm, inner sep=0pt, outer sep=0pt] at
                    ($0.333*(X1-c)+0.333*(X2-c)+0.333*(X3-c)$) {}; }
      \bg{ \draw[gprop] (X1-3) -- (X2-4); }
      \bg{ \draw[gprop] (X2-3) -- (X3-4); }
      \bg{ \draw[gprop] (X3-3) -- (X1-4); }
    \end{scope}

    \node[anchor=center] at (1.1,0) {$=$};
    \node[anchor=center] at (4.1,0) {$\displaystyle{}-\frac13$};

  \end{scope}
\end{tikzpicture}
}

\newcommand{\newRPAdiagramsImplement}{
\begin{tikzpicture}
  \pgfdeclarelayer{bg}
  \pgfsetlayers{bg,main}
  \pic (Gf) at (0,0) {Vff};
  \node[bdiv2-6] at (Gf-c) {$\Gamma^f$};
  \node[anchor=center] at (0.75,0) {$=$};

  \pic (V1) at (1.5,0) {A22};
  \pic[rotate=180] (V2) at (3,0) {A22};
  \node[anchor=center, A22color!10!white] at (V1-c) {\scriptsize$A^{2:2}$};
  \node[anchor=center, A22color!10!white] at (V2-c) {\scriptsize$A^{2:2}$};
  \bg{ \node[ellipse, fill=fspacebg, minimum width=1.5cm, minimum height=1.2cm] at (2.25,0) {};
       \draw[gprop] (V1-4) to[out=45, in=135] node[midway, above,gcolor!80!black, yshift=2pt] {$g^{(1)}$} (V2-3);
       \draw[gprop] (V2-4) to[out=225, in=315] (V1-3); }

  \node[anchor=center] at (3.75,0) {$+$};

  \begin{scope}[shift={(0,2.45)}]
    \coordinate (X1) at (4.6,0);
    \coordinate (X2) at (6.1,0);
    \pic[rotate= 45] (V3) at (X1) {A31};
    \pic[rotate=135] (V4) at (X2) {A31};
    \bg{ \node[ellipse, fill=fspacebg, minimum width=1.5cm, minimum height=0.6cm] at ($0.5*(X1)+0.5*(X2)$) {};
         \draw[gprop] (V4-1) -- (V3-3);}
    \node[anchor=center, A31color!50!black] at (V3-c) {\scriptsize$A^{\!3:1}$};
    \node[anchor=center, A31color!50!black] at (V4-c) {\scriptsize$A^{\!3:1}$};

    \node[single arrow, draw=bdiv2-6, fill=bdiv2-6!30!white, rotate=-90,
        minimum width=14pt, single arrow head extend=3pt,
        minimum height=10mm] (arrow) at ($0.5*(X1)+0.5*(X2)-(0,0.8)$) {};
    \node[scale=0.6,anchor=center, rotate=-90, xshift=2pt, bdiv2-6] at (arrow.center)
    {\scriptsize$\Gamma^{(6)}\to\Gamma^{(4)}$};
  \end{scope}

  \coordinate (X1) at (4.6,0);
  \coordinate (X2) at (6.1,0);
  \pic[rotate= 45] (V3) at (X1) {A31};
  \pic[rotate=135] (V4) at (X2) {A31};
  \node[anchor=center, A31color!50!black] at (V3-c) {\scriptsize$A^{\!3:1}$};
  \node[anchor=center, A31color!50!black] at (V4-c) {\scriptsize$A^{\!3:1}$};
  \bg{ \node[ellipse, fill=fspacebg, minimum width=1.5cm, minimum height=0.6cm] at ($0.5*(X1)+0.5*(X2)$) {};
       \draw[gprop] (V4-1) -- (V3-3);
       \draw[fprop] (V3-4) to[out=45, in=135] node[midway, above, yshift=-2.5pt] {$G^f_0$} (V4-3); }

  \coordinate (c5) at ($(V4-c) + (V3-c) - (V2-c) - (1.5,0)$);
  \begin{scope}[shift={(c5)}]
    \node[anchor=center] at (0.75,0) {$+$};
    \pic (V5) at (1.5,0) {A22};
    \pic[rotate=180] (V6) at (3,0) {Cvert};
    \node[anchor=center, HoleColor] at (V6-c) {$C$};
    \pic (V7) at (4.5,0) {A22};
    \bg{ \node[ellipse, fill=fspacebg, minimum width=3.4cm, minimum height=1.4cm] at (3.0,0) {};
         \draw[gprop] (V5-4) to[out=45, in=135]   (V6-3);
         \draw[gprop] (V6-4) to[out=225, in=315]  (V5-3);
         \draw[gprop] (V6-2) to[out=-45, in=-135] (V7-1);
         \draw[gprop] (V6-1) to[out=45, in=135]   (V7-2); }
  \end{scope}
  \node[anchor=center] at ($(V7-c)+(0.75,0)$) {$+$};

  \coordinate (c6) at ($(V7-c) - (V2-c)$);
  \begin{scope}[shift={(c6)}]
  \coordinate (X3) at (4.6,0);
  \coordinate (X4) at (8.1,0);
  \coordinate (X34) at ($0.5*(X3)+0.5*(X4)$);
  \pic[rotate= 45] (V8) at (X3) {A31};
  \pic[rotate=-45] (V9) at (X4) {A31};
  \pic[rotate=180] (V10) at (X34) {A22};
  \bg{
    \fill[fspacebg, draw=fspacebg, line width=2pt]
    (V8-c) to[out=45, in=-180] (V10-c) to[out=0, in=135] (V9-c) to[out=-135, in=-45] (V8-c);
    \fill[fspacebg] ($(V8-c)-(0,0.1)$) rectangle ($(V9-c)+(0,0.02)$);
    \draw[gprop] (V8-3) to[out=-60, in=-135] (V10-4);
    \draw[gprop] (V10-2) to[out=-45, in=-120] (V9-1);
    \draw[fprop] (V8-4) to[out=45, in=120] (V10-3);
    \draw[fprop] (V10-1) to[out=60, in=135] (V9-2);
  }
  \node[anchor=west] at ($(X4)+(0.5,0)$) {${}+\dots$};
  \begin{scope}[shift={(0,2.45)}]
    \coordinate (X3) at (4.6,0);
    \coordinate (X4) at (8.1,0);
    \coordinate (X34) at ($0.5*(X3)+0.5*(X4)$);
    \pic[rotate= 45] (V8) at (X3) {A31};
    \pic[rotate=-45] (V9) at (X4) {A31};
    \pic[rotate=180] (V10) at (X34) {A22};
    \bg{
      \fill[fspacebg, draw=fspacebg, line width=2pt]
      (V8-c) to[out=45, in=-180] (V10-c) to[out=0, in=135] (V9-c) to[out=-135, in=-45] (V8-c);
      \fill[fspacebg] ($(V8-c)-(0,0.1)$) rectangle ($(V9-c)+(0,0.02)$);
      \draw[gprop] (V8-3) to[out=-60, in=-135] (V10-4);
      \draw[gprop] (V10-2) to[out=-45, in=-120] (V9-1);
    }
    \node[single arrow, draw=bdiv2-6, fill=bdiv2-6!30!white, rotate=-90,
        minimum width=14pt, single arrow head extend=3pt,
        minimum height=10mm] (arrow) at ($0.5*(X3)+0.5*(X4)-(0,1.4)$) {};
    \node[scale=0.6,anchor=center, rotate=-90, xshift=2pt, bdiv2-6] at (arrow.center)
    {\scriptsize$\Gamma^{(8)}\to\Gamma^{(4)}$};
  \end{scope}
  \end{scope}

\end{tikzpicture}
}

\newcommand{\crossdiagrams}{
\begin{tikzpicture}
  \pgfdeclarelayer{bg}
  \pgfsetlayers{bg,main}
  \pic (Gf) at (0,0) {Vff};
  \node[bdiv2-6] at (Gf-c) {$\Gamma^f$};
  \coordinate (X1) at (4.6,0);
  \coordinate (X2) at (6.1,0);
  \coordinate (X3) at (1.5,0);
  \coordinate (X4) at (3.0,0);
  \pic (V1) at (1.5,0) {A22};
  \pic[rotate=180] (V2) at (3,0) {Vff};
  \node[bdiv2-6] at (V2-c) {$F$};
  \node[anchor=center, A22color!10!white] at (V1-c) {\scriptsize$A^{2:2}$};
  \bg{ \node[ellipse, fill=fspacebg, minimum width=0.6cm, minimum height=1.2cm] at (1.8,0) {};
       \draw[very thick, B22color] (V1-4) to[out=45, in=135]node[midway, above,B22color!80!black, yshift=0pt] {$g_{cf}^{(1)}$}  (V2-3);
       \draw[very thick, B22color] (V2-4) to[out=225, in=315] (V1-3);}
  
  \begin{scope}[shift={(0,3)}]
  \pic (V1) at (1.5,0) {A22};
  \pic[rotate=180] (T2) at (2.8,-0.3) {A11};
  \pic (T1) at (2.8,0.3) {A11};
  \node[anchor=center, A22color!10!white] at (V1-c) {\scriptsize$A^{2:2}$};
  \node[anchor=center, A22color!50!black] at (2.9, 0.55) {\scriptsize$A^{1:1}$};
  \node[anchor=center, A22color!50!black] at (2.9, -0.55) {\scriptsize$A^{1:1}$};
  \bg{ \node[ellipse, fill=fspacebg, minimum width=1.5cm, minimum height=1.2cm] at (2.2,0) {};
       \draw[gprop] (V1-4) to[out=45, in=135] node[midway, above,gcolor!80!black, yshift=2pt] {$g_{c}^{(1)}$} (T1-1);
       \draw[gprop] (T2-2) to[out=225, in=315] (V1-3);
    }
    
    \node[single arrow, draw=bdiv2-6, fill=bdiv2-6!30!white, rotate=-90,
        minimum width=14pt, single arrow head extend=3pt,
        minimum height=10mm] (arrow) at (1.6,-1.4) {};
    \node[scale=0.6,anchor=center, rotate=-90, xshift=2pt, bdiv2-6] at (arrow.center)
    {$\times$};
    \pic[rotate=180] (V2) at (3,-1.3) {Vff};
    \node[bdiv2-6] at (V2-c) {$F$};
    \draw[fprop] ($(V2-3)-(0.7,0.0)$) to[out=45, in=135] (V2-3);
    \draw[fprop] ($(V2-4)-(0.7,0.0)$)  to[out=-45, in=-135] (V2-4);
    \node[bdiv2-6] at ($(V2-c)-(0.7,0.0)$){$G^f_0$};
  \end{scope}
    \node[] at (2.25,-1)
    {$\propto g_{cf}^{(1)}g_{cf}^{(1)}$};

  \pic (V3) at (X1) {A31};
  \pic (V4) at (X2) {A31};
  \node[anchor=center, A31color!50!black] at (V3-c) {\scriptsize$A^{\!3:1}$};
  \node[anchor=center, A31color!50!black] at (V4-c) {\scriptsize$A^{\!3:1}$};
  \bg{\node[ellipse, fill=fspacebg, minimum width=1.4cm, minimum height=0.5cm, rotate=0] at (5.3,-0.25) {};
         \draw[very thick, B22color] (V3-4) to[out=45, in=135] node[midway, above,B22color!80!black, yshift=2.5pt] {$g_{cf}^{(1)}$} (V4-1);
         \draw[very thick, B22color] (V3-3) to[out=45, in=135]  (V4-2);}
   
  \begin{scope}[shift={(4,3)}]
      \pic (V1) at (0.4,-0.3) {A31};
      \pic[rotate=180] (T1) at ($(V1-3) + (0.8,0.0)$) {A11};
      \pic[rotate=180] (V2) at (2, 0.3) {A31};
      \pic (T2) at ($(V2-3) - (0.8,0.0)$) {A11};
      \node[anchor=center, A31color!50!black] at (V1-c) {\scriptsize$A^{\!3:1}$};
      \node[anchor=center, A31color!50!black] at (V2-c) {\scriptsize$A^{\!3:1}$};
      \bg{ 
      \node[ellipse, fill=fspacebg, minimum width=1.0cm, minimum height=0.5cm, rotate=0] at ($(V1-3) + (0.25,0.0)$) {};
      \node[ellipse, fill=fspacebg, minimum width=1.0cm, minimum height=0.5cm, rotate=0] at ($(T2-2) + (0.25,0.0)$) {};
      \draw[gprop] (V1-3) to[out=0, in=180] (T1-1);
      \draw[gprop] (T2-2) to[out=0, in=180] (V2-3);
      }
      \node[single arrow, draw=bdiv2-6, fill=bdiv2-6!30!white, rotate=-90,
            minimum width=14pt, single arrow head extend=3pt,
            minimum height=10mm] (arrow) at (0.3,-1.4) {};
        \node[scale=0.6,anchor=center, rotate=-90, xshift=2pt, bdiv2-6] at (arrow.center)
        {$\times$};
        \draw[fprop] (0.9,-1.1) to[out=-90, in=90] (0.9,-1.8);
        \draw[fprop] (1.5,-1.1) to[out=-90, in=90] node[midway, xshift=10pt] {$G^f_0$} (1.5,-1.8);
  \end{scope}
    \node at (5.25,-1)
    {$\propto g_{cf}^{(1)}g_{fc}^{(1)}$};
  
  \begin{scope}[shift={(7.5,0)}]
  \pic (V1) at (0.0,0.0) {A31};
  \node[anchor=center, A31color!50!black] at (V1-c) {\scriptsize$A^{\!3:1}$};
  \pic[rotate=180] (V2) at (1.5,0) {Vff};
  \node[bdiv2-6] at (V2-c) {$F$};
  \bg{ 
    \node[ellipse, fill=fspacebg, minimum width=0.5cm, minimum height=0.5cm, rotate=0] at (V1-4) {};
    \draw[fprop] (V1-3) to[out=-45, in=-135] (V2-4);
    \draw[very thick, B22color] (V1-4) to[out=45, in=135] (V2-3);
  }
    \node at (0.75,-1)
    {$\propto g_{cf}^{(1)}G_{0}^{f}$};
  \end{scope}
  
  \begin{scope}[shift={(7.5,3)}]
      \pic (V1) at (0.4,0.0) {A31};
      \pic[rotate=180] (T1) at ($(V1-3) + (0.8,0.0)$) {A11};
      \node[anchor=center, A31color!50!black] at (V1-c) {\scriptsize$A^{\!3:1}$};
      \bg{ 
      \node[ellipse, fill=fspacebg, minimum width=1.0cm, minimum height=0.5cm, rotate=0] at ($(V1-3) + (0.25,0.0)$) {};
      \draw[gprop] (V1-3) to[out=0, in=180] (T1-1);
      }
      \node[single arrow, draw=bdiv2-6, fill=bdiv2-6!30!white, rotate=-90,
            minimum width=14pt, single arrow head extend=3pt,
            minimum height=10mm] (arrow) at (0.1,-1.4) {};
        \node[scale=0.6,anchor=center, rotate=-90, xshift=2pt, bdiv2-6] at (arrow.center)
        {$\times$};
        \pic[rotate=180] (V2) at (1.5,-1.3) {Vff};
        \node[bdiv2-6] at (V2-c) {$F$};
        \draw[fprop] ($(V2-3)-(0.7,0.0)$) to[out=45, in=135] (V2-3);
        \draw[fprop] ($(V2-4)-(0.7,0.0)$)  to[out=-45, in=-135] (V2-4);
        \node[bdiv2-6] at ($(V2-c)-(0.7,0.0)$){$G^f_0$};
  \end{scope}
  
  \begin{scope}[shift={(10.5,0)}]
  \pic (V1) at (0.0,0.0) {A31};
  \node[anchor=center, A31color!50!black] at (V1-c) {\scriptsize$A^{\!3:1}$};
  \pic[rotate=180] (V2) at (1.5,0) {A22};
  \node[anchor=center, A22color!10!white] at (V2-c) {\scriptsize$A^{2:2}$};
  \bg{ 
    \node[ellipse, fill=fspacebg, minimum width=1.6cm, minimum height=0.9cm, rotate=20] at (0.85,-0.1) {};
    \draw[gprop] (V1-3) to[out=-45, in=-135] (V2-4);
    \draw[very thick, B22color] (V1-4) to[out=45, in=135] (V2-3);
  }
    \node at (0.75,-1)
    {$\propto g_{cf}^{(1)}g^{(1)}$};
  \end{scope}
  
  \begin{scope}[shift={(10.5,3)}]
        \pic (V1) at (0.0,-0.4) {A31};
      \node[anchor=center, A31color!50!black] at (V1-c) {\scriptsize$A^{\!3:1}$};
      \pic[rotate=180] (V2) at (1.5,-0.4) {A22};
      \node[anchor=center, A22color!10!white] at (V2-c) {\scriptsize$A^{2:2}$};
      \pic[rotate=0] (T2) at ($(V2-3) + (-0.5,0.4)$) {A11};
      \bg{ 
        \node[ellipse, fill=fspacebg, minimum width=1.6cm, minimum height=0.9cm, rotate=20] at      (0.85,-0.5) {};
        \node[ellipse, fill=fspacebg, minimum width=0.7cm, minimum height=0.7cm, rotate=135] at      ($(V2-3) + (-0.15,0.25)$) {};
        \draw[gprop] (V1-3) to[out=-45, in=-135] (V2-4);
        \draw[gprop] (V2-3) to[out=120, in=-60] (T2-2);
        
    \node[single arrow, draw=bdiv2-6, fill=bdiv2-6!30!white, rotate=-90,
        minimum width=14pt, single arrow head extend=3pt,
        minimum height=10mm] (arrow) at (0.1,-1.7) {};
    \node[scale=0.6,anchor=center, rotate=-90, xshift=2pt, bdiv2-6] at (arrow.center)
    {\scriptsize$\Gamma^{(6)}\to\Gamma^{(4)}$};
    \draw[fprop] (0.6,-1.7) to[out=0, in=180] node[midway, above] {$G^f_0$} (1.6,-1.7);
      }
  \end{scope}
         
  \node[anchor=center] at (0.75,0) {$=$};
  \node[anchor=center] at (3.75,0) {$+$};
  \node[anchor=center] at (6.75,0) {$+$};
  \node[anchor=center] at (9.75,0) {$+$};
  \node[anchor=west] at (12.75,0) {${}+\dots$};
\end{tikzpicture}
}

\newcommand{\connectedness}{
\begin{tikzpicture}[x=.8cm, y=.8cm]
  \pgfdeclarelayer{bg}
  \pgfsetlayers{bg,main}

  \node[anchor=north west] at (-3, 1) {(a)};
  \node[anchor=north west] at (-3,-1) {(b)};
  \node[anchor=north west] at (-3,-3) {(c)};
  \draw [dashed, color=gray!50!white] (-2,-0.8) -- (17,-0.8);
  \draw [dashed, color=gray!50!white] (-2,-2.8) -- (17,-2.8);
  
  \begin{scope}[shift={(-1.1,0)}]
  \pic[rotate=-90] (T1) at (-0.6,0) {A11};
  \pic[rotate=90] (T2) at (0,0) {A11};
  \pic[rotate=90] (V1) at (0.8,0) {A22};
      \bg{\node[rectangle, fill=fspacebg, minimum width=2cm, minimum height=0.8cm, rounded corners] at (0.2,0.4) {};
      \coordinate (A) at (-0.6,0.25);
      \coordinate (B) at (0.0,0.25);
      \coordinate (C) at (0.55,0.25);
      \coordinate (D) at (1.05,0.25);
      \draw[fill=gcolor]
        (A) 
        to[out=45, in=135, looseness=0.6] (B)
        to[out=45, in=135, looseness=0.6] (C)
        to[out=45, in=135, looseness=0.6] (D)
        to[out=90, in=90, looseness=0.9] (A)
       -- cycle;
   }
  \end{scope}
  
  \begin{scope}[shift={(-1,-4)}]
  \pic[rotate=-90] (T1) at (-0.55,0) {A11};
  \pic[rotate=-90] (T2) at (-0.1,0) {A11};
  \pic[rotate=90] (T4) at (0.35,0) {A11};
  \pic[rotate=90] (T3) at (0.8,0) {A11};
      \bg{\node[rectangle, fill=fspacebg, minimum width=1.8cm, minimum height=0.8cm, rounded corners] at (0.15,0.4) {};
      \coordinate (A) at (-0.55,0.25);
      \coordinate (B) at (-0.1,0.25);
      \coordinate (C) at (0.35,0.25);
      \coordinate (D) at (0.8,0.25);
      \draw[fill=gcolor]
        (A) 
        to[out=45, in=135, looseness=0.6] (B)
        to[out=45, in=135, looseness=0.6] (C)
        to[out=45, in=135, looseness=0.6] (D)
        to[out=90, in=90, looseness=0.9] (A)
       -- cycle;
   }
  \end{scope}

  \begin{scope}[shift={(-0.9,-2)}]
  \pic[rotate=-45] (V1) at (-0.5,0.2) {A31};
  \pic[rotate=-135] (V2) at (0.5,-0.2) {A31};
      \bg{\node[rectangle, fill=fspacebg, minimum width=2cm, minimum height=0.9cm, rounded corners] at (0.0,0.45) {};
      \coordinate (A) at (-0.85,0.2);
      \coordinate (B) at (-0.5,0.55);
      \coordinate (C) at (-0.15,0.2);
      \coordinate (D) at (0.5,0.15);
      \draw[fill=gcolor]
        (A) 
        to[out=80, in=-180, looseness=0.6] (B)
        to[out=0, in=100, looseness=0.6] (C)
        to[out=70, in=160, looseness=0.6] (D)
        to[out=90, in=90, looseness=1.7] (A)
       -- cycle;
   }
  \end{scope} 
  \node[anchor=west] at (.9,0.0) {$=$};
  \node[anchor=west] at (.9,-4.0) {$=$};
  \node[anchor=west] at (.9,-2.0) {$=$};
  
  \begin{scope}[shift={(3,0)}]
  \pic[rotate=-90] (T1) at (-0.6,0) {A11};
  \pic[rotate=90] (T2) at (0,0) {A11};
  \pic[rotate=90] (V1) at (0.8,0) {A22};
      \bg{\node[rectangle, fill=fspacebg, minimum width=2cm, minimum height=0.8cm, rounded corners] at (0.2,0.4) {};
      \coordinate (A) at (-0.6,0.25);
      \coordinate (B) at (0.0,0.25);
      \coordinate (C) at (0.55,0.25);
      \coordinate (D) at (1.05,0.25);
      \draw[fill=G2c]
        (A) 
        to[out=45, in=135, looseness=0.6] (B)
        to[out=45, in=135, looseness=0.6] (C)
        to[out=45, in=135, looseness=0.6] (D)
        to[out=90, in=90, looseness=0.9] (A)
       -- cycle; }
  \end{scope}
  
  \begin{scope}[shift={(3,-4)}]
  \pic[rotate=-90] (T1) at (-0.55,0) {A11};
  \pic[rotate=-90] (T2) at (-0.1,0) {A11};
  \pic[rotate=90] (T4) at (0.35,0) {A11};
  \pic[rotate=90] (T3) at (0.8,0) {A11};
      \bg{\node[rectangle, fill=fspacebg, minimum width=1.8cm, minimum height=0.8cm, rounded corners] at (0.15,0.4) {};
      \coordinate (A) at (-0.55,0.25);
      \coordinate (B) at (-0.1,0.25);
      \coordinate (C) at (0.35,0.25);
      \coordinate (D) at (0.8,0.25);
      \draw[fill=G2c]
        (A) 
        to[out=45, in=135, looseness=0.6] (B)
        to[out=45, in=135, looseness=0.6] (C)
        to[out=45, in=135, looseness=0.6] (D)
        to[out=90, in=90, looseness=0.9] (A)
       -- cycle;
   }
  \end{scope}
  
  \node[anchor=west] at (4.7,0.0) {$-$};
  \node[anchor=west] at (4.7,-2.0) {$-$};
  \node[anchor=west] at (8.3,-2.0) {$+$};
  
  \begin{scope}[shift={(6.6,0)}]
  \pic[rotate=-90] (T1) at (-0.6,0) {A11};
  \pic[rotate=90] (T2) at (0,0) {A11};
  \pic[rotate=90] (V1) at (0.8,0) {A22};
      \bg{\node[rectangle, fill=fspacebg, minimum width=2cm, minimum height=0.8cm, rounded corners] at (0.2,0.4) {};
      \coordinate (A) at (-0.6,0.25);
      \coordinate (B) at (0.0,0.25);
      \coordinate (C) at (0.55,0.25);
      \coordinate (D) at (1.05,0.25);
      \draw[gprop] (A) to[out=90, in=90] (C);
      \draw[gprop] (B) to[out=90, in=90] (D);
    }
  \end{scope}

  \begin{scope}[shift={(3.2,-2)}]
  \pic[rotate=-45] (V1) at (-0.5,0.2) {A31};
  \pic[rotate=-135] (V2) at (0.5,-0.2) {A31};
      \bg{\node[rectangle, fill=fspacebg, minimum width=2cm, minimum height=0.9cm, rounded corners] at (0.0,0.45) {};
      \coordinate (A) at (-0.85,0.2);
      \coordinate (B) at (-0.5,0.55);
      \coordinate (C) at (-0.15,0.2);
      \coordinate (D) at (0.5,0.15);
      \draw[fill=G2c]
        (A) 
        to[out=80, in=-180, looseness=0.6] (B)
        to[out=0, in=100, looseness=0.6] (C)
        to[out=70, in=160, looseness=0.6] (D)
        to[out=90, in=90, looseness=1.7] (A)
       -- cycle;
       \node[anchor=center, G2c!10!white] at (0.05,0.48) {\scriptsize$G^{(2)}_c$};
   }
  \end{scope}
  
  \begin{scope}[shift={(6.8,-2)}]
  \pic[rotate=-45] (V1) at (-0.5,0.2) {A31};
  \pic[rotate=-135] (V2) at (0.5,-0.2) {A31};
      \bg{\node[rectangle, fill=fspacebg, minimum width=2cm, minimum height=0.9cm, rounded corners] at (0.0,0.45) {};
      \coordinate (A) at (-0.85,0.2);
      \coordinate (B) at (-0.5,0.55);
      \coordinate (C) at (-0.15,0.2);
      \coordinate (D) at (0.5,0.15);
      \coordinate (E) at (-0.5, 0.82);
      \draw[gprop] (A) to[out=90, in=180, looseness=2] (E);
      \draw[gprop] (E) to[out=0, in=90, looseness=2] (C);
      \draw[gprop] (B) to[out=90, in=90] (D);
   }
  \end{scope}
  \begin{scope}[shift={(10.3,-2)}]
  \pic[rotate=-45] (V1) at (-0.5,0.2) {A31};
  \pic[rotate=-135] (V2) at (0.5,-0.2) {A31};
      \bg{\node[rectangle, fill=fspacebg, minimum width=2cm, minimum height=0.9cm, rounded corners] at (0.0,0.45) {};
      \coordinate (A) at (-0.85,0.2);
      \coordinate (B) at (-0.5,0.55);
      \coordinate (C) at (-0.15,0.2);
      \coordinate (D) at (0.5,0.15);
      \draw[gprop] (A) to[out=90, in=90] (B);
      \draw[gprop] (C) to[out=90, in=90] (D);
   }
  \end{scope}

  \begin{scope}[shift={(19,0)}]
  \bg{\node[rectangle, draw=notconn, ultra thick, minimum width=4cm, minimum height=5cm, rounded corners] (Rl) at (-3.95,-1.7) {};}
  \node[anchor=center, notconn] (i1) at (Rl.center) {\begin{minipage}{0.3\textwidth}\scshape Disconnected \\ not contained in $g^{(2)}$\end{minipage}};
  \begin{scope}[shift={(-4.1,0)}]
  \pic[rotate=-90] (T1) at (-0.6,0) {A11};
  \pic[rotate=90] (T2) at (0,0) {A11};
  \pic[rotate=90] (V1) at (0.8,0) {A22};
      \bg{\node[rectangle, fill=fspacebg, minimum width=2cm, minimum height=0.8cm, rounded corners] at (0.2,0.5) {};
      \coordinate (A) at (-0.6,0.25);
      \coordinate (B) at (0.0,0.25);
      \coordinate (C) at (0.55,0.25);
      \coordinate (D) at (1.05,0.25);
      \draw[gprop] (A) to[out=90, in=90] node[midway, above, gcolor!20!black, yshift=-2pt]{$g^{(1)}$} (B);
      \draw[gprop] (C) to[out=90, in=90] (D);
    }
  \end{scope}

  \begin{scope}[shift={(-5.1,-4)}]
  \pic[rotate=-90] (T1) at (-0.55,0) {A11};
  \pic[rotate=-90] (T2) at (-0.1,0) {A11};
  \pic[rotate=90] (T4) at (0.35,0) {A11};
  \pic[rotate=90] (T3) at (0.8,0) {A11};
      \bg{\node[rectangle, fill=fspacebg, minimum width=1.8cm, minimum height=0.8cm, rounded corners] at (0.15,0.4) {};
      \coordinate (A) at (-0.55,0.25);
      \coordinate (B) at (-0.1,0.25);
      \coordinate (C) at (0.35,0.25);
      \coordinate (D) at (0.8,0.25);
      \draw[gprop] (A) to[out=90, in=90] (C);
      \draw[gprop] (B) to[out=90, in=90] (D);
   }
  \end{scope}

  \begin{scope}[shift={(-3.1,-4)}]
  \pic[rotate=-90] (T1) at (-0.55,0) {A11};
  \pic[rotate=-90] (T2) at (-0.1,0) {A11};
  \pic[rotate=90] (T4) at (0.35,0) {A11};
  \pic[rotate=90] (T3) at (0.8,0) {A11};
      \bg{\node[rectangle, fill=fspacebg, minimum width=1.8cm, minimum height=0.8cm, rounded corners] at (0.15,0.4) {};
      \coordinate (A) at (-0.55,0.25);
      \coordinate (B) at (-0.1,0.25);
      \coordinate (C) at (0.35,0.25);
      \coordinate (D) at (0.8,0.25);
      \draw[gprop] (A) to[out=90, in=90] (D);
      \draw[gprop] (B) to[out=90, in=90] (C);
   }
  \end{scope}
  \end{scope}
   
\end{tikzpicture}
}

\newcommand{\constructSign}{
\begin{tikzpicture}
  \pgfdeclarelayer{bg}
  \pgfsetlayers{bg,main}
    
  \begin{scope}[shift={(0,0)}, very thick]
      \node[anchor=center] (it1) at (-0.8,-1) {$\tilde i_1$};
      \node[anchor=center] (it2) at (-0.2,-1) {$\tilde i_2$};
      \node[anchor=center] (i2) at (0.4,-1) {$i_2$};
      \node[anchor=center] (i1) at (1,-1) {$i_1$};
      \bg{\node[rectangle, rounded corners, fill=fspacebg, minimum width=2.5cm, minimum height=0.6cm] at (0.15,1.3) {};}
      \pic[rotate=90] (V1) at (1,1) {A22};
      \pic[rotate=-90] (V2) at (0,1) {A11};
      \pic[rotate=90] (V3) at (-0.8,1) {A11};
      \node[anchor=north,yshift=-2.5pt] (pi1) at (V3-1) {$i_1$};
      \node[anchor=north,yshift=-2.5pt] (pti1) at (V2-2) {$\tilde{i}_1$};
      \node[anchor=north,yshift=-2.5pt] (pti2) at (V1-2) {$\tilde{i}_2$};
      \node[anchor=north,yshift=-2.5pt] (pi2) at (V1-1) {$i_2$};
      \node[anchor=south] at (V1-4) {1};
      \node[anchor=south] at (V1-3) {2};
      \node[anchor=south] at (V2-1) {3};
      \node[anchor=south] at (V3-2) {4};
      \node[anchor=west] at (-1, 2) {$\times \langle \bar c_1 c_2 c_3 \bar c_4 \rangle$};

      \draw[] (pi1.south) -- (i1.north);
      \draw[] (pi2.south) -- (i2.north);
      \draw[] (pti1.south) -- (it1.north);
      \draw[] (pti2.south) -- (it2.north);
      
      \node[anchor=west] at (-2,1) {(a1)};
      
      \node[anchor=west] at (-1, 4.5) {$\# \text{perm}$ = 4};
      \node[anchor=west] at (-1, 4) {$\# \text{vertices}$ = 3};
      \node[anchor=west] at (-1, 3.5) {$\# \tilde{A}^{1:1}$ = 1};
      \node[anchor=west] at (-1, 3) {$\mathcal{P}$ = 3};
      \node[anchor=west] at (-1, 2.5) {$\# A^{-:\text{odd}}$ = 2};
  \end{scope}
  
  \begin{scope}[shift={(0,-3.5)}, very thick]
  \bg{\node[rectangle, rounded corners, fill=fspacebg, minimum width=2.5cm, minimum height=0.6cm] at (0.15,1.3) {};}
      \node[anchor=west] at (-2,1) {(a2)};
      \pic[rotate=90] (V1) at (1,1) {A22};
      \pic[rotate=-90] (V2) at (0,1) {A11};
      \pic[rotate=90] (V3) at (-0.8,1) {A11};
      \node[anchor=north,yshift=-2.5pt] (pi1) at (V3-1) {$i_2$};
      \node[anchor=north,yshift=-2.5pt] (pti1) at (V2-2) {$\tilde{i}_1$};
      \node[anchor=north,yshift=-2.5pt] (pti2) at (V1-2) {$\tilde{i}_2$};
      \node[anchor=north,yshift=-2.5pt] (pi2) at (V1-1) {$i_1$};
  \end{scope}
  \begin{scope}[shift={(0,-5)}, very thick]
  \bg{\node[rectangle, rounded corners, fill=fspacebg, minimum width=2.5cm, minimum height=0.6cm] at (0.15,1.3) {};}
      \node[anchor=west] at (-2,1) {(a3)};
      \pic[rotate=90] (V1) at (1,1) {A22};
      \pic[rotate=-90] (V2) at (0,1) {A11};
      \pic[rotate=90] (V3) at (-0.8,1) {A11};
      \node[anchor=north,yshift=-2.5pt] (pi1) at (V3-1) {$i_2$};
      \node[anchor=north,yshift=-2.5pt] (pti1) at (V2-2) {$\tilde{i}_2$};
      \node[anchor=north,yshift=-2.5pt] (pti2) at (V1-2) {$\tilde{i}_1$};
      \node[anchor=north,yshift=-2.5pt] (pi2) at (V1-1) {$i_1$};
  \end{scope}
  \begin{scope}[shift={(0,-6.5)}, very thick]
  \bg{\node[rectangle, rounded corners, fill=fspacebg, minimum width=2.5cm, minimum height=0.6cm] at (0.15,1.3) {};}
      \node[anchor=west] at (-2,1) {(a4)};
      \pic[rotate=90] (V1) at (1,1) {A22};
      \pic[rotate=-90] (V2) at (0,1) {A11};
      \pic[rotate=90] (V3) at (-0.8,1) {A11};
      \node[anchor=north,yshift=-2.5pt] (pi1) at (V3-1) {$i_1$};
      \node[anchor=north,yshift=-2.5pt] (pti1) at (V2-2) {$\tilde{i}_2$};
      \node[anchor=north,yshift=-2.5pt] (pti2) at (V1-2) {$\tilde{i}_1$};
      \node[anchor=north,yshift=-2.5pt] (pi2) at (V1-1) {$i_2$};
  \end{scope}

  \begin{scope}[shift={(4,0)}, very thick]
      \node[anchor=center] (it1) at (-0.8,-1) {$\tilde i_1$};
      \node[anchor=center] (it2) at (-0.2,-1) {$\tilde i_2$};
      \node[anchor=center] (i2) at (0.4,-1) {$i_2$};
      \node[anchor=center] (i1) at (1,-1) {$i_1$};
      \node[anchor=west] at (-2,1) {(b1)};
      \bg{\node[rectangle, rounded corners, fill=fspacebg, minimum width=2.5cm, minimum height=0.6cm] at (0.15,1.3) {};}
      \pic (V1) at (1,1) {B22};
      \pic[rotate=-90] (V2) at (0,1) {A11};
      \pic[rotate=-90] (V3) at (-0.8,1) {A11};
      \node[anchor=north,yshift=-2.5pt] (pti1) at (V3-2) {$\tilde{i}_1$};
      \node[anchor=north,yshift=-2.5pt] (pti2) at (V2-2) {$\tilde{i}_2$};
      \node[anchor=north,yshift=-2.5pt] (pi2) at (V1-1) {$i_2$};
      \node[anchor=north,yshift=-2.5pt] (pi1) at (V1-3) {$i_1$};
      \node[anchor=south] at (V1-2) {1};
      \node[anchor=south] at (V1-4) {2};
      \node[anchor=south] at (V2-1) {3};
      \node[anchor=south] at (V3-1) {4};
      \node[anchor=west] at (-1, 2) {$\times \langle \bar c_1 \bar c_2 c_3 c_4 \rangle$};

      \draw[] (pi1.south) -- (i1.north);
      \draw[] (pi2.south) -- (i2.north);
      \draw[] (pti1.south) -- (it1.north);
      \draw[] (pti2.south) -- (it2.north);
      
      \node[anchor=west] at (-1, 4.5) {$\# \text{perm}$ = 2};
      \node[anchor=west] at (-1, 4) {$\# \text{vertices}$ = 3};
      \node[anchor=west] at (-1, 3.5) {$\# \tilde{A}^{1:1}$ = 0};
      \node[anchor=west] at (-1, 3) {$\mathcal{P}$ = 0};
      \node[anchor=west] at (-1, 2.5) {$\# A^{-:\text{odd}}$ = 2};
  \end{scope}
  \begin{scope}[shift={(4,-3.5)}, very thick]
  \bg{\node[rectangle, rounded corners, fill=fspacebg, minimum width=2.5cm, minimum height=0.6cm] at (0.15,1.3) {};}
      \node[anchor=west] at (-2,1) {(b2)};
      \pic (V1) at (1,1) {B22};
      \pic[rotate=-90] (V2) at (0,1) {A11};
      \pic[rotate=-90] (V3) at (-0.8,1) {A11};
      \node[anchor=north,yshift=-2.5pt] (pti1) at (V3-2) {$\tilde{i}_2$};
      \node[anchor=north,yshift=-2.5pt] (pti2) at (V2-2) {$\tilde{i}_1$};
      \node[anchor=north,yshift=-2.5pt] (pi2) at (V1-1) {$i_2$};
      \node[anchor=north,yshift=-2.5pt] (pi1) at (V1-3) {$i_1$};
      \draw[red] (-1,1.5) -- (1.25,0.5);
      \draw[red] (-1,0.5) -- (1.25,1.5);
      \node[fill=red!10!white, rounded corners, anchor=south] at (0.125,1.1) {= (b1)};
  \end{scope}
  \begin{scope}[shift={(4,-5)}, very thick]
  \bg{\node[rectangle, rounded corners, fill=fspacebg, minimum width=2.5cm, minimum height=0.6cm] at (0.15,1.3) {};}
      \node[anchor=west] at (-2,1) {(b3)};
      \pic (V1) at (1,1) {B22};
      \pic[rotate=-90] (V2) at (0,1) {A11};
      \pic[rotate=-90] (V3) at (-0.8,1) {A11};
      \node[anchor=north,yshift=-2.5pt] (pti1) at (V3-2) {$\tilde{i}_1$};
      \node[anchor=north,yshift=-2.5pt] (pti2) at (V2-2) {$\tilde{i}_2$};
      \node[anchor=north,yshift=-2.5pt] (pi2) at (V1-1) {$i_1$};
      \node[anchor=north,yshift=-2.5pt] (pi1) at (V1-3) {$i_2$};
  \end{scope}
  \begin{scope}[shift={(4,-6.5)}, very thick]
  \bg{\node[rectangle, rounded corners, fill=fspacebg, minimum width=2.5cm, minimum height=0.6cm] at (0.15,1.3) {};}
      \node[anchor=west] at (-2,1) {(b4)};
      \pic (V1) at (1,1) {B22};
      \pic[rotate=-90] (V2) at (0,1) {A11};
      \pic[rotate=-90] (V3) at (-0.8,1) {A11};
      \node[anchor=north,yshift=-2.5pt] (pti1) at (V3-2) {$\tilde{i}_2$};
      \node[anchor=north,yshift=-2.5pt] (pti2) at (V2-2) {$\tilde{i}_1$};
      \node[anchor=north,yshift=-2.5pt] (pi2) at (V1-1) {$i_1$};
      \node[anchor=north,yshift=-2.5pt] (pi1) at (V1-3) {$i_2$};
      \draw[red] (-1,1.5) -- (1.25,0.5);
      \draw[red] (-1,0.5) -- (1.25,1.5);
      \node[fill=red!10!white, rounded corners, anchor=south] at (0.125,1.1) {= (b3)};
  \end{scope}
  
  \begin{scope}[shift={(8,0)}, very thick]
      \node[anchor=west] at (-2,1) {(c1)};
      \node[anchor=center] (it1) at (-0.8,-1) {$\tilde i_1$};
      \node[anchor=center] (it2) at (-0.2,-1) {$\tilde i_2$};
      \node[anchor=center] (i2) at (0.4,-1) {$i_2$};
      \node[anchor=center] (i1) at (1,-1) {$i_1$};
      \bg{\node[rectangle, rounded corners, fill=fspacebg, minimum width=2.5cm, minimum height=0.6cm] at (0.15,1.3) {};}
      \pic[rotate=-90] (V1) at (-0.8,1) {A11};
      \pic[rotate=90] (V2) at (-0.2,1) {A11};
      \pic[rotate=-90] (V3) at (0.4,1) {A11};
      \pic[rotate=90] (V4) at (1.0,1) {A11};
      
      \node[anchor=north,yshift=-2.5pt] (pti1) at (V1-2) {$\tilde{i}_1$};
      \node[anchor=north,yshift=-2.5pt] (pti2) at (V3-2) {$\tilde{i}_2$};
      \node[anchor=north,yshift=-2.5pt] (pi2) at (V4-1) {$i_2$};
      \node[anchor=north,yshift=-2.5pt] (pi1) at (V2-1) {$i_1$};
      \node[anchor=south] at (V4-2) {1};
      \node[anchor=south] at (V3-1) {2};
      \node[anchor=south] at (V2-2) {3};
      \node[anchor=south] at (V1-1) {4};
      \node[anchor=west] at (-1, 2) {$\times \langle \bar c_1 c_2 \bar c_3 c_4 \rangle$};

      \draw[] (pi1.south) -- (i1.north);
      \draw[] (pi2.south) -- (i2.north);
      \draw[] (pti1.south) -- (it1.north);
      \draw[] (pti2.south) -- (it2.north);
      
      \node[anchor=west] at (-1, 4.5) {$\# \text{perm}$ = 1};
      \node[anchor=west] at (-1, 4) {$\# \text{vertices}$ = 4};
      \node[anchor=west] at (-1, 3.5) {$\# \tilde{A}^{1:1}$ = 2};
      \node[anchor=west] at (-1, 3) {$\mathcal{P}$ = 2};
      \node[anchor=west] at (-1, 2.5) {$\# A^{-:\text{odd}}$ = 4};
  \end{scope}
  \begin{scope}[shift={(8,-3.5)}, very thick]
  \bg{\node[rectangle, rounded corners, fill=fspacebg, minimum width=2.5cm, minimum height=0.6cm] at (0.15,1.3) {};}
      \node[anchor=west] at (-2,1) {(c2)};
      \pic[rotate=-90] (V1) at (-0.8,1) {A11};
      \pic[rotate=90] (V2) at (-0.2,1) {A11};
      \pic[rotate=-90] (V3) at (0.4,1) {A11};
      \pic[rotate=90] (V4) at (1.0,1) {A11};
      
      \node[anchor=north,yshift=-2.5pt] (pti1) at (V1-2) {$\tilde{i}_2$};
      \node[anchor=north,yshift=-2.5pt] (pti2) at (V3-2) {$\tilde{i}_1$};
      \node[anchor=north,yshift=-2.5pt] (pi2) at (V4-1) {$i_2$};
      \node[anchor=north,yshift=-2.5pt] (pi1) at (V2-1) {$i_1$};
      \draw[red] (-1,1.5) -- (1.25,0.5);
      \draw[red] (-1,0.5) -- (1.25,1.5);
      \node[fill=red!10!white, rounded corners, anchor=south] at (0.125,1.1) {= (c1)};
  \end{scope}
  \begin{scope}[shift={(8,-5)}, very thick]
  \bg{\node[rectangle, rounded corners, fill=fspacebg, minimum width=2.5cm, minimum height=0.6cm] at (0.15,1.3) {};}
      \node[anchor=west] at (-2,1) {(c3)};
      \pic[rotate=-90] (V1) at (-0.8,1) {A11};
      \pic[rotate=90] (V2) at (-0.2,1) {A11};
      \pic[rotate=-90] (V3) at (0.4,1) {A11};
      \pic[rotate=90] (V4) at (1.0,1) {A11};
      
      \node[anchor=north,yshift=-2.5pt] (pti1) at (V1-2) {$\tilde{i}_1$};
      \node[anchor=north,yshift=-2.5pt] (pti2) at (V3-2) {$\tilde{i}_2$};
      \node[anchor=north,yshift=-2.5pt] (pi2) at (V4-1) {$i_1$};
      \node[anchor=north,yshift=-2.5pt] (pi1) at (V2-1) {$i_2$};
      \draw[red] (-1,1.5) -- (1.25,0.5);
      \draw[red] (-1,0.5) -- (1.25,1.5);
      \node[fill=red!10!white, rounded corners, anchor=south] at (0.125,1.1) {= (c1)};
  \end{scope}
  \begin{scope}[shift={(8,-6.5)}, very thick]
  \bg{\node[rectangle, rounded corners, fill=fspacebg, minimum width=2.5cm, minimum height=0.6cm] at (0.15,1.3) {};}
      \node[anchor=west] at (-2,1) {(c4)};
      \pic[rotate=-90] (V1) at (-0.8,1) {A11};
      \pic[rotate=90] (V2) at (-0.2,1) {A11};
      \pic[rotate=-90] (V3) at (0.4,1) {A11};
      \pic[rotate=90] (V4) at (1.0,1) {A11};
      
      \node[anchor=north,yshift=-2.5pt] (pti1) at (V1-2) {$\tilde{i}_2$};
      \node[anchor=north,yshift=-2.5pt] (pti2) at (V3-2) {$\tilde{i}_1$};
      \node[anchor=north,yshift=-2.5pt] (pi2) at (V4-1) {$i_1$};
      \node[anchor=north,yshift=-2.5pt] (pi1) at (V2-1) {$i_2$};
      \draw[red] (-1,1.5) -- (1.25,0.5);
      \draw[red] (-1,0.5) -- (1.25,1.5);
      \node[fill=red!10!white, rounded corners, anchor=south] at (0.125,1.1) {= (c1)};
  \end{scope}

  \begin{scope}[shift={(12,0)}, very thick]
      \node[anchor=center] (it1) at (-0.8,-1) {$\tilde i_1$};
      \node[anchor=center] (it2) at (-0.2,-1) {$\tilde i_2$};
      \node[anchor=center] (i2) at (0.4,-1) {$i_2$};
      \node[anchor=center] (i1) at (1,-1) {$i_1$};
      \bg{\node[rectangle, rounded corners, fill=fspacebg, minimum width=2.5cm, minimum height=0.6cm] at (0.15,1.3) {};}
      \pic[rotate=90] (V1) at (-0.5,1) {A22};
      \pic[rotate=90] (V2) at (0.8,1) {A22};
      
      \node[anchor=north,yshift=-2.5pt] (pti1) at (V1-2) {$\tilde{i}_1$};
      \node[anchor=north,yshift=-2.5pt] (pti2) at (V2-2) {$\tilde{i}_2$};
      \node[anchor=north,yshift=-2.5pt] (pi2) at (V1-1) {$i_2$};
      \node[anchor=north,yshift=-2.5pt] (pi1) at (V2-1) {$i_1$};
      \node[anchor=south] at (V2-4) {1};
      \node[anchor=south] at (V2-3) {2};
      \node[anchor=south] at (V1-4) {3};
      \node[anchor=south] at (V1-3) {4};
      \node[anchor=west] at (-1, 2) {$\times \langle \bar c_1 c_2 \bar c_3 c_4 \rangle$};

      \draw[] (pi1.south) -- (i1.north);
      \draw[] (pi2.south) -- (i2.north);
      \draw[] (pti1.south) -- (it1.north);
      \draw[] (pti2.south) -- (it2.north);
      \node[anchor=west] at (-2,1) {(d1)};
      
      \node[anchor=west] at (-1, 4.5) {$\# \text{perm}$ = 2};
      \node[anchor=west] at (-1, 4) {$\# \text{vertices}$ = 2};
      \node[anchor=west] at (-1, 3.5) {$\# \tilde{A}^{1:1}$ = 0};
      \node[anchor=west] at (-1, 3) {$\mathcal{P}$ = 1};
      \node[anchor=west] at (-1, 2.5) {$\# A^{-:\text{odd}}$ = 0};
  \end{scope}

\begin{scope}[shift={(12,-3.5)}, very thick]
  \bg{\node[rectangle, rounded corners, fill=fspacebg, minimum width=2.5cm, minimum height=0.6cm] at (0.15,1.3) {};}
      \node[anchor=west] at (-2,1) {(d2)};
      \pic[rotate=90] (V1) at (-0.5,1) {A22};
      \pic[rotate=90] (V2) at (0.8,1) {A22};
      
      \node[anchor=north,yshift=-2.5pt] (pti1) at (V1-2) {$\tilde{i}_2$};
      \node[anchor=north,yshift=-2.5pt] (pti2) at (V2-2) {$\tilde{i}_1$};
      \node[anchor=north,yshift=-2.5pt] (pi2) at (V1-1) {$i_2$};
      \node[anchor=north,yshift=-2.5pt] (pi1) at (V2-1) {$i_1$};
  \end{scope}
  \begin{scope}[shift={(12,-5)}, very thick]
  \bg{\node[rectangle, rounded corners, fill=fspacebg, minimum width=2.5cm, minimum height=0.6cm] at (0.15,1.3) {};}
      \node[anchor=west] at (-2,1) {(d3)};
      \pic[rotate=90] (V1) at (-0.5,1) {A22};
      \pic[rotate=90] (V2) at (0.8,1) {A22};
      
      \node[anchor=north,yshift=-2.5pt] (pti1) at (V1-2) {$\tilde{i}_1$};
      \node[anchor=north,yshift=-2.5pt] (pti2) at (V2-2) {$\tilde{i}_2$};
      \node[anchor=north,yshift=-2.5pt] (pi2) at (V1-1) {$i_1$};
      \node[anchor=north,yshift=-2.5pt] (pi1) at (V2-1) {$i_2$};
      \draw[red] (-1,1.5) -- (1.25,0.5);
      \draw[red] (-1,0.5) -- (1.25,1.5);
      \node[fill=red!10!white, rounded corners, anchor=south] at (0.125,1.1) {= (d2)};
  \end{scope}
  \begin{scope}[shift={(12,-6.5)}, very thick]
  \bg{\node[rectangle, rounded corners, fill=fspacebg, minimum width=2.5cm, minimum height=0.6cm] at (0.15,1.3) {};}
      \node[anchor=west] at (-2,1) {(d4)};
      \pic[rotate=90] (V1) at (-0.5,1) {A22};
      \pic[rotate=90] (V2) at (0.8,1) {A22};
      
      \node[anchor=north,yshift=-2.5pt] (pti1) at (V1-2) {$\tilde{i}_2$};
      \node[anchor=north,yshift=-2.5pt] (pti2) at (V2-2) {$\tilde{i}_1$};
      \node[anchor=north,yshift=-2.5pt] (pi2) at (V1-1) {$i_1$};
      \node[anchor=north,yshift=-2.5pt] (pi1) at (V2-1) {$i_2$};
      \draw[red] (-1,1.5) -- (1.25,0.5);
      \draw[red] (-1,0.5) -- (1.25,1.5);
      \node[fill=red!10!white, rounded corners, anchor=south] at (0.125,1.1) {= (d1)};
  \end{scope}
  
\end{tikzpicture}
}

\newcommand{\makeauthor}[2]{\newcommand{#1}[1]{{%
  \protect%
  \color{#2}{%
    \bfseries\begingroup%
        \escapechar=-1\edef\x{\endgroup\string#1}\x:%
  }\itshape{} ##1}}%
  \MakeRobustCommand#1}
\newcommand{\bvec}[1]{\bm{#1}}
\newcommand{\vdagger}{{\vphantom{\dagger}}}
\newcommand{\dd}{\mathrm{d}}
\newcommand{\DD}{\mathcal{D}}
\newcommand{\Tr}{\mathrm{Tr}}
\newcommand{\matrixOne}{\mathds{1}}
\newcommand{\angstrom}{\text{\AA}}

\makeauthor{\lk}{purple}
\makeauthor{\jp}{ForestGreen}
\makeauthor{\rv}{brown}
\makeauthor{\jv}{red}
\makeauthor{\mr}{blue}

\begin{document}

\title{Exact downfolding and its perturbative approximation}

\author{Jonas B.~Profe}
\email{Profe@itp.uni-frankfurt.de}
\affiliation{Institute for Theoretical Physics, Goethe University Frankfurt,
Max-von-Laue-Straße 1, 60438 Frankfurt a.M., Germany}
\author{Jak\v{s}a Vu\v{c}i\v{c}evi\'{c}}
\affiliation{Scientific Computing Laboratory, Center for the Study of Complex Systems,
Institute of Physics Belgrade, University of Belgrade, Pregrevica 118, 11080 Belgrade, Serbia}
\author{P.~Peter Stavropoulos}
\affiliation{Institute for Theoretical Physics, Goethe University Frankfurt,
Max-von-Laue-Straße 1, 60438 Frankfurt a.M., Germany}
\author{Malte Rösner}
\affiliation{Institute for Molecules and Materials, Radboud University, Nijmegen, The Netherlands}
\affiliation{Faculty of Physics, Bielefeld University, 33501 Bielefeld, Germany}
\author{Roser Valentí}
\affiliation{Institute for Theoretical Physics, Goethe University Frankfurt,
Max-von-Laue-Straße 1, 60438 Frankfurt a.M., Germany}
\author{Lennart Klebl}
\affiliation{Institut für Theoretische Physik und Astrophysik and Würzburg-Dresden Cluster
of Excellence ct.qmat, Universität Würzburg, 97074 Würzburg, Germany}

\date{\today}
\begin{abstract}
Solving the many-electron problem, even approximately, is one of the most challenging and simultaneously most important problems in contemporary condensed matter physics with various connections to other fields. The standard approach is to follow a \emph{divide and conquer} strategy that combines various numerical and analytical techniques. A crucial step in this strategy is the derivation of an effective model for a subset of degrees of freedom by a procedure called \emph{downfolding}, which often corresponds to integrating out energy scales far away from the Fermi level. In this work we present a rigorous formulation of this downfolding procedure, which complements the renormalization group picture put forward by Honerkamp [\href{https://doi.org/10.1103/PhysRevB.85.195129}{PRB~{\bf{}85},~195129~(2012)}]. We derive an exact effective model in an arbitrarily chosen target space (e.g.~low-energy degrees of freedom) by explicitly integrating out the the rest space (e.g.~high-energy degrees of freedom). Within this formalism we state conditions that justify a perturbative truncation of the downfolded effective interactions to just a few low-order terms. Furthermore, we utilize the exact formalism to formally derive the widely used constrained random phase approximation (cRPA), uncovering underlying approximations and highlighting relevant corrections in the process. Lastly, we detail different contributions in the material examples of fcc Nickel and the infinite-layer cuprate SrCuO\textsubscript2. 
Our results open up a new pathway to obtain effective models in a controlled fashion and to judge whether a chosen target space is suitable. 
\end{abstract}

\maketitle

\section{Introduction}
Central to the design of new functional materials is a deep microscopic understanding of their various physical properties. In general, we might differentiate between  weakly correlated materials describable on a mean-field level, such as density functional theory~\cite{DFT1, DFT2}, and strongly correlated materials, for which effective mean-field descriptions fail. 
Fortunately, to model and eventually understand strongly correlated materials we can make use of the hierarchical nature of physics: Relevant degrees of freedom can be distilled from complex microscopic equations, amounting to the formulation of effective, interacting (lattice) theories. In recent decades, a plethora of analytical~\cite{Bethe1931,AMI_env,kopietz2010} and numerical~\cite{Hedin, gros1989,valenti1992,metzner1989,RevModPhys.73.33,  becca2017quantum,Schollw_ck_2011,PhysRevB.101.075113, PhysRevResearch.3.023082,vilk1997,profe2024multi,full_cell_dmft2, fullcell_DMFT, PhysRevLett.111.036601} techniques that aid the formulation and solution of corresponding effective models have been developed. In practice, the typical workflow for the study of strongly correlated materials consists of a multi-step procedure starting from an \emph{ab initio} characterization~\cite{DFT1,DFT2,Hedin,aryasetiawan1998,GW2,Lee2023,Mushkaev_2024}. Based on this calculation a model describing the relevant degrees of freedom is derived by a downfolding scheme. This model can then be solved with many-body techniques~\cite{ gros1989,valenti1992,metzner1989,RevModPhys.73.33,  becca2017quantum,Schollw_ck_2011,PhysRevB.101.075113, PhysRevResearch.3.023082,vilk1997,profe2024multi,full_cell_dmft2, fullcell_DMFT}.
Some examples for such hierarchical calculations can be found in the fields of (unconventional) superconductors~\cite{Misawa_2012,PhysRevB.87.024505,glasbrenner2015,Deng2019,vanLoon2018, Chen2022, Hirayama_2018,Kitatani_2020,Tazai2022,ferreira2023search, PhysRevB.108.L201121, Witt_2024, Cui2025}, quantum spin-liquid candidates and frustrated magnets
~\cite{yamaji2014,winter2017models,riedl2019,eichstaedt2019,hering2022,razpopov2023}
and Mott insulators~\cite{Sutter2017, PhysRevB.91.125122, PhysRevLett.118.086401, PhysRevLett.117.056402, grytsiuk2023nb3cl8prototypicallayeredmotthubbard}.

Each of these three building blocks, namely the \emph{ab initio} starting point, the downfolding procedure and finally the (approximate) solutions to the downfolded models, comes with its own challenges. However, there has been tremendous advancements in recent years, especially on the model-solving side. In particular, it has become possible to achieve ``handshakes''{} for ground state properties derived from different numerical algorithms for (strongly correlated) single-band models~\cite{PhysRevX.10.031016, Thomas_2021_footprints, Qin2022,imkovic2024}. This showcases that, out of the three building blocks, solving the effective model might not pose the most significant source of error. Simultaneously, the necessary initial \emph{ab initio} calculations can nowadays be performed on several, well defined, and well controllable levels. Thus, in the hunt for fully predictive calculations for real (correlated) materials we have to turn our attention to the downfolding technique with which we formulate effective target-space models.

There are two main downfolding strategies one can employ to derive a target-space model: One can either directly start from the Schrödinger equation and require that the spectrum of the target-space model is (approximately) identical to a part of the spectrum of the full model~\cite{Lwdin1951, Lwdin1964}, or alternatively, one starts from the partition function or the density matrix and performs a partial trace over the rest subsystem~\cite{2012_cfrg}.  The first strategy gave rise to a plethora of downfolding techniques, such as active space downfolding~\cite{Tenno2013, Eskridge_2019, doi:10.1021/ct4006486}, coupled cluster downfolding~\cite{Bauman2019, bauman2023coupledclusterdownfoldingtechniques}, canonical transformations~\cite{Chao1977, White2002,Yanai2006,PhysRevA.111.042825}, operator perturbation theory~\cite{Kato1949, Kato1995,Canestraight2025}, density-matrix renormalization group~\cite{PhysRevB.108.L161111}, density-matrix downfolding~\cite{Changlani_2015, Zheng_2018, PhysRevB.110.195103} unitary circuit-based downfolding~\cite{jiang2025groundstatebasedmodelreductionunitary}, techniques mapping to spin models~\cite{Sharma2023, RevModPhys.95.035004} and constrained local density approximation (cLDA)~\cite{cLDA1,cLDA2,cLDA3,carta2025bridgingconstrainedrandomphaseapproximation}. The second strategy leads to the (stochastic) constrained random phase approximation (cRPA)~\cite{crpa_2008,hcRPA_1,hcRPA_2,PhysRevB.86.165105,PhysRevLett.132.076401, carta2025bridgingconstrainedrandomphaseapproximation,10.21468/SciPostPhys.16.2.046,shinaoka2015accuracy, chang2024downfolding, reddy2025unveilingcrpacomparativeanalysis,Romanova2022, Romanova2023} and constrained functional renormalization group (cFRG)~\cite{2012_cfrg,2015_cfrg,2018_cfrg}. cRPA was initially introduced as an \emph{ad-hoc} approximation~\cite{aryasetiawan1998} to obtain screened two-particle interactions within the target space. Together with further approximations, e.g. on the so-called double counting corrections~\cite{dc_1_dmft, dc_2_dmft, dc_3_dmft, dc_4_dmft,PhysRevB.97.201116} for single particle properties, it has been shown to be a good approximation for sufficiently gapped systems~\cite{shinaoka2015accuracy, loon2021random, chang2024downfolding}. cFRG on the other hand was introduced as an exact theory for downfolding~\cite{2012_cfrg}, but no applications beyond model calculations were performed so far~\cite{2015_cfrg,2018_cfrg,han2021investigation}. 
Thus, there is an inherent need for an in-depth analysis of an exact downfolding theory and the corresponding effective action that is obtained for different real materials.

In this work, starting from the partition function we construct an exact downfolding approach based on splitting the degrees of freedom into a rest space and a target space in an arbitrary way. The rest space degrees of freedom are integrated out using the path integral formalism expressed in terms of an effective action, cf.~\cref{sec::deriv_main}. Such a procedure is reminiscent of the cavity construction in dynamical mean-field theory (DMFT)~\cite{DMFT}. From the exact target-space action we develop a diagrammatic theory for the effective target-space model and discuss controlled approximations of it, cf.~\cref{ssec::howtoapprox}. The diagrammatic representation allows us to recover cRPA by resummation of specific diagrams highlighting and understanding the approximations required by the method, cf.~\cref{ssec::crpa}. Next, we calculate simple test cases analytically that showcase the structure of the resulting contributions to the target-space action, cf.~\cref{sec:anastruct}. In addition we perform \emph{ab initio} simulations to calculate the various terms appearing within the formalism for two material examples: fcc Nickel and the infinite layer cuprate SrCuO\textsubscript2, cf.~\cref{sec:paramreal}. Our approach makes it possible to establish the relevance of each of the contributing terms allowing to directly gauge the quality of the derived effective target-space model.

\section{Exact downfolding}
\label{sec::deriv_main}
Our starting point is the following generic two-particle interacting Hamiltonian:
\begin{equation}
    \label{eq:hamiltonian}
    \def\cop{\mathsf{c}}
    H = T_{12} \mathsf d^\dagger_2 \mathsf d^\vdagger_1 + \frac12 U_{1234} \mathsf d^\dagger_3 \mathsf d^\dagger_4 \mathsf d^\vdagger_2 \mathsf d^\vdagger_1 \,.
\end{equation}
The operator $\mathsf{d}^{(\dagger)}_1$ annihilates (creates) an electron in state $\ket{\psi_1}$. The indices are (implicitly) summed over and they combine all the relevant degrees of freedom, e.g., site, orbital, spin, etc. The matrix elements $T_{12}$ and $U_{1234}$ are determined in a given basis by
\begin{equation}
    T_{12} = \braket{\psi_2|\hat H_\mathrm{sp}|\psi_1} \,, \quad
    U_{1234} = \braket{\psi_3\psi_4|\hat H_\mathrm{int}|\psi_1\psi_2} \,,
\end{equation}
with $\hat H_\mathrm{sp}$ and $\hat H_\mathrm{int}$ the single-particle contributions (encompassing kinetic energy as well as potential energy) and the Coulomb interaction, respectively. From here on, we will work in the path integral formalism. The exact partition function of the system is given by~\cite{fetter2012quantum}
\begin{equation}
    \label{eq:partition}
    \mathcal Z = \int \DD d\DD\bar d\, e^{-S[d,\bar d]} \,,
\end{equation}
where $S[d,\bar d]$ is the action corresponding to the Hamiltonian \cref{eq:hamiltonian} and $d, \bar d$ are the fermionic (Grassman) fields~\footnote{We assume the interaction to be anti-symmetrized and prefactors absorbed back into the antisymmetric tensor}. We still use numbers as multi-indices, which in the action formalism also carry the Matsubara frequency arguments. We do not explicitly write out all Kronecker-$\delta$'s that arise from (imaginary) time translational invariance.

The next step is to separate the single-particle basis \cref{eq:partition} into target space ($\cal T$) and rest space ($\cal R$) states and subsequently integrate out the rest space. Often the target space is spanned by a few selected low-energy states and the rest space is spanned by high-energy states. However, this is not a requirement for our construction. At this point we can separate the system into two subspaces in an arbitrary way. The resulting integrand can be expressed into the form of an action which we refer to as the ``downfolded model''. On this formal level, no approximations have been made.  All correlation functions within the target space are fully equivalent between the full and the downfolded action, see~\cref{App:simple} for an explicit demonstration of this in a simple toy-model. 

In practice, the separation into target and rest space 
amounts to splitting the sums over multi-indices into two parts in the following way (say, for the field $d_1$):
\begin{equation}
    \sum_1 d_1 = \sum_{1 \in {\mathcal T}} d_1 + \sum_{1 \in {\mathcal R}}d_1 \equiv \sum_{1} f_1 + \sum_{1}c_1 \,.
\end{equation}
In the last step, for the sake of brevity, we introduced a short-hand notation: the fields $f$ and $c$ correspond to the fields $d$ in the target and the rest space, respectively. We implicitly assume that the multi-index $1$ in the subscript of $f$ belongs to the set of target space degrees of freedom $\mathcal T$ and in the subscript of $c$ belongs to the set of rest space degrees of freedom $\mathcal R$. We now apply this separation to each of the fields in all terms in the (full) action. For example, the interaction term becomes
\begin{widetext}
\begin{equation}
    \frac12\sum_{1234} U_{1234} \bar{d}_3\bar{d}_4 d_2 d_1 = \frac12\sum_{1234} U_{1234} (\bar{f}_3 + \bar{c}_3)(\bar{f}_4 + \bar{c}_4) (f_2 + c_2) (f_1+c_1) \,.
\end{equation}
Multiplying the summands yields $2^4=16$ terms. Due to anticommutation of Grassman fields and the antisymmetry of the tensor $U$ (e.g. $U_{1234}=-U_{1243}$), some of the terms turn out to be equal, and we can group them together with additional combinatorial prefactors in front, see \cref{App:split}. 

The action can now be split into parts in the following way:

\begin{equation}
    \mathcal Z = \int \DD f \DD \bar f \int \DD c \DD \bar c ~ e^{-S_f[f,\bar f]} \, e^{-S_c[c, \bar c]} \, e^{-\mathcal A[f,\bar f,c,\bar c]} \,,
\end{equation}
where we defined the target-space action $S_f[f,\bar f]$, the rest-space action $S_c[c,\bar c]$, and the coupling action $\mathcal A[f,\bar f,c,\bar c]$. These different contributions read
\begin{align}
    S_f[f,\bar f] &{}= \sum_{12 \in \mathcal{T}} \bar f_2 (\delta_{12}(i\omega-\mu) + T_{12}) f_1 +  \sum_{1234\in \mathcal{T}} \underbrace{\frac12 U_{1234}}_{F_{1234}} \bar f_3 \bar f_4 f_2 f_1 \,, \\
    S_c[c,\bar c] &{}= \sum_{12 \in \mathcal{R}} \bar c_2 (\delta_{12}(i\omega-\mu) + T_{12}) c_1 + \sum_{1234\in \mathcal{R}} \underbrace{\frac12 U_{1234}}_{C_{1234}} \bar c_3 \bar c_4 c_2 c_1 \,, \\
    \mathcal A[f,\bar f,c,\bar c] &{}= \begin{multlined}[t]
    \label{eq:coupling-action}
     \bar f_2 \underbrace{T_{12}}_{A^{1:1}_{12}} c_1 + \bar c_2 \underbrace{T_{12}}_{\tilde{A}^{1:1}_{12}} f_1  + 
    \underbrace{U_{1234}}_{{A}^{1:3}_{1234}}    \bar f_3 \bar c_4 c_2 c_1 +
    \underbrace{U_{1234}}_{\tilde{A}^{1:3}_{1234}}    f_2 \bar c_3 \bar c_4 c_1 +
    \underbrace{U_{1234}}_{{A}^{3:1}_{1234}}      \bar f_3 \bar f_4 f_2 c_1 +
    \underbrace{U_{1234}}_{\tilde{A}^{3:1}_{1234}} \bar f_3 f_2 f_1 \bar c_4 
     \\ +
        \underbrace{\frac{1}{2}U_{1234}}_{{B}^{2:2}_{1234}} \bar f_3 \bar f_4 c_2 c_1+
        \underbrace{\frac{1}{2}U_{1234}}_{\tilde{B}^{2:2}_{1234}} f_2 f_1  \bar c_3 \bar c_4 +
        \underbrace{2U_{1234}}_{{A}^{2:2}_{1234}} \bar f_4 f_2 \bar c_3 c_1 
     \,,
    \end{multlined}
\end{align}
where we introduced new vertices for the different terms in the coupling action. The superscripts denote the number of $f$ and $c$ operators associated to the respective vertex, i.e., the number of target and rest space fields attached to it. The vertices with and without tilde are related by complex conjugation.
Whenever visually representing contributions throughout the paper, these will not be explicitly differentiated, but which vertex is required follows from the context.
Each of these terms can be associated with a physical process: $A^{1:1}$ is a hopping process between target and rest space, $A^{2:2}$ encompasses both density-density and spin-spin Coulomb like interactions between the spaces. $B^{2:2}$ describes pair-hopping processes between target and rest space. Lastly, both $A^{1:3}$ and $A^{3:1}$ describe assisted hopping processes between target and rest space.

\begin{figure*}
    \centering
    \includegraphics{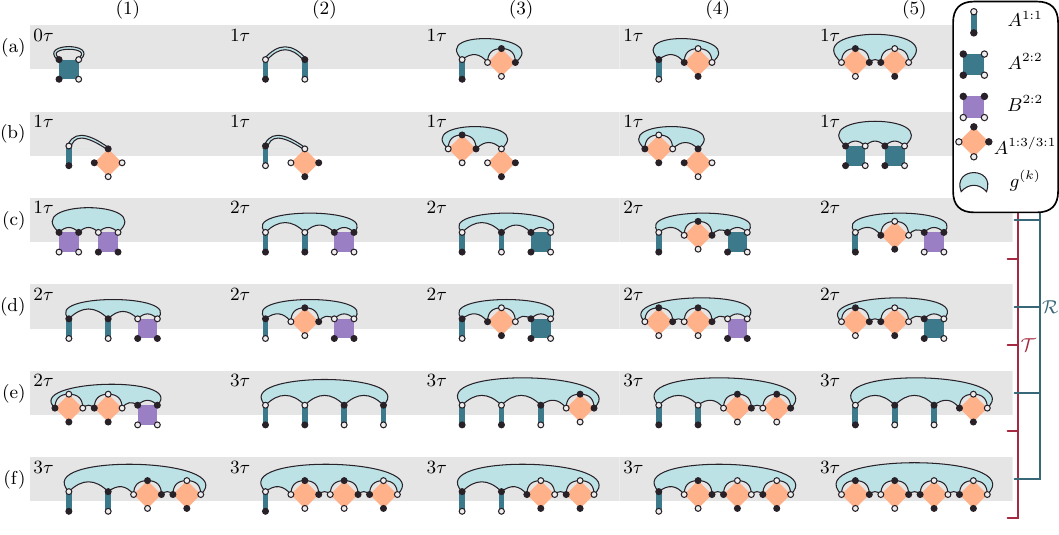}
    \caption{Diagrammatic representation of $G^{(1)}$ (a) and $G^{(2)}$ (b--f). The gray background indicates rest space ($\color{bdiv2-3}\mathcal R$) variables which are summed over, while the white background indicates target space ($\color{bdiv2-6}\mathcal T$). White circles indicate ingoing and black circles indicate outgoing legs. Light cyan filled shapes in the rest space $\mathcal R$ indicate $g^{(k)}$ (see main text). The one-particle vertex $A^{1:1}$ is represented with a dark cyan line, while the two-particle vertices $A^{2:2}$, $A^{3:1/1:3}$, $B^{2:2}$ are given as dark cyan, orange, and purple squares, respectively. We only draw topologically distinct vertices, reducing the number of contributions to 25 for the two-particle case. The number of (imaginary) times that each diagram depends on is denoted in the top left corner as $N\tau$, under the assumption of instantaneous vertices.}
    \label{fig:diagrams}
\end{figure*}

With these definitions at hand, we restructure the path integral into
\begin{equation}
    \mathcal Z = \int \DD f \DD \bar f \, e^{-S_f[f,\bar f]} \Braket{e^{-\mathcal A[f,\bar f, c, \bar c]}}_{S_c} = \int \DD f \DD \bar f \, \exp\Big[-S_f[f, \bar f] + \underbrace{\log\Big( \Braket{\exp\big[
    -\mathcal A[f,\bar f,c,\bar c]
    \big]}_{S_c}\Big)}_{\mathcal G[f, \bar f]}\Big] \,.
    \label{eq:func}
\end{equation}
The resulting action $(S_f - \mathcal{G})$ is what we will refer to as the ``effective model''{} or ``downfolded model''{}.
We recognize the expectation value over the rest space fields as a generalized generating functional for Green's functions~\cite{fetter2012quantum} of the rest space with the target space fields as sources which are \emph{not} one-to-one coupled to the rest space ones. This is, in contrast to, e.g., the usual derivation of DMFT~\cite{DMFT} where the fields are one-to-one coupled via the hopping between the impurity and the bath. We Taylor-expand the functional $\mathcal G$ around $f = 0 = \bar{f}$. Due to charge conservation in the subspace $\mathcal{T}$, terms with unequal number of $\bar{f}$ and $f$ fields will ultimately drop out. What remains has the form of $n$-particle interactions, i.e.,
\begin{equation}
    \mathcal G[f, \bar f] = \sum_{n=1}^\infty \sum_{\bar{i}_1,...,\bar{i}_n}\sum_{i_1,...,i_n}\underbrace{\frac{1}{(n!)^2} \frac{\delta^{2n} \mathcal G[f, \bar f]}
    {\delta f_{i_1}\cdots\delta f_{i_n}\delta \bar{f}_{\bar i_n}\cdots\delta \bar{f}_{\bar i_1} 
     }\Bigg|_{\{f,\bar f\}=0}}_{G^{(n)}_{i_1,\dots,i_n;\bar i_n,\dots,\bar i_1}} 
      \bar{f}_{\bar i_1}\cdots \bar{f}_{\bar i_n}  f_{i_n}\cdots f_{i_1}
     \,.
    \label{eq:Gff}
\end{equation}

Inserting the Taylor expansion, \cref{eq:Gff}, into the restructured path integral, \cref{eq:func}, we obtain the following effective action, i.e., the model action, for the target space ($f$) degrees of freedom
\begin{equation}
\label{eq:Seff}
    S_{\mathrm{eff}}[f,\bar f] = S_{f}[f,\bar f] - 
    \sum_{n=1}^\infty G^{(n)}_{1,\dots,n;\bar n,\dots,\bar1} \, 
      \bar{f}_{\bar i_1}\cdots \bar{f}_{\bar i_n} \,  f_{i_n} \cdots f_{i_1}\,.
\end{equation}
\end{widetext}
At this point, we arrived at an exact theory for the target space degrees of freedom which formally is equivalent to the cFRG downfolding approach~\cite{2012_cfrg}. For a comparison to Hamiltonian-based downfolding approaches, see \cref{App::Hambaseddown}.

Individual terms in the effective action $S_{\mathrm{eff}}[f,\bar f]$ can be expressed in terms of vertex amplitudes ($A$ and $B$) and the correlators within the rest space of the form $\langle \bar{c}_1 \bar{c}_2 ... c_{\bar{2}} c_{\bar{1}}\rangle_{S_c}$.
These terms can be compactly illustrated as shown in \cref{fig:diagrams}. We stress that, in general, all orders of $G^{(n)}$ have to be considered; in the figures and the derivations, we mostly constrain ourselves to the single- and two-particle terms ($n\leq2$) to keep the presentation manageable.

Up to the two-body level, we find 30 topologically distinct terms (5 single-body, 25 two-body). We classify the contributions into terms depending on zero, one, two, or three imaginary times, as denoted in the top left corner of each diagram in \cref{fig:diagrams}.

\begin{figure*}
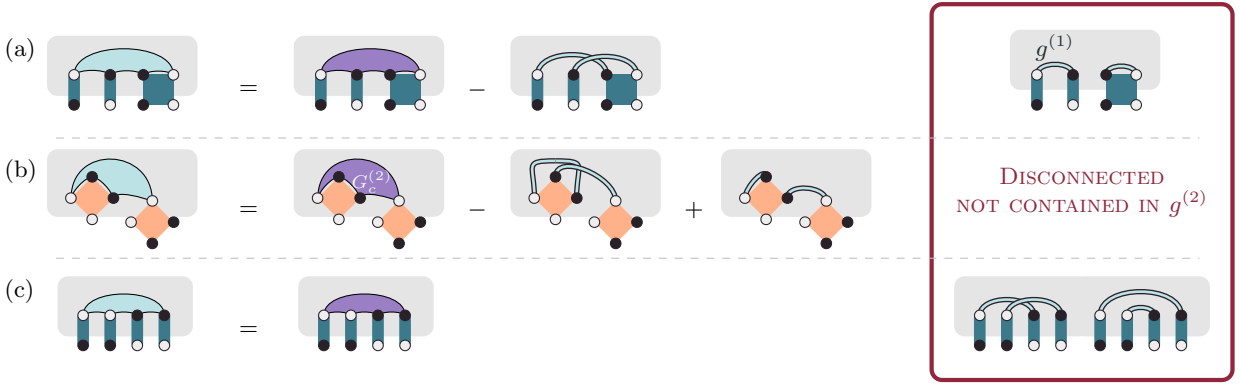

    \centering
    \connectedness
    \caption{Graphical definition of connectedness for diagrams (b3), (c3) and (e2) (cf.~\cref{fig:diagrams}) in terms of connected single-particle $g^{(1)}$ and two-particle $G^{(2)}_c$ Green's functions. We depict $G^{(2)}_c$ as a purple filled shape. The light cyan filled shape ($g^{(k)}$) is defined such that it encompasses only the set of possible decompositions in terms of connected Green's functions which lead to a connected diagram. The disconnected contributions (right, red box) are \emph{not} included in $g^{(k)}$, as they cancel by virtue of the linked cluster theorem.}
    \label{fig:cancellation}
\end{figure*}

The light cyan filled shapes introduced in the diagrams (\cref{fig:diagrams}) denote
$g^{(k)}=\langle c_1...c_k\bar{c}_{\bar{k}}...\bar{c}_{\bar{1}} \rangle_{S_c}^{\text{conn}}$, where the superscript `conn'{} indicates that the $n$-particle expectation value will be replaced by all combinations of $k_i\leq n$-particle connected Green's functions (with $\sum_i k_i=n$) of the rest-space action ($G^{(k)}_c$)  which lead to a connected diagram, cf.~\cref{fig:cancellation}. The cancellation of the disconnected diagrams is a direct consequence of the linked cluster theorem in combination with the generalized Wick theorem~\cite{Helias_2020}.

Now that we have laid out the individual terms in the effective action, let us discuss their physical meaning. The single-particle terms are depicted in row (a) of \cref{fig:diagrams}. We find five different contributions: diagram (a1) is a simple Hartree-like term which can be absorbed into the tight-binding matrix of the effective model: orbital-diagonal terms are chemical potential shifts, while the off-diagonal terms are hopping renormalizations. All of the remaining terms are retarded and are formally analogous to the hybridization terms in (cluster) impurity problems~\cite{Hewson1993, DMFT, RevModPhys.83.349}. 
The second diagram is due to electrons hopping to the rest space, propagating there, and hopping back. 
The last three contributions occur due to the $A^{1:3}$ couplings (assisted hoppings between the target and the rest spaces). 
It is important to note that, depending on the choice of basis, the kinetic term connecting the target and the rest space can vanish. All the diagrams featuring $A^{1:1}$ then immediately drop out. However, one can still have an effective hybridization due to assisted hopping processes ($A^{3:1}$), cf.~diagram~(a5).

On the two-particle level, the diagram~(b5) is the one that is most commonly kept in the effective model as it encompasses retarded density-density and spin-spin interactions. Retarded pair-hopping is captured by the diagram~(c1).

\begin{widetext}
To translate the meaning of the diagrams, we need to follow the diagrammatic rules stated in \cref{App:diagrules}. As examples, we show the expressions for diagram~(a1) of \cref{fig:diagrams}:
\begin{equation}
    \begin{tikzpicture}[x=0.8cm, y=0.8cm, baseline=(base)]
        \coordinate (base) at (0,-0.1);
        \pgfdeclarelayer{bg}
        \pgfsetlayers{bg,main}
        \bg{\node[rectangle, fill=fspacebg, minimum width=0.9cm, minimum height=0.6cm, rounded corners] at (0.0,0.4) {};}
        \pic[rotate=90] (V) at (0,0) {A22};
        \bg{\draw[gprop] (V-4) to[in=35, out=145, looseness=3] (V-3);}
    \end{tikzpicture} \!\!\!
    = -\sum_{1,3\in\mathcal R} g^{(1)}_{1,3} A^{2:2}_{1i_13\bar i_1} \bar{f}_{\bar i_1} f_{i_1}
     \,,
\end{equation}
and diagram~(d3) of \cref{fig:diagrams}:
\begin{equation}
    \begin{tikzpicture}[x=0.8cm, y=0.8cm, baseline=(base)]
        \coordinate (base) at (0,-0.1);
        \pgfdeclarelayer{bg}
        \pgfsetlayers{bg,main}
        \bg{\node[rectangle, fill=fspacebg, minimum width=2.3cm, minimum height=0.8cm, rounded corners] at (0.1,0.4) {};}
        \pic[rotate=90] (V1) at (-1,0) {A11};
        \pic[rotate=135] (V2) at (0,0.1) {A31};
        \pic[rotate=90] (V3) at (1,0) {A22};
        \bg{
          \coordinate (A) at (-1,0.25);
          \coordinate (B) at (-0.35,0.1);
          \coordinate (C) at (0.0,0.45);
          \coordinate (D) at (0.35,0.1);
          \coordinate (E) at (0.75,0.25);
          \coordinate (F) at (1.25,0.25);
          \draw[fill=gcolor]
            (A) 
            to[out=45, in=135, looseness=0.5] (B)
            to[out=90, in=180, looseness=0.5] (C)
            to[out=0, in=90, looseness=0.5] (D)
            to[out=45, in=180, looseness=0.5] (E)
            to[out=45, in=135, looseness=0.5] (F)
            to[out=90, in=90, looseness=0.7] (A)
           -- cycle;}
      \end{tikzpicture}~
    = \sum_{1,2,3,4,5,6 \in \mathcal{R}}4 g^{(3)}_{1,2,3,4,6,5} 
    \tilde{A}^{1:1}_{i_2,4}A^{1:3}_{2,1,\bar{i}_2,5}A^{2:2}_{3,i_1,6,\bar{i}_1} \bar{f}_{\bar i_1}  \bar{f}_{\bar i_2} f_{i_2}f_{i_1}
    \,.
\end{equation}

Collecting all terms in \cref{fig:diagrams}, we arrive at the following expressions for $G^{(1)}$ and $G^{(2)}$:
\begin{align}
    \label{eq:geff1}
     G^{(1)}_{i_1,\bar{i}_1} &{}= 
     \left(A^{2:2}_{1{i_1}3\bar{i}_1} + A_{1,\bar{i}_1}^{1:1} \tilde{A}^{1:1}_{i_1{3}}\right)  \underbrace{\braket{ c_1\bar{c}_{3}}^{\text{conn}}_{S_c}}_{g^{(1)}_{1,3}} + \left(\tilde{A}^{1:1}_{i_1{3}} A^{1:3}_{12\bar{i}_14}  + A_{2,\bar{i}_1}^{1:1}\tilde{A}^{1:3}_{1i_134} \right) \underbrace{\braket{c_1 c_{2}\bar c_{3} \bar c_{4}}^{\text{conn}}_{S_c}}_{g^{(2)}_{1,2,3,4}} 
     \\ &\nonumber
     + A^{1:3}_{12\bar{i}_14}\tilde{A}^{1:3}_{3i_165} \underbrace{\braket{c_1 c_{2}c_3\bar c_{4} \bar c_{5}\bar c_{6}}^{\text{conn}}_{S_c}}_{g^{(3)}_{1,2,3,4,5,6}}\,, 
\end{align}
\begin{align}
     \label{eq:geff2}
     G^{(2)}_{i_1,i_2,\bar{i}_2,\bar{i}_1} &{}= \begin{multlined}[t]
     \frac{1}{4} \left(4A^{1:1}_{1,\bar{i}_2}\tilde{A}^{3:1}_{i_1,i_2,\bar{i}_1,2}+4\tilde{A}^{1:1}_{i_2,2}A^{3:1}_{1,i_1,\bar{i}_2,\bar{i}_1} \right)g^{(1)}_{1,2} \\
     \frac{1}{4}\Big( A^{1:1}_{2,{\bar{i}_1}} \tilde{A}^{1:1}_{i_1,3} A^{1:1}_{1,{\bar{i}_2}} \tilde{A}^{1:1}_{i_2,4} + 2 A^{2:2}_{2,i_2,4,\bar{i}_1}A^{2:2}_{1,i_1,3,\bar{i}_2} + 
        4\tilde{A}^{1:1}_{i_1,3}A^{1:1}_{1,\bar{i}_2}A^{2:2}_{2,i_2,4,\bar{i}_1} 
        + 2\tilde{A}^{1:1}_{i_1,3}\tilde{A}^{1:1}_{i_2,4}B^{2:2}_{1,2,\bar{i}_1,\bar{i}_2}
        \\
        {} +2A^{1:1}_{1,\bar{i}_1}A^{1:1}_{2,\bar{i}_2}\tilde{B}^{2:2}_{i_2,i_1,4,3} + 4B^{2:2}_{1,2,\bar{i}_1,\bar{i}_2}\tilde{B}^{2:2}_{i_2,i_1,3,4}
        +4A^{1:3}_{1,2,\bar{i}_2,3}\tilde{A}^{3:1}_{i_2,i_1,\bar{i}_1,4}+4\tilde{A}^{1:3}_{1,i_2,4,3}A^{3:1}_{2,i_1,\bar{i}_1,\bar{i}_2}  
    \Big)\,g^{(2)}_{1,2,3,4} \\
    {}
    + \frac{1}{4} \Big( 2\tilde{A}^{1:1}_{i_1,4}\tilde{A}^{1:1}_{i_2,5}A^{1:1}_{2,\bar{i}_2}A^{1:3}_{3,1,\bar{i}_1,6} + 2\tilde{A}^{1:1}_{i_2,6}\tilde{A}^{1:3}_{2,i_1,5,4}A^{1:1}_{3,\bar{i}_1}A^{1:1}_{1,\bar{i}_2} + 
4\tilde{A}^{1:1}_{i_2,4}A^{1:3}_{2,1,\bar{i}_2,5}A^{2:2}_{3,i_1,6,\bar{i}_1} +\\
{}4 A^{1:1}_{2,\bar{i}_2}A^{1:3}_{1,i_2,5,4}A^{2:2}_{3,i_1,6,\bar{i}_1} +4\tilde{A}^{1:1}_{i_2,6}\tilde{A}^{1:3}_{1,i_1,5,4}B^{2:2}_{3,2,\bar{i}_1,\bar{i}_2}+4A^{1:1}_{1,\bar{i}_2}A^{1:3}_{3,2,\bar{i}_1,4}\tilde{B}^{2:2}_{i_2,i_1,5,6} \Big) g^{(3)}_{1,2,3,4,6,5}  \\
    {} + \frac{1}{4} \Big(
\tilde{A}^{1:1}_{i_1,5}\tilde{A}^{1:1}_{i_2,6}A^{1:3}_{4,3,\bar{i}_1,8}A^{1:3}_{2,1,\bar{i}_2,7} +A^{1:1}_{4,\bar{i}_1}A^{1:1}_{3,\bar{i}_2}A^{1:3}_{1,i_1,6,5}A^{1:3}_{2,i_2,8,7}  + 4 A^{1:1}_{2,\bar{i}_2}\tilde{A}^{1:1}_{i_2,7}A^{1:3}_{4,3,\bar{i}_1,8}\tilde{A}^{1:3}_{1,i_1,6,5}
\\{}+4 A^{1:3}_{4,3,\bar{i}_1,8}\tilde{A}^{1:3}_{1,i_2,6,5}A^{2:2}_{2,i_1,7,\bar{i}_2} + 2\tilde{A}^{1:3}_{1,i_1,6,5}\tilde{A}^{1:3}_{2,i_2,7,8}B^{2:2}_{4,3,\bar{i}_1,\bar{i}_2}+2A^{1:3}_{4,3,\bar{i}_1,6}A^{1:3}_{2,1,\bar{i}_2,5}\tilde{B}^{2:2}_{i_2,i_1,7,8} \Big) g^{(4)}_{1,2,3,4,5,6,7,8} \\
+{} \left( 2A^{1:1}_{3,\bar{i}_2}A^{1:3}_{5,4,\bar{i}_1,10}\tilde{A}^{1:3}_{1,i_1,6,7}\tilde{A}^{1:3}_{2,i_2,9,8} + 2\tilde{A}^{1:1}_{i_1,6}A^{1:3}_{5,4,\bar{i}_1,10}A^{1:3}_{3,2,\bar{i}_2,7}\tilde{A}^{1:3}_{1,i_2,9,8}  \right) g^{(5)}_{1,2,3,4,5,6,7,8,10,9}\\ 
   {} A^{1:3}_{6,5,\bar{i}_1,12}A^{1:3}_{1,i_1,8,7}A^{1:3}_{4,3,\bar{i}_2,11}A^{1:3}_{2,i_2,10,9} \, g^{(6)}_{1,2,3,4,5,6,7,8,9,10,11,12} \,,
    \end{multlined}
\end{align}
\end{widetext}
which can be either derived from diagrams or by performing the functional derivatives, see \cref{App:GFGF}.

\subsection{How to approximate?}
\label{ssec::howtoapprox}
As mentioned above, the effective action \cref{eq:Seff} represents the \emph{exact downfolding}: Observables involving only $f$ fields are identical to those calculated using the full partition function \cref{eq:partition}. In practical applications, however, the summation over $n$ most likely has to be truncated. Any $n>2$ will generate six- and higher point vertices and thereby exclude the application of most of the established quantum-many-body techniques to solve fermionic lattice models. Hence we define that a faithful target-space model exists whenever contributions beyond $n=2$ become irrelevant. Whether this is the case is dictated by the rest-space action and the coupling amplitudes between rest and target space. The following rules can be formulated:
\begin{enumerate}    
    \item The rest space itself has to have a rapidly convergent expansion of its Green's function generating functional ($\cal G$). That is, a perturbative solution of the rest space has to be well approximated by the lowest two orders, $G^{(1)}_c$ and $G^{(2)}_c$. If this is not the case, we do not expect a closed form for the target-space model to be derivable. In other words, all connected Green's functions of the rest space involving more than $2$ particles have to be negligibly small.
    \label{rule:one}
    \item There has to be a hierarchy of the $A$ and $B$ couplings [cf.~\cref{eq:coupling-action}] which guarantees that no significant three-fermion term is generated. Explicitly, $A^{3:1}$ has to be negligible as otherwise $3n$ particle interactions are generated from an $n$-particle expectation value in the rest space.
\end{enumerate}
The first condition can be readily checked by employing a perturbative expansion for the connected Green's functions of the rest space (for which we have to check convergence of the coefficient tensors). The second condition can be checked by calculating all couplings from the {\it ab initio} wavefunctions, as we exemplify in \cref{sec:paramreal}. This hierarchy can then be used to drop certain terms from \cref{eq:geff1,eq:geff2}.

These two rules have two important implications. First, the hierarchy between A and B tensor elements is decisively controlled by the basis, such that the reliability of the downfolded theory is intimately connected to the target space basis set choice. Second, the larger the target space basis, the easier it is to derive a faithful target space theory~\cite{Mushkaev_2024, full_cell_dmft2}. However, since the subsequent handling of the target-space theory becomes more and more involving with increasing target space size, the aim is to find an optimal target basis set size that balances accuracy and feasibility.

Beyond approximating the exact theory, typically, we want to have a Hamiltonian instead of an action, i.e., keep only the instantaneous terms on both the single- and the two-particle level. How to approximate a retarded interaction by an instantaneous one
is a matter we do not address in this paper, but we note that there are optimized schemes for this procedure~\cite{Casula_2012, PhysRevLett.132.076401,malte_inprep}.

In addition  to the described formalism, there is a second potential hurdle that we have not yet addressed: the double-counting problem~\cite{dc_1_dmft, dc_2_dmft, dc_3_dmft, dc_4_dmft}. It refers to the fact that both downfolding and model calculations contain contributions which were already included in the \emph{ab initio} starting point. To correct for this, we have to subtract those contributions which appear in both approaches. For example, considering an ab initio GW approximation~\cite{GW2} as a starting point, we have to drop those diagrams in the downfolding which were already part of the GW calculation~\footnote{Note that by construction there is no double counting between the downfolding and the model solution side.}.

In the following, we will consider several instructive approximations to \cref{eq:geff1,eq:geff2}. We start with a derivation of cRPA in the present formalism.

\subsection{Reproducing cRPA}
\label{ssec::crpa}
\begin{figure*}
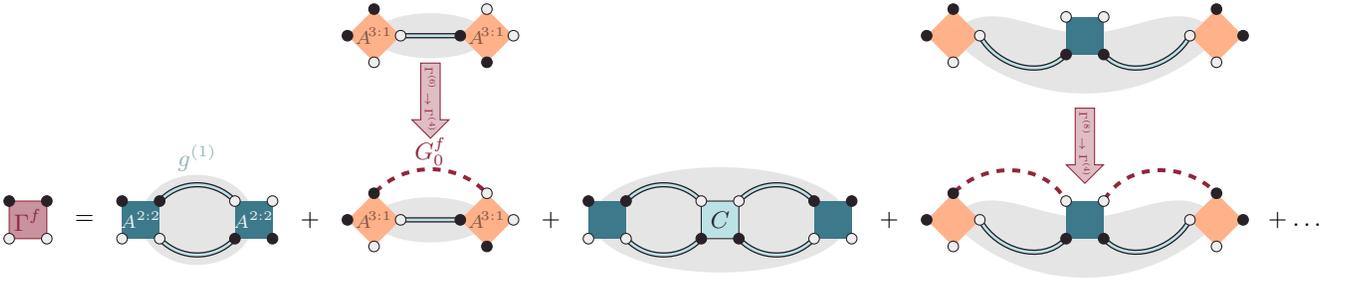

    \centering%
    \newRPAdiagramsImplement%
    \caption{Diagrams contained in cRPA in orders $(A^{n:m})^2$ and $(g^{(1)})^2$ and $(A^{n:m})^3$ and $(g^{(1)})^4$ under the assumption of orthogonal basis sets of the rest- and target space. Light blue lines denote rest space single-particle propagators, vertices are colored according to \cref{fig:diagrams}. The red dashed line represents a bare target space propagator $G^f_0$ that is required to construct a four-point vertex from the six-point contribution $A^{3:1} g^{(1)} A^{3:1}$ (indicated by the red arrow). The rest space vertex is defined by the pale-blue four-point vertex. }
    \label{fig:cRPA_diag}
\end{figure*}

A widely used approach for the extraction of effective two-particle interactions in the target space is the cRPA~\cite{PhysRevB.70.195104}. Within this approximation, the target space interaction parameters include charge screening processes from the rest space. More specifically, the screened Coulomb interaction kernel is obtained through restricted RPA resummation, where the target space two-particle propagator is subtracted from the full two-particle propagator:
\begin{equation}
    \Gamma^f = \frac{V}{\matrixOne + (\chi - \chi_f) V}\,,
\end{equation}
where $\Gamma^f$ is the effective cRPA target space interaction, $\chi$ and $\chi_{f}$ are the non-interacting two-particle propagators in the full and target space, respectively, and $V$ is the bare Coulomb interaction.
By subtracting the target space propagator $\chi_f$ from the full two-particle propagator $\chi$ we end up with a pure rest space propagation $\chi_c = g^{(1)} g^{(1)}$ and a mixed propagation between rest and target space, $\chi_{cf}$ (assuming orthogonal target and rest spaces, i.e., $A^{1:1} = 0$):
\begin{equation}
\label{eq:crpaseparation}
\begin{aligned}
    \chi - \chi_f &{}= (g^{(1)} + G_0^f)(g^{(1)} + G_0^f) - G_0^f G_0^f \\
    &{}= \underbrace{g^{(1)}g^{(1)}}_{\chi_c} + \underbrace{G_0^fg^{(1)} + g^{(1)}G_0^f}_{\chi_{cf}} \,.
\end{aligned}
\end{equation}
To obtain the cRPA diagrams in this setting from the exact starting point \cref{eq:Seff}, we need to reconstruct both diagram types visualized in \cref{fig:cRPA_diag}. 
To recover the first and the third diagram of \cref{fig:cRPA_diag} we approximate the rest space two-particle propagator by a charge channel RPA which partially encompasses diagram (b5) in \cref{fig:diagrams}: This induces a leading order correction of the form $A^{2:2} \chi_c A^{2:2}$ and $A^{2:2} \chi_c C \chi_c A^{2:2}$ in the next to leading order, where $C$ is the rest space vertex. The second and fourth diagrams in \cref{fig:cRPA_diag} are constructed analogously to Ref.~\cite{2012_cfrg}: The diagrams are obtained by a resummation of higher order vertices arising in the target-space action due to the coupling $A^{3:1}$. To make this more explicit, we again consider the lowest order of this type which is a six point interaction of the form $A^{3:1} g^{(1)} A^{3:1}$ (second diagram in \cref{fig:cRPA_diag}). cRPA approximates their contribution by closing bare target space lines in a way similar to the rest-space RPA resulting in $A^{3:1} \chi_{cf} A^{3:1}$. The next-to-leading order of such mixed propagation terms is $A^{3:1} \chi_{cf} A^{2:2} \chi_{cf} A^{3:1}$ and corresponds to an eight point vertex in the target space-action. So the main differences to the pure rest-space series are (i) the smaller energy denominators due to the introduction of the target space propagators $G_0^f$ and (ii) the inclusion of $A^{3:1}$ to connect to the target space. Combining the two classes generates the standard cRPA series for orthogonal basis sets. Notably, the second term implicitly renders the screened interaction dependent on the target space non-interacting Green's function $G_0^f$. As a consequence, the self-energy in the target-space model becomes a functional of not only the fully dressed Green's function but also of the bare Green's function.

To discuss the relevance of the two diagrammatic classes, as an example let us assume interaction values similar to the ones found below for Nickel (cf.~\cref{table:1}).  We assume one band at the Fermi level (being the target space) and two remote bands at distance $\pm\Delta$ from the Fermi level (being the rest-space), with $\Delta$ being much larger than the bandwidth. Under these circumstances, the full (target and rest space) non-interacting, static two-particle propagator reads
\begin{equation}
\begin{aligned}
    \chi_{nm}(\bvec{q}) &{}= \sum_{\bvec{k},\nu}G_{nn}(\bvec{k}-\bvec{q},\nu) \, G_{mm}(\bvec{k},\nu) \\ &{}= \sum_{\bvec{k}} \frac{n_{f}(\epsilon_n(\bvec{k}-\bvec{q}))-n_{f}(\epsilon_m(\bvec{k}))}{\epsilon_n(\bvec{k}-\bvec{q}) - \epsilon_m(\bvec{k})}\,,
\end{aligned}
\end{equation}
where $n,m$ is an orbital/band index, 
$\epsilon_n$ is the corresponding energy of the non-interacting Hamiltonian, $n_{f}$ is the Fermi function (still considering orthogonal orbitals) and $G$ is the full Green's function. In this setting the pure rest space two-particle propagator ($n,m\in \mathcal{R}$) has a denominator bound by $1/(2\Delta)$ (for $m\neq n$).
\begin{figure*}
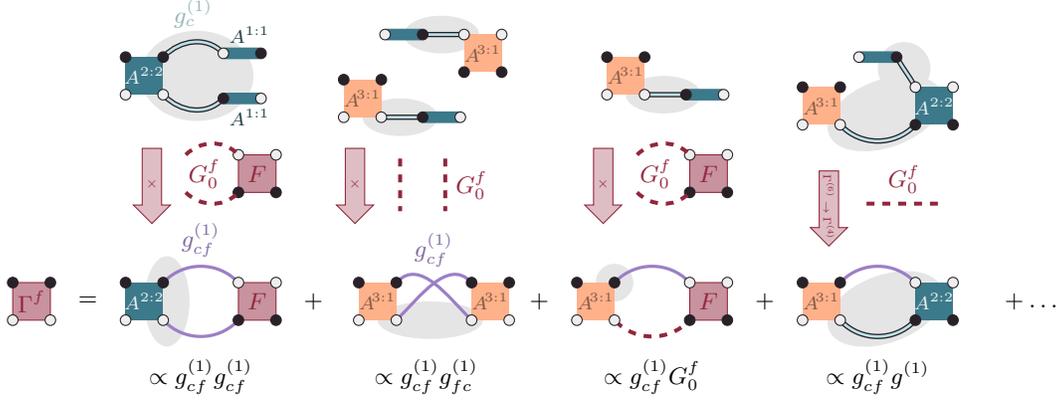

    \centering
    \crossdiagrams
    \caption{Additional diagrams of order $(A^{n:m})^2$ in the case of non-orthogonal basis functions of the target and rest spaces. Light purple lines denote the rest-to-target propagator. Vertices are colored according to \cref{fig:diagrams}. The lowest row visualizes the diagrammatic content of cRPA, while the upper two rows graphically construct these starting from the exact downfolding.}
    \label{fig:cRPA_cross}
\end{figure*}
 The mixed ($n\in \mathcal{R}$ and $m \in \mathcal{T}$) two-particle propagator's denominator is bound by $1/\Delta$
 , so it is roughly twice as large as the pure rest space two-particle propagator. From this standpoint, it appears that mixed propagation terms are more relevant and consequently three particle vertices can never be neglected. However, we have to keep in mind that these diagrams require different outer interaction vertices. For example, we find that $A^{3:1}/A^{2:2} \approx 10^{-2}$ for Nickel~(cf.~\cref{sec:paramreal}). Therefore, we conclude that effects stemming from mixed propagation in the cRPA series should be negligible as long as the target-space model is accurately described by two- and four-point vertices and we are in a perturbative regime of the rest space. Again, note that this depends on the chosen basis for the separation of target and rest spaces.

Going back to the general case, in many instances we do not necessarily have orthogonal basis functions of the target  and rest spaces. In this case the two-particle propagator is a rank four tensor, such that we arrive at the following, more complex [compared to \cref{eq:crpaseparation}], structure of the projected propagator
\begin{equation}
\begin{aligned}
    \chi - \chi_{f} &{}= \left(g^{(1)} + G_0^f + g_{cf}^{(1)} + g_{fc}^{(1)}\right)^2 - G_0^f G_0^f  \\
    &{}= \begin{multlined}[t]\chi_c + \chi_{cf} + (g^{(1)} + G_0^f )( g_{cf}^{(1)} + g_{fc}^{(1)}) \\
    + ( g_{cf}^{(1)} + g_{fc}^{(1)})(g^{(1)}+G_0^f ) \\ +g_{cf}^{(1)}g_{fc}^{(1)} + g_{fc}^{(1)}g_{cf}^{(1)}\,. \end{multlined}
\end{aligned}
\end{equation}
The first two terms in the second line correspond to the contributions discussed before. The other terms generate new classes of diagrams in the screening. In \cref{fig:cRPA_cross} we visualize one representative case of each of these classes of diagrams and construct it graphically from the exact downfolding formalism. 
To reproduce these contributions we first have to identify how the exact formalism captures effects of the mixed single-particle propagator $g^{(1)}_{cf}$. As we show in the next section the kinetic coupling results in a single-particle propagator of the target-space action of the form $(g^{(1)}_{f})^{-1} = (G_0^f)^{-1} + A^{1:1} g_0^{c}\tilde{A}^{1:1}$ such that we arrive at $g^{(1)}_{cf} = g_0^{c}A^{1:1}g^{(1)}_{f}$~\footnote{This follows from a block matrix inversion.}. Thus, some contributions stemming from $g_{cf}^{(1)}$ are captured by inclusion of the hybridization corrections in the target space, i.e., the first three diagrams in the summation (lower row) in \cref{fig:cRPA_cross}
are contained in the downfolded model upon performing a non-self-consistent perturbation theory in the target space. Only the last diagram in \cref{fig:cRPA_cross}, which is proportional to $g_{cf}^{(1)}g^{(1)}$ requires perturbative closing of selected legs of a six-point vertex.

\subsection{Comparison to other downfolding approaches}
There are many other approaches tackling the downfolding problem or at least some aspects of it~\cite{riedl2019, PhysRevLett.111.036601, RevModPhys.95.035004,Sharma2023}. In the following we briefly compare our approach to some of the previously suggested ones.

\subsubsection{constrained FRG} In the initially proposed form~\cite{2012_cfrg}, cFRG is formally equivalent to the approach proposed here.
In contrast to our approach, cFRG has been formulated in the band basis and the majority of cFRG studies did not analyze the basis dependence of their results. Furthermore, cFRG has been derived in terms of Polchinski's effective actions~\cite{Polchinski1984}, which is not as widely known as the applied perturbation theory here.
In practice, cFRG performs an additional approximation~\cite{2015_cfrg, 2018_cfrg} of closing single target-space propagator lines perturbatively within the RG flow leading to cRPA-like equations for effective interactions in the target space. Therefore, it is hampered by the same conceptual issue of the Luttinger-Ward functional depending on both $G$ and $G_0$.

\subsubsection{Coupled-Cluster based downfolding} The different variants of Coupled-Cluster (CC) downfolding, see Ref.~\cite{Bauman2019}, are based on exploiting subsystem embedding sub-algebras (SBS). These algebras are spanned by the particle-hole operators defined for selected empty and occupied single particle states of the reference solution. Coupled cluster downfolding is exact for the ground-state energy and excited states~\cite{Bauman2019excited}, by generating state specific effective Hamiltonians for the chosen SBS. In contrast, within our approach the total energy is not accessible directly anymore, instead we have the guarantee that observables defined purely within the target space are exactly reproduced upon convergence.

\subsubsection{Renormalized Density matrix downfolding} Unrenormalized density matrix downfolding (DMD)~\cite{Changlani_2015} determines the effective Hamiltonian by matching its ground-state energy, excited states and two-body density matrices within a selected set of (potentially) eigenstates of the ab-initio Hamiltonian. Renormalized DMD~\cite{PhysRevB.110.195103} expands on this by ensuring that the model eigenstates and the ab-initio eigenstates map one-to-one. Therefore, if the reference has high accuracy, the derived effective Hamiltonian will be faithful; if the reference is not accurate, the effective Hamiltonian will be a bad representation of the full problem. Therefore, the downfolding was so far mainly performed from a beyond DFT starting point, e.g.~real-space variational Monte Carlo~\cite{PhysRevB.110.195103}.
Depending only on the convergence of the rest-space expansion, the approach suggested here allows, in principle, to start from a lower-level ab initio calculation.

\subsubsection{Canonical transformation} The aim of canonical transformations is to bring the Hamiltonian into a block diagonal form by applying unitary transformations~\cite{White2002}. This is done by subsequently removing off-diagonal elements in the many-body Hamiltonian which generates new higher order terms within the mutually diagonal blocks~\cite{PhysRevA.111.042825}. 
The downfolded basis is thus spanned by many-body eigenstates, which are potentially of multi-reference character. As such, a clear identification of single- and two-particle interaction terms within the target-space is not as straightforward as in our suggested approach.

\subsubsection{Perturbative approaches} There exists a number of operator based perturbative approaches which are in a sense close in spirit to our proposed approach---noteworthy approaches include Schrieffer-Wolff downfolding~\cite{Bravyi_2011} and operator perturbation theory~\cite{Kato1949, Kato1995}. While these approaches are widely used in deriving spin-models~\cite{PhysRevB.110.195103}, applications for direct ab-initio Hamiltonians are rare. Since standard operator perturbation theory is based on a resolvent expansion of the stationary Schrödinger equation, the approach guarantees (if converged) the correct many-body eigenenergies, in contrast to our approach which guarantees correct observables within the targeted subspace. 

\section{Analytic structure}
\label{sec:anastruct}
Having shown that cRPA can be recovered from diagrams contained in the proposed downfolding procedure by perturbatively closing target-space lines and adding target-space vertices,
in this section we construct simplified cases that allow to understand the implications of different contributions to the coupling action \cref{eq:coupling-action}.
We start with a purely kinetic coupling (only relevant for a non-diagonal basis, e.g., the orbital basis) in \cref{sec:kincpl} and continue with adding particle-hole coupling $A^{2:2}$ in \cref{sec:ph-interactions}.

\subsection{Kinetic coupling}
\label{sec:kincpl}
One of the simplest cases one can envision is a non-interacting rest space that is kinetically coupled to the target space, i.e.,
\begin{align}
    \mathcal A[f, \bar f,c, \bar c] &{}= \bar c_2 \tilde{A}^{1:1}_{12} f_1 + \bar f_2 A^{1:1}_{12} c_1\,,\\
    S[c, \bar c] &{}= \bar c_2 \big[ \underbrace{ \delta_{12}(i\omega - \mu) + T^c_{12} }_{\tilde{T}^c_{12}} \big] c_1 \,,\\
    S[f, \bar f] &{}= \bar f_2 \big[ \underbrace{ \delta_{12}(i\omega - \mu) + T^f_{12} }_{\tilde{T}^f_{12}} \big] f_1 +\bar f_3 f_1 U_{1234} \bar f_4 f_2 \,.
\end{align}
Such a scenario is relevant for cases in which target and rest space are entangled. In the absence of entanglement we have $A^{1:1}=0$.
The case of purely kinetic coupling has been discussed in depth in the context of, e.g., impurity models~\cite{PhysRev.124.41} and transport through macroscopic systems~\cite{lee1981anderson, groth2014kwant}. 
In this limit, the rest space induces a retarded hopping (retarded hybridization) in the target space, see the diagram (a2) in \cref{fig:diagrams}:
\begin{equation}
\begin{aligned}
    \mathcal{G}[f, \bar f] &{}= \log\Big( \Braket{\exp\big[
    \mathcal A[f,\bar f,c,\bar c]
    \big]}_{S_c}\Big) \\
    &{}= \tilde{A}^{1:1}_{12}A^{1:1}_{34} \bar{f}_4f_1 \braket{c_3\bar{c}_2} = \tilde{A}^{1:1}_{12}A^{1:1}_{34} \bar{f}_4f_1 \, g^{(1)}_{2,3} \,.
\end{aligned}
\end{equation}
The exact target-space action is then given by 
\begin{equation}
    S[f, \bar f] = \bar f_2 (\tilde{T}^f_{12} + A^{1:1}_{42}g^{(1)}_{3,4} \tilde{A}^{1:1}_{13} ) f_1 +\bar f_3 f_1 U_{1234} \bar f_4 f_2 \,.
\end{equation}
We find the standard result that the kinetic coupling to remote bands induces an effective ``retarded hopping'' due to propagation in the rest space. This kind of term is also known as (retarded) hybridization. In the present example, this term is sufficient to ensure that the spectral function calculated from the effective target-space model is identical to the spectral function obtained by the corresponding calculation in the full model. Importantly, we will get formally identical contributions in the target-space action even in the absence of $A^{1:1}$ due to $A^{1:3}$, cf. diagram (a5) in \cref{fig:diagrams}.  

\begin{figure*}
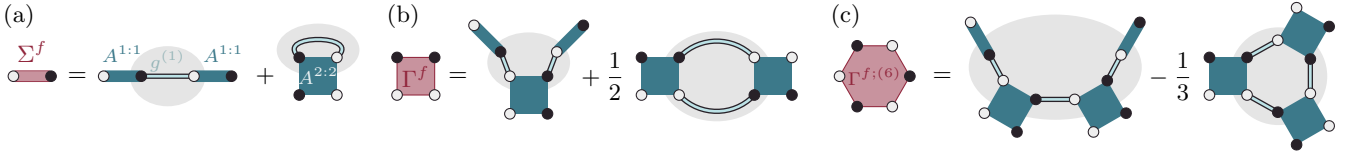
%
    \centering%
    \newDoubleDiagramsImplement%
    \caption{Diagrams in the target space effective theory up to four external $f$ fields under the approximations detailed in \cref{sec:ph-interactions}. (a)~Single-particle term (two-point vertex) $\Sigma^f$, (b)~Two-particle term (four-point vertex) $\Gamma^f$, (c)~Six-point vertex $\Gamma^{f;(6)}$. The $c$ propagators are denoted with light cyan lines, and the $cf$ vertices with dark cyan rectangles/squares (as in \cref{fig:diagrams}).}
    \label{fig:density}
\end{figure*}

\subsection{Particle-hole interactions}
\label{sec:ph-interactions}
\begin{figure}
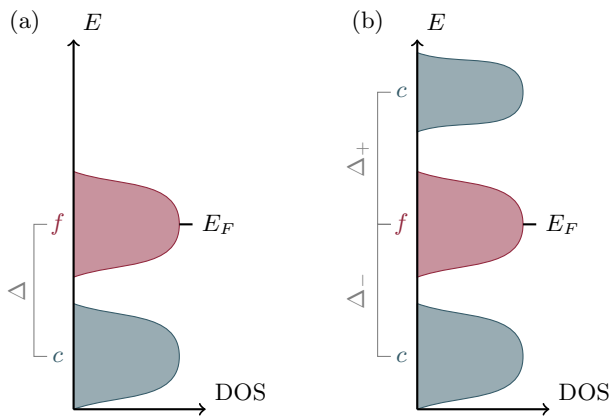

    \centering
    \dosExamplesImplement
    \caption{Sketch of the two different energetic setups in the examples considered in the main text, with panels~(a) and~(b) corresponding to \cref{sssec:above,sssec:above-and-below}, respectively.}
    \label{fig:sketch}
\end{figure}

To extend the discussion above, let us add an $A^{2:2}$ contribution to the coupling between the sectors, while keeping the rest space non-interacting. To obtain the target-space action, we explicitly perform the path integral over $c$ fields and then identify the diagrams that correspond to the leading order corrections. Suppose the coupling 
action and the rest-space action are given by
\begin{align}
    \mathcal A[f, \bar f,c, \bar c] &{}= \bar c_3 c_1 A^{2:2}_{1234} \bar f_4 f_2 + \bar c_2 \tilde{A}^{1:1}_{12} f_1 + \bar f_2 A^{1:1}_{12} c_1\,,\\
    S[c, \bar c] &{}= \bar c_2 \tilde{T}^c_{12} c_1 \,,
\end{align}
i.e., only kinetic and particle-hole coupling with particle-hole pairs in the target space as well as the rest space. Note that this simple form is basis dependent: Under a global change of basis, finite contributions of all other types of vertices may be generated. Further, we can also differentiate from context whether $\tilde{A}$ or $A$ is required by inspecting the attached fields. 
To integrate out the rest space fields we rearrange all quadratic terms in the $\bar c,c$ fields:
\begin{widetext}
\begin{equation}
\begin{aligned}
    \bar c \tilde{T}^c c + \bar c A^{2:2} \bar f f c + \bar c \tilde{A}^{1:1} f + \bar f A^{1:1} c &{}= \bar c_2 \underbrace{\big( \tilde{T}^c_{12} + A^{2:2}_{1i2j}\bar f_j f_i \big)}_{M_{12}} c_1 + \bar c_2 \tilde{A}^{1:1}_{j2} f_j + \bar f_i A^{1:1}_{1i} c_1 \\
    &{}= \underbrace{\big(\bar c_2 + \bar f_i A^{1:1}_{\tilde 1 i} M^{-1}_{\tilde 1 2}\big)}_{\bar \kappa_2} M_{12} \underbrace{\big(c_1 + M^{-1}_{1 \tilde2 } \tilde{A}^{1:1}_{j\tilde 2} f_j\big)}_{\kappa_1} - \bar f_i A^{1:1}_{\tilde 1 i} M^{-1}_{\tilde 1 \tilde 2} \tilde{A}^{1:1}_{j\tilde 2} f_j \,,
\end{aligned}
\end{equation}
which allows us to shift the integration and then perform the Gaussian integral, i.e.,
\begin{equation}
\begin{aligned}
    \mathcal Z &{}= \int\DD f\DD\bar f \exp\Big\{-S_f[f,\bar f] + \bar f_i A^{1:1}_{1 i} M^{-1}_{1 2} \tilde{A}^{1:1}_{j2} f_j \Big\} \int\DD\kappa\DD\bar\kappa \exp\Big\{-(\bar\kappa,M\kappa)\Big\} \\
    &{}=\int\DD f\DD\bar f \exp\Big\{-S_f[f,\bar f] + \bar f_i A^{1:1}_{1 i} M^{-1}_{1 2} \tilde{A}^{1:1}_{j2} f_j + \Tr\log\big[M\big] \Big\} \,.
\end{aligned}
\end{equation}
We recognize that the $M$-matrix contains a rest space propagator and the interaction. If we choose to expand $M$ in powers of the rest space propagator $g^{(1)}$, we obtain
\begin{align}
    \label{eq:exactcouplings1}
    M^{-1} &{}= \frac{g^{(1)}}{\matrixOne + g^{(1)} A^{2:2}_{ji} \bar f_i f_j} = g^{(1)} - g^{(1)} A^{2:2}_{ji} g^{(1)}\bar f_i f_j + \mathcal O\Big(\big[g^{(1)} A^{2:2} \bar f f\big]^2\Big)\,,\\
    \label{eq:exactcouplings2}
    \Tr\log\big[M\big]\stackrel{\sim}= \Tr\log\big[\matrixOne + g^{(1)}A^{2:2} \bar f f\big]  &{}= \Tr\Big[ g^{(1)} A^{2:2} \bar f f -\frac12 \big(g^{(1)} A^{2:2} \bar f f\big)^2 + \mathcal O\Big(\big[g^{(1)} A^{2:2} \bar f f\big]^3\Big)\Big] \,.
\end{align}
\end{widetext}
\Cref{fig:density} visualizes the leading order contributions (up to the six-point vertex) diagrammatically. In principle all $2n$-point vertices are generated, but the higher-order ($n>2$) ones are suppressed by factors of $(g^{(1)}A^{2:2})^{n-1}$, as each pair of $\bar ff$ comes with this factor. Using the diagrams for the six-point vertex [cf.~\cref{fig:density}~(c)] as an example, the simple diagrammatic rules to obtain all $2n$-point vertices [cf.~\cref{eq:exactcouplings1,eq:exactcouplings2}] can be explained: All diagrams that contain $A^{1:1}$ are strings of $n-1$ vertices $A^{2:2}$ that end in $A^{1:1}$ (with $2n$ external $f$ lines and $n$ internal $c$ lines). Diagrams without $A^{1:1}$ are circles of $n$ vertices $A^{2:2}$ connected by $n$ internal $c$ lines (also resulting in $2n$ external $f$ lines). For an improved understanding of the relevance of different diagrammatic contributions, let us consider different rest spaces: One where the $c$ electrons are located only below the Fermi level (cf.~\cref{fig:sketch}~(a), \cref{sssec:above}) and one where the $c$ electrons have spectral weight both below and above the Fermi level (cf.~\cref{fig:sketch}~(b), \cref{sssec:above-and-below}). We will focus on the leading order diagrams, i.e.~neglect the six point vertex.

\subsubsection{Bands below the Fermi level}
\label{sssec:above}
In the first case [cf.~\cref{fig:sketch}~(a)], the rest space is exclusively on \emph{one} side of the Fermi level (here: below). As a consequence, the second diagram in \cref{fig:density}~(b) is suppressed exponentially with $T/\Delta$, where $T$ denotes temperature and $\Delta$ the gap magnitude. The first diagrams of \cref{fig:density}~(a,b) on the other hand only come with an algebraic suppression by $\Delta$---each target space Green's function is bounded from above by its zero frequency value. Assuming narrow $c$-electron bandwidth, we find
\begin{align}
    \Sigma^f(\omega) \lesssim{}& (A^{1:1})^2 \frac{1}{i\omega - \Delta}  \,, \\
    \Gamma^f(\omega_1,\omega_2,\omega_3) \lesssim{}& (A^{1:1})^2 A^{2:2} \frac{1}{i\omega_1 - \Delta}\frac{1}{i\omega_3 - \Delta}\,.
    \label{eq:screening-int}
\end{align}
As a reminder, we denote a degree of freedom as gapped when it has negligible partial DOS near the Fermi level, which does not exclude entanglement with the target space bands, as  the $4s$ orbitals in Nickel (see \cref{ssec:nickel}). As explained above, for such systems, the self-energy acts as a hybridization between target and rest space. There is also  a non-negligible frequency-dependent screening correction to the interaction [cf.~\cref{eq:screening-int}]. Material examples of the energetic setup sketched in \cref{fig:sketch}~(a) include noble gases and $s$-electron systems.

\subsubsection{Bands above and below the Fermi level}
\label{sssec:above-and-below}
In the case of $c$ bands above \emph{and} below the Fermi level [cf.~\cref{fig:sketch}~(b)], the second diagram of \cref{fig:density}~(b) is no longer exponentially suppressed because particle-hole excitations become available. This is the diagram type resummed in the cRPA for the target space two-particle interaction. At low temperatures $T/\Delta_\pm \ll 1$ (i.e., the gaps $\Delta_\pm$ as dominant energy scale), its contribution to $\Gamma^f$ is bounded by 
\begin{equation}
    \Gamma^f(\omega_1,\omega_2,\omega_3) \lesssim \frac{\big(A^{2:2}\big)^2}{i(\omega_3-\omega_1) - (\Delta_+ +\Delta_-)} \,.
    \label{eq:rpa-contrib}
\end{equation}
Comparing to \cref{eq:screening-int} (and assuming $\Delta_\pm\approx\Delta$) reveals an enhancement of a factor $A^{2:2}/A^{1:1} \cdot \Delta/A^{1:1}$ in the static limit $\omega_i \to 0$. The scenario of \cref{fig:sketch}~(b) is relevant in most $d^n$ electron systems, especially in strongly correlated materials~\cite{Bednorz1986,sachdev2003,medici2017hundsmetalsexplained,checkelsky2024}: These systems typically have a filled $s^n$-shell (and sometimes $p^{n-1}$) below the correlated subspace and empty $p^{n}$ or $s^{n+1}$ orbitals above the correlated subspace. The gap sizes $\Delta_\pm$ to these subspaces are often small, such that substantial screening is encountered. Thus cRPA naturally captures the leading order corrections to the target space two-particle interactions for these materials explaining its success \emph{a posteriori}~\cite{crpa_2008, Ni_crpa_2009, chang2024downfolding}. 

In general, we observe that even in this simple example we find a competition between different terms in the downfolding, \cref{eq:rpa-contrib} and \cref{eq:screening-int}. Further, we showed that their hierarchy is detail-dependent and while arguments can be made about which is the larger
coupling, the other couplings are not {\it a priori} negligible.

Beyond the gap magnitude $\Delta_{(\pm)}$, $A^{1:1}$ and $A^{2:2}$ also influence the relevance of different contributions to the target-space model [cf.~\cref{eq:screening-int,eq:rpa-contrib}]. To obtain a hierarchy of contributions, we have to obtain all coupling matrix elements. In the following section, this procedure is carried out for two example materials: fcc Nickel (cf.~\cref{ssec:nickel}) and the infinite layer Cuprate SrCuO\textsubscript2 (cf.~\cref{ssec:cuprate}).

\section{Ab-initio parameters}
\label{sec:paramreal}

After discussing the analytic structure of exactly down-folded models of simplified test cases, we now turn our attention to actual materials and the resulting ab initio parameter estimates. Specifically, we discuss in the following face centered cubic nickel as well as the infinite layer cuprate SrCuO$_2$ and calculate the bare (full model) parameters $A$ and $B$ in different Wannier basis sets. These will allow us to judge on the applicability of the rules defined in \cref{ssec::howtoapprox}.

The DFT calculations we present in the following were all performed with the FPLO code~\cite{FPLO1,FPLO2}. For all of these calculations we employ the local density approximation (LDA) exchange-correlation functional~\cite{PhysRevB.45.13244} and include scalar-relativistic corrections. 

After the DFT simulation, we wannierize all valence and semi-core orbitals considered in the DFT calculation into an atomic orbitals basis. To check for convergence of the real-space mesh we verify that each of the resulting Wannier functions is approximately normalized to one on the chosen grid.

From the obtained Wannier orbitals we calculate the (local) interaction tensor via
\begin{equation}
    \begin{aligned}
    U_{1234} &{}=  \braket{\psi_3 \psi_4 |\hat{H}_\mathrm{pot}|\psi_2 \psi_1} \\
            &{}= \int \dd\bvec{r} \dd\bvec{r}' \, V(|\bvec{r}-\bvec{r}'|) \psi_3(\bvec{r})^* \psi_4(\bvec{r}')^*\psi_2(\bvec{r}') \psi_1(\bvec{r}) \,,
    \end{aligned}
    \label{eq:utensor}
\end{equation}
where $V(|\bvec{r}-\bvec{r}'|)$ is the Coulomb interaction and the indices label both orbital index $o$ and position of the Wannier center $\bvec{R}$. 
The numerical implementation of \cref{eq:utensor} follows the Fourier procedure of Ref.~\cite{schnell2002ab}, see \cref{App:coulombmat} for more details. The integration always has an intrinsic error due to the discretization employed to represent the Wannier functions.

\begin{figure}
    \centering
    \includegraphics{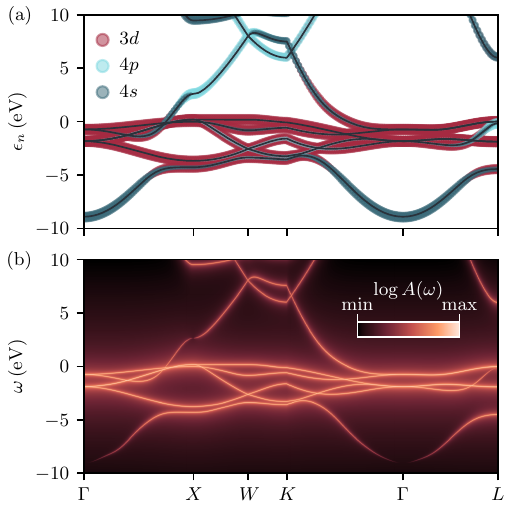}
    \caption{(a)~Band structure of fcc Nickel with colored weights of the $3d$ (bordeaux), $4p$ (light cyan) and $4s$ (dark cyan) orbitals. These three orbital families make up most of the weight near the Fermi level. (b)~Non-interacting spectral function $A(\bvec k,\omega)$ of the downfolded $d$-orbital model for fcc Nickel. As expected, the poles lie on top of the bands of the full system [cf.~panel~(a)]---no additional poles appear or vanish. We set the broadening to $10\,\mathrm{meV}$.}
    \label{fig:nickel}
\end{figure}

\subsection{Face centered cubic nickel}
\label{ssec:nickel}
We start with fcc Nickel which has been extensively studied in the literature~\cite{Ni_2001, Ni_2007, crpa_2008, Ni_crpa_2009, Ni_2012}. To this end, we perform a DFT calculation in which we initialize the lattice in the $Fm\bar3m$ space group (no.~225) with lattice constant $a=3.524\,\angstrom$~\cite{Voronin2016}. The calculations are performed on a $16\times 16\times 16$ momentum mesh. From the converged calculation we obtain a tight-binding model for all bands included in the DFT calculation and the corresponding Wannier functions. \Cref{fig:nickel}~(a) shows the band structure near the Fermi level and color-code the relevant orbital contributions. We define the $3d$ manifold as the target space, which  is responsible for  most of the spectral weight near the Fermi-level. We further observe that there is a hybridization with both $4p$ and $4s$ orbitals close to the Fermi level. This is indicated by bands changing from mainly $3d$ character to $4p$ or $3s$ character, implying $A^{1:1}$ in the Wannier basis to be non-vanishing. Orthogonality of the Wannier orbitals suggests $A^{3:1}$ and $A^{1:3}$ to be smaller than $A^{2:2}$ because of the occurrence of the scalar product-like expression $\dd \bvec r\,\psi_3(\bvec r)^* \psi_1(\bvec r)$ under the integral in \cref{eq:utensor}.

\begin{table}
    \caption{Leading couplings between the rest space states ($4s$ and $4p$) and the target space manifold of the $3d$ states for fcc Nickel in the atomic orbital basis provided by FPLO. All elements are given in~$\mathrm{eV}$. Elements which are forbidden by point group symmetries of the system are marked with ``(sym)''{}.}
    \label{table:1}
    \vspace{0.3em}
    \begin{ruledtabular}
    \def\arraystretch{1.2}
    \begin{tabular}{ccc} 
        coupling & $3d\rightarrow4s$ & $3d\rightarrow4p$  \\[0.2em]\hline
        $A^{1:1}$ & $1.44$ & $-1.19$ \\ 
        $A^{2:2}$ & $10.83$ & $16.35$ \\
        $A^{3:1}$ & $0.0$ & $0.0$ (sym)\\
        $A^{1:3}$ & $0.06$ & $0.0$ (sym)\\
        $B^{2:2}$ & $0.25$ & $1.12$
    \end{tabular}
    \end{ruledtabular}
\end{table}

In \cref{table:1} we summarize the largest magnitude tensor components of the different couplings in the Wannier basis.
As expected $A^{2:2}$ is by far the leading coupling by an order of magnitude and the hybridization $A^{1:1}$ has also relevant contributions. Both $A^{3:1}$ and $A^{1:3}$ are essentially zero, which for some of these elements is expected by symmetry and for others follows from the orthogonality arguments outlined above. In contrast, $B^{2:2}$ is not symmetry forbidden and indeed has a sizable but small contribution.

The influence of the $A^{1:1}$ coupling to the single-particle terms in our target space becomes clear by integrating out all orbitals but the $3d$ shell. \Cref{fig:nickel}~(b) displays the resulting non-interacting target space spectral function. Comparing the spectral function with the DFT bandstructure, we observe that all bands in which the $3d$ manifold had nonzero weight are observable as sharp features in the spectral function. Only those (parts of) bands which have no orbital contributions from that manifold are truly integrated out.
The comparably large $A^{2:2}$ yields on the single-particle side a chemical potential shift (as rendered by diagram (a1) in \Cref{fig:diagrams}) and a charge-screening to the two-particle terms (as rendered by diagram (b5) in \Cref{fig:diagrams}).
The small, but still finite $B^{2:2}$ leads to an additional renormalization due to pair-hopping (as rendered by diagram (c1) in \Cref{fig:diagrams}).
Finally, the combined effects of $A^{1:1}$ and $A^{2:2}$ as well as $A^{1:1}$ and $B^{2:2}$ yield further renormalizations to the screened interactions (diagrams (c3), (c2) and (d1) in \Cref{fig:diagrams}).
All other contributions are negligible due to the vanishingly small $A^{3:1}$ and $A^{1:3}$ terms. 

Thus, if the rest space is perturbatively treatable, fcc nickel fulfills the conditions outlined in \cref{ssec::howtoapprox} and thereby can be captured by an action encompassing single- and two-particle terms.

Finally, we stress that this analysis and conclusion is strongly basis-dependent and as such only valid in the atomic orbital basis we chose---our conclusion relies on $A^{2:2}$ and $B^{2:2}$ being much larger than $A^{3:1}$ and $A^{1:3}$. This hierarchy is affected by any basis transformation (cf.~\cref{App::BvO}) as different vertex components are transformed into each other. More precisely, if we rotate away $A^{1:1}$ we will generate larger $A^{3:1}$ and $A^{1:3}$ contributions appearing due to  $A^{2:2}$ contributions of the orbital space. Therefore, the choice of basis is intimately intertwined with the form and convergence of the low-energy model.

\begin{figure}[!hbt]
    \includegraphics{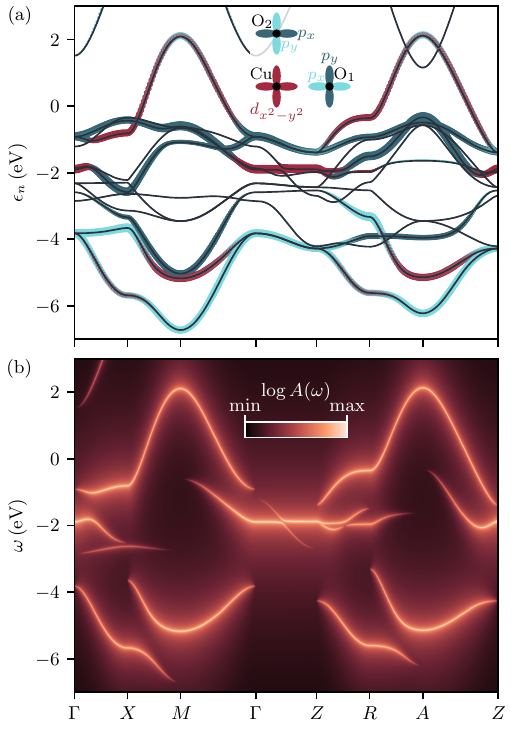}
    \caption{(a) Band structure of SrCuO\textsubscript2 with colored weights of the Copper $3d_{x^2-y^2}$ (bordeaux) and the Emery (light cyan) and non-Emery type (dark cyan) Oxygen $2p$ orbitals as visualized in the inset. The other Copper d orbitals have weight close to the Fermi level as well. (b)~~Non-interacting spectral function $A(\bvec k,\omega)$ of the downfolded single $d_{x^2-y^2}$-orbital model. We observe that the Hubbard like band is essentially completely integrated out along $\Gamma$ to $Z$ as along this path segment the $d_{x^2-y^2}$-orbital does not contribute any weight. The broadening is set to $10\,\mathrm{meV}$.}
    \label{fig:cuprate}
\end{figure}

\subsection{Infinite layer Cuprate SrCuO\textsubscript2}
\label{ssec:cuprate}
We proceed with a more complex material, the infinite-layer cuprate SrCuO\textsubscript{2}~\cite{Siegrist1988, Azuma1992}, which belongs to the intensively studied family of copper oxide superconductors that have been a focal point in condensed matter research since their discovery in 1986~\cite{Bednorz1986}.
We model the crystal by the $P4/mmm$ spacegroup (no.~123) with lattice constants $a=3.927\,\angstrom$ and $c=3.435\,\angstrom$~\cite{Zhou1994}. The calculations were performed on a $20\times 20 \times 10$ momentum mesh.

Often, this class of materials is described by a single band Hubbard model resembling the Copper $d_{x^2-y^2}$ orbital which contributes most weight at the Fermi level. Alternatively, a three band Emery model~\cite{PhysRevLett.58.2794} encompassing $3d_{x^2-y^2}$ from Cu and $2p_x$ and $2p_y$  from the two neighboring in-plane oxygens can be used to describe this class of materials. 
To emphasize the coupling between oxygen $2p$ orbitals, classified as `Emery'-type in light cyan and rest in dark cyan, we encode their respective weight in \cref{fig:cuprate}~(a). We remark that the non-Emery type orbitals are much stronger coupled to the $d_{x^2-y^2}$ orbital than in standard cuprates due to the missing apical oxygen in the infinite layer compound.
We follow the procedure described above to obtain the Wannier orbitals and the DFT bandstructure. In agreement with the single-band Hubbard model picture, we chose the $d_{x^2-y^2}$ orbital as target space, but extending the discussion to the three band Emery model~\cite{PhysRevLett.58.2794} is straightforward.

\Cref{table:cup} summarizes the different couplings of the $d_{x^2-y^2}$ orbital to the rest of the Cu-$3d$ shell as well as to the O-$2p$ shell. The interaction couplings to the Cu-$3d$ shell follow the expected hierarchy of $A^{2:2} > B^{2:2} > A^{1:3}$, while $A^{3:1}$ is symmetry forbidden. Considering the Cu-$3d_{x^2-y^2}$ to O-$2p$ couplings, all classes are symmetry allowed. Again $A^{2:2}$ is the largest by an order of magnitude, and $A^{3:1} > B^{2:2}$. Most notably, the kinetic coupling to the Emery-type O-$2p$ sector is large, see light cyan bands in \cref{fig:cuprate} (a). This large kinetic coupling implies that when downfolding onto the $d_{x^2-y^2}$ orbital a large hybridization arises. This hybridization then ensures that the projected spectral function of the full model is exactly recovered (by construction). Most notably, the downfolded action spectral function cannot be mapped to a single-band Hamiltonian, as it has discontinuities in the weight distribution. 

\begin{table}
    \caption{Leading couplings between the rest space states (Cu-$3d$ and O-$2p$) and the target space $d_{x^2-y^2}$ orbital for the infinite-layer cuprate SrCuO\textsubscript2. All contributions are given in~$\mathrm{eV}$. Contributions which are forbidden by point group symmetries of the system are marked with ``(sym)''{}.}
    \label{table:cup}
    \vspace{0.3em}
    \begin{ruledtabular}
    \def\arraystretch{1.2}
    \begin{tabular}{ccc}
        coupling & $3d_{x^2-y^2}\rightarrow$ Cu-$3d$ & $3d_{x^2-y^2}\rightarrow$ O-$2p$ \\[0.2em]\hline
        $A^{1:1}$ & $-0.09$& $2.8$  \\ 
        $A^{2:2}$ & $25.29$ & $8.04$ \\
        $A^{3:1}$ & $0.0$ (sym) & $0.79$ \\
        $A^{1:3}$ & $0.49$ & $0.45$  \\
        $B^{2:2}$ & $1.09$ & $0.01$
    \end{tabular}
    \end{ruledtabular}
\end{table}

As for the nickel case, we found that the density-density type coupling in $A^{2:2}$ is the largest one throughout the two examined cases. Since this coupling is native to the charge channel, the observed hierarchy justifies cRPA to obtain $\Gamma^{f}$ to some degree. The significance of $A^{1:1}$ on the other hand draws attention to corrections beyond the commonly used downfolding approach of wannierization in combination with cRPA to obtain the screened two-particle interaction due to the correction in the hybridization being missed and corrections due to $B^{2:2}$ or mixed vertex diagrams. 

\begin{figure*}[!hbt]
    \includegraphics{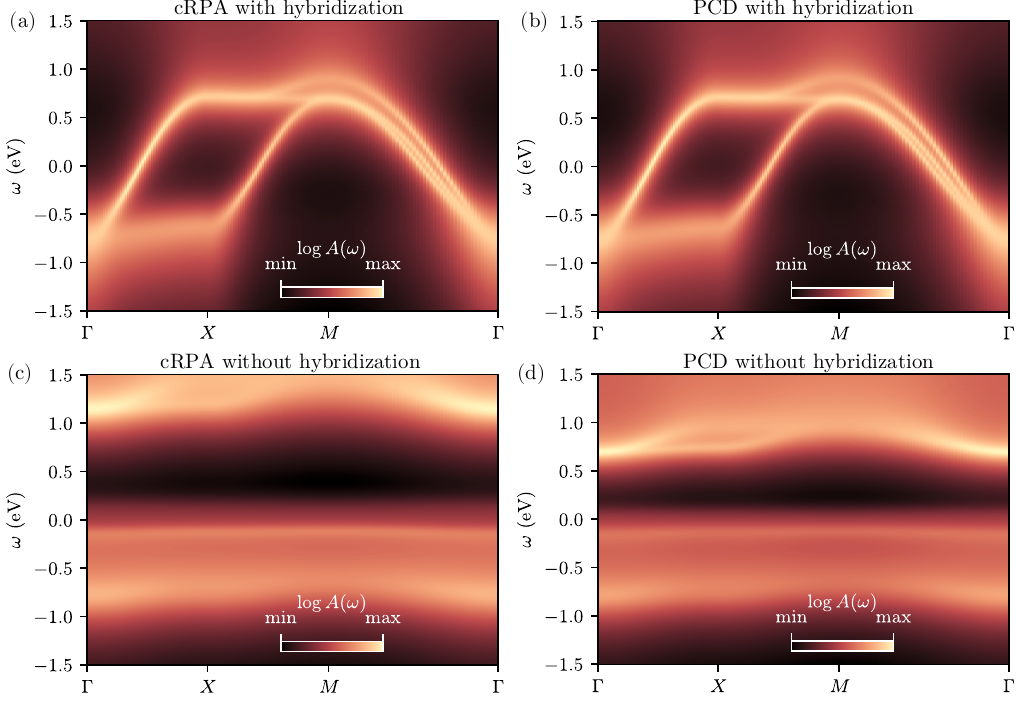}
    \caption{Panels (a) and (c) visualize the spectral function of the $t_{2g}$ subspace from DMFT starting with the downfolded action from cRPA with/without inclusion of the hybridization due to the p-orbitals in (a)/(c). Panels (b) and (d) visualize the spectral function of the $t_{2g}$ subspace from DMFT starting with the downfolded action from PCD with/without inclusion of the hybridization due to the p-orbitals in (b)/(d). We observe that depending on the chosen downfolding method the system is either in a Fermi-liquid metallic state or in a Mott insulating state.}
    \label{fig:SVO}
\end{figure*}

\section{Downfolding in prototypical multi-orbital metals}
In the following final example, we discuss the full impact of our downfolding protocol ranging from the derivation of target-space interactions to their impact upon solving the resulting target-space model and compare to the conventional cRPA scheme. To this end, we focus on correlated multi-orbital metals.

Multi-orbital metals are a class of materials in which the low-energy manifold consists of multiple bands. As a consequence, the Coulomb tensor does not only contain density-density type terms but also pair-hopping terms and spin-flip contributions~\cite{Yin2011, Georges_2013}. The strength of those interaction terms is typically given by the Hund's coupling $J$ following a Hubbard-Kanamori type parametrization of the interaction. In the following we compare different downfolding approaches and the respectively resulting Hund's coupling values and investigate how these in combination with the single-particle terms in the exact down-folding affect the target-space solutions. 

\subsection{The full model and the applied downfolding schemes}
The subsequent calculations are performed for a 14-band Wannier model for SrVO$_3$. SrVO$_3$ is a widely studied multi-orbital metal, which shows metallic behavior down to low temperatures~\cite{Brahlek2024, PhysRevB.58.4372, f4p8-5xb8}. The model consists of five V-$d$ orbitals split in the $t_{2g}$ and ${e_g}$ subspaces (which are orthogonal) as well as nine O-$p$ orbitals~\cite{Mushkaev_2024}. The hopping parameters are extracted from a VASP~\cite{Kresse1996,Hafner2008} calculation using Wannier90~\cite{Pizzi2020}. 
We include a Hubbard-Kanamori interaction on each atom, and a nearest neighbor density-density interaction between the Oxygen $p$-orbitals and the Vanadium $d$-orbitals. For details about the model we refer the reader to App.~\cref{App:model}. For the downfolding, we define the V $t_{2g}$-orbitals as the target and the V $e_g$- and O $p$-orbitals as the rest space. With this choice of model and spaces, $t_{2g}$ and $e_g$ orbitals couple via $B^{2:2}$ and $A^{2:2}$ while the $t_{2g}$ subspace couples to the $p$ orbitals mainly by the $A^{2:2}$ and $A^{1:1}$ coupling. This implies that at second order we only require up to $G^{(2)}$ in the rest space.

We downfold the model to the $t_{2g}$ manifold utilizing cRPA, cFRG in the form of Ref.~\cite{han2021investigation} and our proposed downfolding scheme, which we dub Perturbatively Controlled Downfolding (PCD). For PCD, we consider both a charge-channel RPA truncation of the diagrams and a version that retains all second-order contributions without truncation. Further, we focus on the static component of the resulting two-particle interactions.

All variants of the downfolding approaches are implemented utilizing the divERGe library~\cite{hauck2023diverge} implying only inclusion of all diagrams ending and starting in two-particle interactions. The missing diagrams from the full downfolding scheme are added by hand in a post-processing step. As indicated above, we carried out the truncation of the diagrams in two ways: one using a cRPA-style truncation (denoted as PCD@cRPA) and the other without any truncation (denoted as PCD). 

\subsection{Testing the rules}
\begin{figure}
    \centering
    \includegraphics[width=0.9\linewidth]{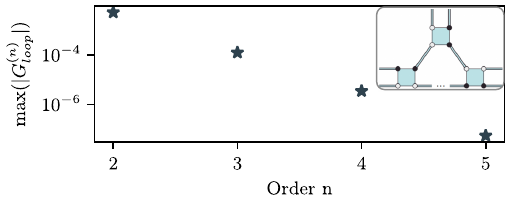}
    \caption{Maximal element of the $n$-point Green's function from loopless diagrams, as indicated in the inset. We observe a power law decay in the log plot as expected from the diagrammatic arguments and the analytic bound.}
    \label{fig:proxy}
\end{figure}

Before performing the downfolding calculations, we need to check whether the perturbative truncation of the downfolding hierarchy is justified. To this end, we need to verify that the rules outlined in \cref{ssec::howtoapprox} are fulfilled. By design, rule 2 is fulfilled, as we have neither $A^{3:1}$ nor $A^{1:3}$ couplings in the full model. Testing rule 1 technically requires  numerically calculating  $n$-point Green's functions of the effectively gapped rest-space. To avoid this numerical heavy calculation, we suggest estimating the size of these contributions by calculating selected contributions to the $n$-particle Green's functions. Specifically, we calculate the series of leading order ring diagrams in the static and local limit. Here, we exploit the fact that in an insulator the presence of a gap leads to exponential localization of the rest-space propagators in real space~\cite{loon2021random}, which renders such an approximate resummation feasible and insightful. To this end, we focus on the class of diagrams consisting of an $n$ two-particle vertices connected in a loopless fashion which translate into $\left(G \cdot V \right)^n$ (we suppressed internal matrix multiplications here). This is converted into a connected $n$-point Green's function by dressing dangling legs of the vertices by in total $2n$ Green's functions at order $n$. Thus, order by order we add a factor of $G \cdot V \cdot G^2$. Note that a general bound on perturbation theory can be proven in the weak coupling limit~\cite{Pedra_2008} and has a similar form. We summarize the results of our convergence test in \cref{fig:proxy}. We observe that already the two-particle connected Green's function is fairly small, supporting the truncation of the diagrammatic expansion at second order. 

\begin{table}
    \caption{Comparison of the local and static target space Hubbard interaction tensor components from different methods for obtaining screened two-particle quantities. All elements are given in eV.}
    \label{tab:int_vals}
    \begin{ruledtabular}
    \def\arraystretch{1.2}
    \begin{tabular}{cccccc}
                & bare& cRPA  & cFRG  & PCD@cRPA & PCD\\\hline
        $U$     & 5.00   & 3.96   & 3.70  & 3.31    & 3.52   \\
        $V_{dd}$& 3.80   & 2.76   & 2.47  & 2.11    & 2.34   \\
        $J_{pp}$& 0.60   & 0.60   & 0.53  & 0.60    & 0.55   \\
        $J_{ss}$& 0.60   & 0.60   & 0.67  & 0.60    & 0.63   \\
    \end{tabular}
    \end{ruledtabular}
\end{table}

\subsection{Comparing downfolding methods}
We summarize the results for downfolded two-particle interactions in \cref{tab:int_vals}. We observe that cRPA leads to substantial charge screening, while leaving the Hund's couplings unchanged. This however is not a general feature but a consequence of the model we picked. 
Adding the pairing and spin-channels, as done in cFRG, further reduces the local Coulomb interaction due to the screening nature of the pairing channel. More importantly, including these channels discriminates between the different Hund's couplings---the pairing channel acts as screening channel, while the spin channel is anti-screening~\cite{honerkamp2018limitations}. Therefore, the spin-flip contribution, which is mainly renormalized by the spin channel, is enhanced in the downfolding process. Further, the pair-hopping contribution, which is mainly renormalized by the pairing channel, is reduced. We stress that this differentiation is not a material specific feature, but instead we expect it to be systematically present due to the inclusion of the spin and pairing screening channels.

Comparing the cRPA calculation to its PCD@cRPA counterpart, i.e., removing the first diagram type in \cref{fig:cRPA_cross} from cRPA, we observe a decrease of the on-site Coulomb tensor. This diagram type is proportional to an off-diagonal propagator (since the propagators traverse between rest and target space), resulting in a case-specific overall-sign of the corresponding contribution. Thus, the contribution can be either screening or anti-screening depending on details of the calculation. Here, we find a  reduction the effective interaction by $\sim 0.7\,\mathrm{eV}$ by excluding this diagram. Including all other diagrams, we find that the on-site $U$ increases due to diagram (c3) in \cref{fig:diagrams}, which is essentially the precursor of the diagram responsible for the increase of the on-site interaction in cRPA. Comparing the PCD result to the PCD@cRPA approximation, we observe the same dichotomy of the Hund's coupling as in the cRPA/cFRG comparison, though slightly weaker due to the absence diagrams containing target space propagators and interactions.

\subsection{Solving the target space model}
To better understand how these differences in the target space models alter the predicted behavior we solve the resulting t$_{2g}$ three band models, once with cRPA-derived interactions and once with the interactions from PCD. In both cases we once exclude and once include the effective hybridization due to the kinetic coupling of the $p$ orbitals in the calculations. To this end, we perform dynamical mean-field calculations using the continuous-time hybridization expansion impurity solver implemented in \texttt{w2dynamics}~\cite{Gull_2011,Wallerberger_2019} at $\beta=20$, with the Anisimov double counting correction~\cite{Anisimov1997}. For the analytical continuation we utilize the maximal entropy method implemented in \texttt{ana\_cont}~\cite{Kaufmann2023}. The spectral functions for the two cases are visualized in \cref{fig:SVO}. We observe that depending on the chosen downfolding method to generate the t$_{2g}$ three band models, we either find a Fermi-liquid metallic state (PCD and cRPA with hybridization) or a Mott insulating state (cRPA and PCD without hybridization). Importantly, the hybridization arising from the kinetic coupling to the rest space enhances the kinetic energy within the target subspace. This effectively promotes delocalization, resulting in a more resilient metal. Thus, the treatment of the hybridization is crucial to obtain the correct physics. Beyond the pronounced impact of including or neglecting hybridization, switching the two-particle treatment between cRPA and PCD leads to comparatively minor changes. This reflects a partial error cancellation: In this specific example cRPA overestimates $U$ compared to PCD, which leads to enhanced correlations. On the other hand, the pair-hopping and spin-flip Hund's coupling are not split, which leads to a reduction of correlations. Thus overall, the two effects cancel out partially, resulting in only a minor change on the two-particle level.

\section{Conclusion}
We derived and discussed a general and exact formalism for downfolding to effective target-space models. Starting from a generic many-body Hamiltonian, we derived a diagrammatic expression for the effective action of the target-space model including interactions up to arbitrary orders. Within the formalism, we specified criteria that help in judging the quality of effective target-space models, focusing on whether such a model represents a faithful approximation to the real material:
First, the rest space has to have a rapidly convergent perturbative solution, as for example is the case in wide-gap rest spaces~\cite{loon2021random}. Second, there has to be a hierarchy of the different coupling contributions which ensures that three- and more-body interactions are suppressed. 
The former can be analyzed by solving the rest-space action by appropriate means ensuring that its $n>3$-particle connected Green's functions are negligible.
The latter can be checked by analyzing the $A$ and $B$ coefficients appearing in the coupling action, which can be calculated using available tools. This downfolding procedure we call perturbatively controlled downfolding (PCD).
While the physical observables themselves are independent of the choice of basis, we stress that the form of these couplings, and thereby the form and convergence properties of the target-space model can be strongly basis dependent (cf.~\cref{App::BvO}).

We further discussed both single- and two-particle terms arising in the downfolded model and compared to the well established cRPA on the level of the two-particle interaction. Notably, the proposed approach naturally resolves the problem of constructing effective models for cases in which target and rest space are entangled~\cite{PhysRev.126.413, PhysRevB.83.121101, PhysRevB.70.195104, Kltak} by the inclusion of all different couplings between the spaces.
Furthermore, it removes the ambiguity of the target space Luttinger-Ward functional depending on the non-interacting Green's function as it is the case when target space interactions are derived within cRPA.

Next, we calculated the different couplings from first principles in two prototypical examples: pure fcc nickel and an infinite layer cuprate. In both, we observed that one has to expect substantial retarded hoppings due to kinetic couplings between target and rest space. These findings indicate that for either of the two materials no faithful target-space model exists which includes only screened interactions. Furthermore, for both materials we found $A^{1:3}$ to be small. Finally, we demonstrated in a prototypical multi-orbital metal the significance of employing a perturbatively controlled downfolding scheme to accurately capture screening effects across the different channels.

In conclusion, to construct reliable effective target-space models, it is essential to carefully treat single-particle hybridization effects and to ensure a well chosen basis such that the model itself can be truncated to include only single- and two-particle terms. Otherwise a theoretical description of the discussed materials purely from the perspective of the corresponding target spaces can suffer from uncontrolled approximations.

As the next step, we envision building a post-\emph{ab initio} tool which automatizes the proposed downfolding scheme incorporating the convergence checks outlined in this manuscript. The starting point for such an implementation could be an orthonormal atomic orbital basis~\cite{FPLO1,FPLO2,buongiornonardelli2018paoflow, cerasoli2021advanced} exploiting symmetries~\cite{PhysRevB.108.115137}, but in principle any basis can be used~\cite{jolly2024tensorizedorbitalscomputationalchemistry}. For the perturbative treatment of the rest space we can rely on a plethora of highly efficient \emph{ab initio} many-body perturbation theory~\cite{GW2,hauck2023diverge} and \emph{ab initio} quantum chemistry codes~\cite{Friesner2005, Motta_2018}. 

\bigskip

\section*{Data availability}
The DFT simulations and wannierizations are publicly available under Ref.~\cite{NOMAD_dataset}.

\section*{Code availability}
The Simulation codes are either available upon reasonable request from the authors or licensed software packages (FPLO).

\begin{acknowledgments}
The authors are grateful to R.~da Silva Souza and T.~O.~Wehling for valuable discussions.
J.~P., R.~V., P.~P.~S. and L.~K. acknowledge support by the Deutsche Forschungsgemeinschaft (DFG, German Research Foundation) for funding through projects QUAST-FOR5249 (449872909) (projects P3 and P4). L.~K. acknowledges support by the Deutsche Forschungsgemeinschaft (DFG, German Research Foundation) for funding through SFB~1170 (258499086), and through the Würzburg-Dresden Cluster of Excellence on Complexity and Topology in Quantum Matter, \emph{ct.qmat} (EXC~2147, 390858490). R.~V. acknowledges support by the Deutsche Forschungsgemeinschaft (DFG, German Research Foundation) for funding through work was funded by the Deutsche Forschungsgemeinschaft (DFG, German Research Foundation) – CRC 1487,
“Iron, upgraded!” – project number 44370300.
J.~V. acknowledges funding provided by the Institute of Physics Belgrade, through the grant by the Ministry of Science, Technological Development and Innovation of the Republic of Serbia, as well as funding by the European Research Council, grant ERC-2022-StG: 101076100.
M.R. acknowledges support from the Vidi ENW research programme of the Dutch Research Council (NWO) [Grant DOI: 10.61686/YDRHT18202] with file number VI.Vidi.233.077.
\end{acknowledgments}

\section*{Author Contributions}
JBP conceived the idea. JBP, LK and JV performed the analytic calculations and JBP performed the DFT calculations. JBP and PPS wrote the code for the calculation of the interaction parameters. RV and LK supervised the work. JBP, JV, PPS, MR, RV and LK contributed in writing the manuscript. 

\section*{Competing Interest}
The Authors declare no Competing Financial or Non-Financial Interests.

\bibliography{bib}

\appendix
\clearpage
\onecolumngrid

\section{Splitting the interaction}
\label{App:split}
After defining target and rest space we have to restructure the interaction accordingly. For this, we start from
\begin{equation}
\label{eq:U}
    \sum_{1234} U_{1234} \bar{d}_3\bar{d}_4 d_2 d_1 = \sum_{1234} U_{1234} (\bar{f}_3 + \bar{c}_3)(\bar{f}_4 + \bar{c}_4) (f_2 + c_2) (f_1+c_1) \,,
\end{equation}
and multiply out the brackets. The terms can then be classified according to \cref{tab:interactionclasses}.
\begin{table}[h]
\caption{Classification of the different contributions arising in \cref{eq:U} and sorting into the different actions.}
\label{tab:interactionclasses}
\centering
\begin{ruledtabular}
\def\arraystretch{1.5}
\begin{tabular}{lcr}
$U_{1234} \bar{d}_3\bar{d}_4 d_2 d_1$ & representative & term in $\mathcal{A}$ (or $\color{bdiv2-6}S_f$, $\color{bdiv2-3}S_c$) \\ \hline
\arrayrulecolor{gray!60!white}
$\frac{1}{2}U_{1234}\bar{f}_3 \bar{f}_4 f_2 f_1$ & $\frac{1}{2}U_{1234}\bar{f}_3 \bar{f}_4 f_2 f_1$
                                                 & \color{bdiv2-6}$F_{1234}\bar{f}_3 \bar{f}_4 f_2 f_1$ \\ \hline
$\frac{1}{2}U_{1234}\bar{c}_3 \bar{f}_4 f_2 f_1$ & \multirow{2}{*}{$U_{1234}\bar{f}_3 \bar{c}_4 f_2 f_1$}
                                                 & \multirow{2}{*}{$\bar{f}_3 f_2 f_1 A^{3:1}_{1234}\bar{c}_4 $} \\ 
$\frac{1}{2}U_{1234}\bar{f}_3 \bar{c}_4 f_2 f_1$ &
                                                 & \\ \hline
$\frac{1}{2}U_{1234}\bar{f}_3 \bar{f}_4 c_2 f_1$ & \multirow{2}{*}{$U_{1234}\bar{f}_3 \bar{f}_4 f_2 c_1$}
                                                 & \multirow{2}{*}{$\bar{f}_3 \bar{f}_4 f_2 \tilde{A}^{3:1}_{1234} c_1$} \\ 
$\frac{1}{2}U_{1234}\bar{f}_3 \bar{f}_4 f_2 c_1$ &
                                                 & \\ \hline
$\frac{1}{2}U_{1234}\bar{c}_3 \bar{c}_4 f_2 f_1$ & $\frac{1}{2}U_{1234}\bar{c}_3 \bar{c}_4 f_2 f_1$
                                                 & $ f_2 f_1 \tilde{B}^{2:2}_{1234} \bar{c}_3 \bar{c}_4 $ \\ \hline
$\frac{1}{2}U_{1234}\bar{c}_3 \bar{f}_4 c_2 f_1$ & \multirow{4}{*}{$2U_{1234}\bar{c}_3 \bar{f}_4 f_2 c_1$}
                                                 & \multirow{4}{*}{$\bar{f}_4 f_2 A^{2:2}_{1234} \bar{c}_3 c_1 $} \\ 
$\frac{1}{2}U_{1234}\bar{c}_3 \bar{f}_4 f_2 c_1$ &
                                                 & \\ 
$\frac{1}{2}U_{1234}\bar{f}_3 \bar{c}_4 c_2 f_1$ &
                                                 & \\ 
$\frac{1}{2}U_{1234}\bar{f}_3 \bar{c}_4 f_2 c_1$ &
                                                 & \\ \hline
$\frac{1}{2}U_{1234}\bar{f}_3 \bar{f}_4 c_2 c_1$ & $\frac{1}{2}U_{1234}\bar{f}_3 \bar{f}_4 c_2 c_1$
                                                 & $\bar{f}_3 \bar{f}_4 {B}^{2:2}_{1234} c_2 c_1$ \\ \hline
$\frac{1}{2}U_{1234}\bar{c}_3 \bar{c}_4 c_2 f_1$ & \multirow{2}{*}{$U_{1234}\bar{c}_3 \bar{c}_4 f_2 c_1$}
                                                 & \multirow{2}{*}{$f_2 A^{1:3}_{1234} \bar{c}_3  \bar{c}_4 c_1 $} \\ 
$\frac{1}{2}U_{1234}\bar{c}_3 \bar{c}_4 f_2 c_1$ &
                                                 & \\ \hline
$\frac{1}{2}U_{1234}\bar{c}_3 \bar{f}_4 c_2 c_1$ & \multirow{2}{*}{$U_{1234}\bar{f}_3 \bar{c}_4 c_2 c_1$}
                                                 & \multirow{2}{*}{$\bar{f}_3 \tilde{A}^{1:3}_{1234} \bar{c}_4 c_2 c_1$} \\ 
$\frac{1}{2}U_{1234}\bar{f}_3 \bar{c}_4 c_2 c_1$ &
                                                 & \\ \hline
\arrayrulecolor{black}
$\frac{1}{2}U_{1234}\bar{c}_3 \bar{c}_4 c_2 c_1$ & $\frac{1}{2}U_{1234}\bar{c}_3 \bar{c}_4 c_2 c_1$
                                                 & \color{bdiv2-3}$C_{1234}\bar{c}_3 \bar{c}_4 c_2 c_1$
\end{tabular}
\end{ruledtabular}
\end{table}

\section{Derivation of the form of the Green's function generating functional}
\label{App:GFGF}
In \cref{eq:func} we found that the nontrivial contribution to the effective target-space action ($f$ Fermions) is given by a generalized Green's function generating functional. In the following we derive the leading orders of its Taylor expansion, see \cref{eq:Gff}. The functional to expand in powers of $f, \bar f$ is
\begin{equation}
    \mathcal G[f,\bar f] =\log\Big( \Braket{\exp\big[
    -\mathcal A[f,\bar f,c,\bar c]
    \big]}_{S_c}\Big) \equiv \log A[f,\bar f]\,.
\end{equation}
We recall that its Taylor coefficients are defined as
\begin{equation}
    \label{eq:taylorG}
    \mathcal G[f, \bar f] = \sum_{n=1}^\infty \underbrace{\frac{1}{(n!)^2} \frac{\delta^{2n} \mathcal G[f, \bar f]}
    {\delta f_{i_1}\cdots\delta f_{i_n}\delta \bar{f}_{\bar i_n} \cdots \delta \bar{f}_{\bar i_1} 
     }\Bigg|_{\{f=0\}}}_{G^{(n)}_{i_1,\dots,i_n;\bar{i}_n,\dots,\bar{i}_1}}
      \bar{f}_{\bar i_1}\cdots \bar{f}_{\bar i_n}  f_{i_n}\cdots f_{i_1}
     \,,
\end{equation}
We note that $\mathcal{A}$ is Grassmann even and so is $A$. Thus a singly differentiated $A$ is Grassmann-odd. 
Keeping this in mind, a formal consideration of the second derivative yields
\begin{equation}
     G^{(1)}_{i_1,\bar{i}_1} = \frac{\delta^2\log(A)}{\delta f_{i_1}\delta\bar f_{\bar{i}_1}}\Bigg|_{\{f=0\}} = \frac{\delta}{\delta f_{i_1}} \frac1{A} \frac{\delta A}{\delta\bar f_{\bar{i}_1}}\Bigg|_{\{f=0\}} = -\frac1{A^2} \frac{\delta A}{\delta f_{i_1}} \frac{\delta A}{\delta \bar f_{\delta \bar{i}_1}}\Bigg|_{\{f=0\}} 
     + \frac1{A} \frac{\delta^2 A}{\delta f_{i_1}\delta \bar f_{\bar{i}_1}}\Bigg|_{\{f=0\}} = \frac1{A} \frac{\delta^2 A}{\delta f_{i_1}\delta \bar f_{\bar{i}_1}}\Bigg|_{\{f=0\}} \,.
\end{equation}
The first summand either becomes zero when setting the sources to zero or contains an odd number of rest space fields, which is zero due to the charge conservation of the rest space theory. As a consequence, only the second summand remains, where the derivative can be taken \emph{inside} the expectation value (because the expectation value integrates over $c,\bar c$). In the same fashion, the fourth and higher derivatives can be obtained:
\begin{equation}
\begin{aligned}
\label{eq::G4}
     4 G^{(2)}_{i_1,i_2,\bar{i}_2,\bar{i}_1} &{}= \frac{\delta^4\log(A)}{\delta f_{i_2}\delta\bar f_{\bar{i}_2} \delta f_{i_1}\delta\bar f_{\bar{i}_1}}\Bigg|_{\{f=0\}} \\
     &= \begin{multlined}[t] \frac\delta{\delta f_{i_2}}  
     \Bigg[ 
     \frac2{A^3} \frac{\delta A}{\delta \bar f_{\bar{i}_2}}\frac{\delta A}{\delta f_{i_1}} \frac{\delta A}{\delta \bar f_{\delta \bar{i}_1}} 
     - 
     \frac1{A^2} \frac{\delta^2 A}{\delta \bar f_{\bar{i}_2}\delta f_{i_1}} \frac{\delta A}{\delta \bar f_{\delta \bar{i}_1}} 
     +
     \frac1{A^2} \frac{\delta A}{\delta f_{i_1}} \frac{\delta^2 A}{\delta \bar f_{\bar{i}_2} \delta \bar f_{\delta \bar{i}_1}} 
     +
     \frac1{A} \frac{\delta^3 A}{\delta \bar f_{\bar{i}_2}\delta f_{i_1}\delta \bar f_{\bar{i}_1}}
     -  {} \\
     \frac1{A^2}\frac{\delta A}{\delta \bar f_{\bar{i}_2}}  \frac{\delta^2 A}{\delta f_{i_1}\delta \bar f_{\bar{i}_1}}
     \Bigg]\Bigg|_{\{f=0\}}
     \end{multlined}
     \\
     &{} = \frac1A \frac{\delta^4 A}{\delta f_{i_2}\delta\bar f_{\bar{i}_2}\delta f_{i_1}\delta\bar f_{\bar{i}_1}}\Bigg|_{\{f=0\}} 
     - \frac1{A^2}\left[
    \frac{\delta^2 A}{\delta f_{i_2}\delta\bar f_{\bar{i}_2}} \frac{\delta^2 A}{\delta f_{i_1}\delta\bar f_{\bar{i}_1}} -
    \frac{\delta^2 A}{\delta f_{i_1}\delta\bar f_{\bar{i}_2}} \frac{\delta^2 A}{\delta f_{i_2}\delta\bar f_{\bar{i}_1}}
    \right]\Bigg|_{\{f=0\}}  \,.
\end{aligned}
\end{equation}
Here, we observe that the logarithm again implies a subtraction of disconnected parts. So we still generate a connected Green's function in the sense that the external legs have to be connected (even though the internal Green's functions in the rest space do not necessarily have to). We recall that the generalized Wick theorem connecting moment, denoted by single brackets $\braket{...}$, and cumulants, denoted by double brackets $\langle\langle ...\rangle\rangle$,
allows to decompose expectation values in connected Green's functions resulting in~\cite{Helias_2020}
\begin{multline}
\label{eq:wick}
    \braket{\bar{c}_{\bar 1} \cdots \bar{c}_{\bar k}c_k \cdots  c_1}_{S_c} = \sum_{n=1}^{k} \quad 
    \sum_{\substack{1 \leq l_i \leq k \\ \sum_i l_i = k}}\sum_{\substack{1 \leq \bar{l}_i \leq \bar{k} \\ \sum_i \bar{l}_i = \bar k}} \sum_{\sigma \in P(\{l_i,\bar{l}_i\},k)} (-1)^P \langle\langle \bar{c}_{\sigma(\bar{1})} \cdots \bar{c}_{\sigma(\bar l_1)}c_{\sigma({l}_1)} \cdots c_{\sigma(1)} \rangle\rangle \cdots {}\\[-1.7em]
    {} \cdots \langle\langle \bar{c}_{\sigma(\bar k-\bar l_n-1)} \cdots \bar{c}_{\sigma(\bar k)}c_{\sigma(k)}...c_{\sigma(k-l_n-1)} \rangle\rangle \,.
\end{multline}
This expression allows us to rewrite the moments in terms of connected Green's functions of the rest space. Thereby, disconnected graphs constructed from lower order subgraphs are canceled. 

With the above realization that we keep the general form of a connected Green's function, we arrive for the three particle Green's function at
\begin{equation}
\begin{aligned}
    G^{(3)}_{i_1,i_2,i_3,\bar{i}_3,\bar{i}_2,\bar{i}_1} &{}= \frac1A\frac{\delta^6 A}{\delta f_{i_3}\delta\bar f_{\bar{i}_3}\delta f_{i_2}\delta\bar f_{\bar{i}_2}f_{i_1}\delta\bar f_{\bar{i}_1}} \Bigg|_{\{f=0\}} 
    - \frac1{A^2} \mathcal P\left\{ 
    \frac{\delta^2 A}{\delta f_{i_3}\delta\bar f_{\bar{i}_3}}\frac{\delta^4 A}{\delta f_{i_2}\delta\bar f_{\bar{i}_2}f_{i_1}\delta\bar f_{\bar{i}_1}} 
    \right\}\Bigg|_{\{f=0\}}  \\ &{}+ 2\frac1{A^3} \mathcal P \left\{ \frac{\delta^2 A}{\delta f_{i_3}\delta\bar f_{\bar{i}_3}}\frac{\delta^2 A}{\delta f_{i_2}\delta\bar f_{\bar{i}_2}}\frac{\delta^2 A}{\delta f_{i_1}\delta\bar f_{\bar{i}_1}} \right\}\Bigg|_{\{f=0\}} \,
\end{aligned}
\end{equation}
where $\mathcal P$ denotes a short hand for all possible permutations we can reach from the base case which still have the same structure including the number of swaps inducing a minus sign each.

To evaluate the expressions above we require the derivatives of $A$ explicitly. As a reminder, $A$ is in general of the form

\begin{equation}
\begin{aligned}
\label{eq:cplact}
    A[f,\bar{f}] = \Big\langle\exp&\left[-(\bar{f}_2 A^{1:1}_{12} c_1 - f_1 \tilde{A}^{1:1}_{12} \bar c_2 + \bar{f}_3\bar{f}_4 f_2 A^{3:1}_{1234} c_1 + \bar{f}_3{f}_2 f_1 \tilde{A}^{3:1}_{1234} \bar c_4 +  \bar{f}_4f_2 A^{2:2}_{1234} \bar{c}_3c_1\right. \\ &\left.+ \bar{f}_3\bar{f}_4 B^{2:2}_{1234} c_2c_1 + {f}_2{f}_1 \tilde{B}^{2:2}_{1234} \bar{c}_3\bar{c}_4 + \bar{f}_3 A^{1:3}_{1234} \bar c_4 c_2 c_1 + {f}_2 \tilde{A}^{1:3}_{1234} \bar c_3 \bar c_4 c_1 \right]\Big\rangle_{S_c}
\end{aligned}
\end{equation}
The second order derivative reads
\begin{equation}
\begin{aligned}
     \frac{\delta^2 A}{\delta f_{i_1} \delta\bar f_{\bar{i}_1}}\Bigg|_{\{f=0\}} &= \Braket{\left(-\frac{\delta^2 \mathcal A}{\delta f_{i_1} \delta\bar f_{\bar{i}_1}} 
    - \frac{\delta \mathcal A}{\delta\bar f_{\bar{i}_1}} \frac{\delta \mathcal A}{\delta f_{i_1}}  \right) e^{-\mathcal A[f,\bar f,c,\bar c]} }\Bigg|_{\{f=0\}} \\
     &= - \Big\langle \left(-2 \bar{f}_4  A_{1,{i_1},\bar{i}_1,4}^{3:1} c_1 +
     f_1 A_{1,{i_1},\bar{i}_1,4}^{3:1} \bar c_4 + A^{2:2}_{1{i_1}3\bar{i}_1} \bar{c}_3c_1  \right)
     \Big\rangle_{S_c} \Bigg|_{\{f=0\}} \\
     &\quad - \Big\langle \left(A_{1,\bar{i}_1}^{1:1}c_1 + 2 \bar{f}_4 f_2 A_{1,2,\bar{i}_1,4}^{3:1} c_1 +
     f_2 f_1 A_{1,2,\bar{i}_1,4}^{3:1} \bar c_4 + f_2 A^{2:2}_{123\bar{i}_1} \bar{c}_3c_1  \right. 
     \\
     &\quad \quad \quad\quad \left.+ 2\bar{f}_4 B^{2:2}_{12\bar{i}_14} c_2c_1 + A^{1:3}_{12\bar{i}_14} \bar c_4 c_2 c_1  \right) 
     \\
     &\quad \quad \quad \times \left((-\bar{c}_{2'} \tilde{A}^{1:1}_{i_1{2'}} + \bar{f}_{3'}\bar{f}_{4'} A^{3:1}_{1'i_13'4'} c_{1'} + 2\bar{f}_{3'}{f}_{2'} \tilde{A}^{3:1}_{i_1 2'3'4'} \bar c_{4'} -  \bar{f}_{4'} A^{2:2}_{1'i_13'4'} \bar{c}_{3'}c_{1'} \right. \\ 
    &\quad \quad \quad\quad+ \left. 2{f}_{1'} B^{2:2}_{1'i_13'4'} \bar{c}_{3'}\bar{c}_{4'} + \tilde{A}^{1:3}_{1'i_13'4'} \bar c_{3'} \bar c_{4'} c_{1'} )\right) 
     \Big\rangle_{S_c} \Bigg|_{\{f=0\}} \\
    &= A^{2:2}_{1{i_1}3\bar{i}_1} \braket{ c_1\bar{c}_3}_{S_c} + A_{1,\bar{i}_1}^{1:1} \tilde{A}^{1:1}_{i_1{2'}} \braket{ c_1\bar{c}_{2'}}_{S_c} - A_{1,\bar{i}_1}^{1:1}\tilde{A}^{1:3}_{1'i_13'4'}  \braket{\bar c_{3'} \bar c_{4'} c_1 c_{1'}}_{S_c} + \tilde{A}^{1:1}_{i_1{2'}} A^{1:3}_{12\bar{i}_14} \braket{\bar c_4 \bar{c}_{2'} c_2 c_1 }_{S_c} \\
    & \quad - A^{1:3}_{12\bar{i}_14}\tilde{A}^{1:3}_{1'i_13'4'} \braket{ \bar c_4 \bar c_{3'} \bar c_{4'} c_2 c_{1'} c_{1}}_{S_c}
\end{aligned}
\end{equation}
These are all diagrams with which we obtain the correct number of target space fields. Since the rest space theory is typically charge conserving we directly conclude that the insertion of \cref{eq:wick} cannot generate disconnected graphs. 

Due to the large number of different source fields, the number of terms also rapidly increases. We made use of Mathematica to generate the fourth order contribution. The convention used here is that we calculate
\begin{equation}
    \frac{\delta^4 A}{\delta f_{i_2} \delta\bar f_{\bar{i}_2}\delta f_{i_1} \delta\bar f_{\bar{i}_1}}\Bigg|_{\{f=0\}}\,,
\end{equation}
keeping in mind that penultimately $f$ fields will be added and a summation is added. As shown in \cref{fig:diagrams}, the two-particle interaction consists of 25 topologically distinct diagrams, which each can be associated to an analytical term classified according to the different vertices contained. 
First we have a term in which only $A^{1:1}$ appears, cf.~\cref{fig:diagrams} (e2),
\begin{equation}
    \label{eq::4:1}
    \braket{\bar{c}_{4}\bar{c}_{3}c_{2}c_{1}}_{S_c}  A^{1:1}_{2,{\bar{i}_1}} \tilde{A}^{1:1}_{i_1,3} A^{1:1}_{1,{\bar{i}_2}} \tilde{A}^{1:1}_{i_2,4}\,.
\end{equation}
Notably, when utilizing the generalized Wick theorem \cref{eq:wick} we observe that the disconnected parts are canceled as they are contained within the second order.
Next, we can replace any number of $A^{1:1}$($\tilde{A}^{1:1}$) vertices by $A^{1:3}$($\tilde{A}^{1:3}$) vertices without changing the number of external legs in the target space. Introducing a single $A^{3:1}$ we have two different terms 

\begin{equation}
    \begin{aligned}
    &-\langle \bar{c}_{4_a}c_{1_a}c_{2_a}c_{1_d}\bar{c}_{2_e}\bar{c}_{2_b} \rangle_{S_c}\tilde{A}^{1:1}_{i_1,2_b}\tilde{A}^{1:1}_{i_2,2_e}A^{1:1}_{1_d,\bar{i}_2}A^{1:3}_{1_a,2_a,\bar{i}_1,4_a}
-\langle c_{1_a}\bar{c}_{4_d}c_{1_d}c_{2_d}\bar{c}_{2_e}\bar{c}_{2_b} \rangle_{S_c}\tilde{A}^{1:1}_{i_1,2_b}\tilde{A}^{1:1}_{i_2,2_e}A^{1:1}_{1_a,\bar{i}_1}A^{1:3}_{1_d,2_d,\bar{i}_2,4_d} \\ 
&=\langle \bar{c}_{4_a}\bar{c}_{2_e}\bar{c}_{2_b}c_{1_a}c_{1_d}c_{2_a} \rangle_{S_c} \left(\tilde{A}^{1:1}_{i_1,2_b}\tilde{A}^{1:1}_{i_2,2_e}A^{1:1}_{1_d,\bar{i}_2}A^{1:3}_{1_a,2_a,\bar{i}_1,4_a} - \tilde{A}^{1:1}_{i_1,2_b}\tilde{A}^{1:1}_{i_2,2_e}A^{1:1}_{1_d,\bar{i}_1}A^{1:3}_{1_a,2_a,\bar{i}_2,4_a}\right)\,.
\end{aligned}
\end{equation}
Using here, that we can swap $\bar{i}_1$ and $\bar{i}_2$ by changing the attached $f$-field order and relabeling, we find

\begin{equation}
    \begin{aligned}
    2\langle \bar{c}_{4_a}\bar{c}_{2_e}\bar{c}_{2_b}c_{1_a}c_{1_d}c_{2_a} \rangle_{S_c} \tilde{A}^{1:1}_{i_1,2_b}\tilde{A}^{1:1}_{i_2,2_e}A^{1:1}_{1_d,\bar{i}_2}A^{1:3}_{1_a,2_a,\bar{i}_1,4_a}\,.
\end{aligned}
\end{equation}
and analogously for all other terms where we replace $A^{1:1}$ by $A^{3:1}$ (cf.~\cref{fig:diagrams} (e3-5) and (f1-5)) such that in total these diagrams contribute as:
\begin{equation}
\label{eq::4:2}
    \begin{aligned}
    \langle \bar{c}_{6}\bar{c}_{5}\bar{c}_{4}c_{3}c_{2}c_{1} \rangle_{S_c}& \left(2\tilde{A}^{1:1}_{i_1,4}\tilde{A}^{1:1}_{i_2,5}A^{1:1}_{2,\bar{i}_2}A^{1:3}_{3,1,\bar{i}_1,6} + 2\tilde{A}^{1:1}_{i_2,6}\tilde{A}^{1:3}_{2,i_1,5,4}A^{1:1}_{3,\bar{i}_1}A^{1:1}_{1,\bar{i}_2} \right) \\
+\langle \bar{c}_{8}\bar{c}_{7}\bar{c}_{6}\bar{c}_{5}c_{4}c_{3}c_{2}c_{1}\rangle_{S_c}& \Big(\tilde{A}^{1:1}_{i_1,5}\tilde{A}^{1:1}_{i_2,6}A^{1:3}_{4,3,\bar{i}_1,8}A^{1:3}_{2,1,\bar{i}_2,7} +A^{1:1}_{4,\bar{i}_1}A^{1:1}_{3,\bar{i}_2}A^{1:3}_{1,i_1,6,5}A^{1:3}_{2,i_2,8,7} \\ & + 4 A^{1:1}_{2,\bar{i}_2}\tilde{A}^{1:1}_{i_2,7}A^{1:3}_{4,3,\bar{i}_1,8}\tilde{A}^{1:3}_{1,i_1,6,5}\Big)\\  
+\langle \bar{c}_{10}\bar{c}_{9}\bar{c}_{8}\bar{c}_{7}\bar{c}_{6}c_{5}c_{4}c_{3}c_{2}c_{1} \rangle_{S_c}& \left( 2A^{1:1}_{3,\bar{i}_2}A^{1:3}_{5,4,\bar{i}_1,10}\tilde{A}^{1:3}_{1,i_1,6,7}\tilde{A}^{1:3}_{2,i_2,9,8} + 2\tilde{A}^{1:1}_{i_1,6}A^{1:3}_{5,4,\bar{i}_1,10}A^{1:3}_{3,2,\bar{i}_2,7}\tilde{A}^{1:3}_{1,i_2,9,8}  \right)\\ 
+\langle \bar{c}_{12}\bar{c}_{11}\bar{c}_{10}\bar{c}_{9}\bar{c}_{8}\bar{c}_{7}c_{6}c_{5}c_{4}c_{3}c_{2}c_{1} \rangle_{S_c}&A^{1:3}_{6,5,\bar{i}_1,12}A^{1:3}_{1,i_1,8,7}A^{1:3}_{4,3,\bar{i}_2,11}A^{1:3}_{2,i_2,10,9} \\ 
\end{aligned}
\end{equation}

Turing to diagrams containing $A^{2:2}$ (cf.~\cref{fig:diagrams} (b5), (c3), (c4), (d3) and (d5)) we find
\begin{equation}
\label{eq::4:3}
    \begin{aligned}
        \langle \bar{c}_{4}\bar{c}_{3}c_{2}c_{1} \rangle_{S_c}& \left( 2 A^{2:2}_{2,i_2,4,\bar{i}_1}A^{2:2}_{1,i_1,3,\bar{i}_2} + 
        4\tilde{A}^{1:1}_{i_1,3}A^{1:1}_{1,\bar{i}_2}A^{2:2}_{2,i_2,4,\bar{i}_1} \right) \\
+\langle \bar{c}_{6}\bar{c}_{5}\bar{c}_{4}c_{3}c_{2}c_{1} \rangle_{S_c}& \left(4\tilde{A}^{1:1}_{i_2,4}A^{1:3}_{2,1,\bar{i}_2,5}A^{2:2}_{3,i_1,6,\bar{i}_1} +4 A^{1:1}_{2,\bar{i}_2}A^{1:3}_{1,i_2,5,4}A^{2:2}_{3,i_1,6,\bar{i}_1}  \right)\\ 
+\langle \bar{c}_{8}\bar{c}_{7}\bar{c}_{6}\bar{c}_{5}c_{4}c_{3}c_{2}c_{1} \rangle_{S_c}& \left(4 A^{1:3}_{4,3,\bar{i}_1,8}\tilde{A}^{1:3}_{1,i_2,6,5}A^{2:2}_{2,i_1,7,\bar{i}_2}\right)
    \end{aligned}
\end{equation}

Which completes all diagrams constructable without having a $B^{2:2}$ or $A^{3:1}$ vertex. Let us first construct those equations corresponding to diagrams containing $B^{2:2}$ (cf.~\cref{fig:diagrams} (c1), (c2), (c5), (d1), (d2), (d4) and (e1)):
\begin{equation}
\label{eq::4:4}
    \begin{aligned}
    \langle \bar{c}_{4}\bar{c}_{3}c_{2} c_{1}\rangle_{S_c}&\left( 2\tilde{A}^{1:1}_{i_1,3}\tilde{A}^{1:1}_{i_2,4}B^{2:2}_{1,2,\bar{i}_1,\bar{i}_2} +2A^{1:1}_{1,\bar{i}_1}A^{1:1}_{2,\bar{i}_2}\tilde{B}^{2:2}_{i_2,i_1,4,3} + 4B^{2:2}_{1,2,\bar{i}_1,\bar{i}_2}\tilde{B}^{2:2}_{i_2,i_1,3,4} \right)\\
    +\langle \bar{c}_{6}\bar{c}_{5}\bar{c}_{4}c_{3}c_{2}c_{1} \rangle_{S_c}&\left(4\tilde{A}^{1:1}_{i_2,6}\tilde{A}^{1:3}_{1,i_1,5,4}B^{2:2}_{3,2,\bar{i}_1,\bar{i}_2}+4A^{1:1}_{1,\bar{i}_2}A^{1:3}_{3,2,\bar{i}_1,4}\tilde{B}^{2:2}_{i_2,i_1,5,6}\right) \\
\langle \bar{c}_{8}\bar{c}_{7}\bar{c}_{6}\bar{c}_{5}c_{4}c_{3}c_{2}c_{1} \rangle_{S_c}&\left(2\tilde{A}^{1:3}_{1,i_1,6,5}\tilde{A}^{1:3}_{2,i_2,7,8}B^{2:2}_{4,3,\bar{i}_1,\bar{i}_2}+2A^{1:3}_{4,3,\bar{i}_1,6}A^{1:3}_{2,1,\bar{i}_2,5}\tilde{B}^{2:2}_{i_2,i_1,7,8} \right) 
    \end{aligned}
\end{equation}
And finally, we construct the terms containing $A^{3:1}$ (cf.~\cref{fig:diagrams} (b1-b4)):
\begin{equation}
\label{eq::4:5}
    \begin{aligned}
        \langle \bar{c}_{2}c_{1} \rangle_{S_c}& \left(4A^{1:1}_{1,\bar{i}_2}\tilde{A}^{3:1}_{i_2,i_1,\bar{i}_1,2}+4\tilde{A}^{1:1}_{i_2,2}A^{3:1}_{1,i_1,\bar{i}_1,\bar{i}_2} \right)\\  
        +\langle \bar{c}_{4}\bar{c}_{3}c_{2}c_{1} \rangle_{S_c}&\left(4A^{1:3}_{1,2,\bar{i}_2,3}\tilde{A}^{3:1}_{i_2,i_1,\bar{i}_1,4}+4\tilde{A}^{1:3}_{1,i_2,4,3}A^{3:1}_{2,i_1,\bar{i}_1,\bar{i}_2}  \right)
    \end{aligned}
\end{equation}

By combining \crefrange{eq::4:1}{eq::4:5}, we obtain the first term of the Taylor coefficient in \cref{eq::G4}. Additionally we have to subtract the disconnected part constructable from the lower orders, which manifests itself by rendering all contributions constructable by lower order combinations to be connected, cf.~\cref{fig:cancellation}. In second order, we additionally have no way to connect the newly appearing diagram types in a disconnected fashion. Therefore, there are no disconnected diagrams appearing in the expansion in the sense that they can be subdivided into disconnected vertices. We again stress, that this does not mean that the internal rest space propagator has to be a connected Green's function. In most cases this is not the case.  
\subsection{Diagrammatic rules}
\label{App:diagrules}
From \crefrange{eq:cplact}{eq::4:5} we can deduce the rules to translate diagrams into equations with the correct prefactors:
\begin{enumerate}
    \item At order $n$ construct all unique combinations of vertices with $n$ in-going and $n$ out-going legs in the target space.
    \item For each unique combination of vertices, we get a prefactor of $\#\mathrm{perm}/(n!)^2$. $\#\mathrm{perm}$ is the magnitude of the set of all permutations of the assigned target space indices and their connectivity which are are inequivalent under exchanging vertices. 
    \item To get the correct sign, we have to define a representative diagram of the group. We pick one specific permutation and label all in-going legs from $i_n$ to $i_1$ and all out-going legs from $\bar{i}_n$ till $\bar{i}_1$.
    \begin{enumerate}
        \item\label{rule:derivative} From the derivatives of the exponential we get a prefactor of $(-1)^{\# \text{vertices}}\times (-1)^{\# \tilde{A}^{1:1}}$. 
        \item\label{rule:simple_sign} To count the number of permutations of derivatives, we count the permutations to bring the index order of the specific permutation to the index order $\bar{i}_1\dots \bar{i}_n {i}_n \dots {i}_1$. In total we get $(-1)^{\mathcal{P}}$. Additionally, we get a $(-1)^{\# A^{-:\mathrm{odd}}}$ where $\# A^{-:\mathrm{odd}}$ is the number of vertices with an odd number of rest space fields. 
        \item To collect the rest space expectation value, we start with the rightmost vertex and write the corresponding rest space fields leftmost in the expectation-value. Then the vertex to the left of the rightmost diagram contributes the next fields and so on and so forth. 
    \end{enumerate}
\end{enumerate}
Therefore in total we find
\begin{equation}
\label{eq:prefac}
    \frac{\#\mathrm{perm}}{(n!)^2} (-1)^{\# \text{vertices}}\times (-1)^{\# \tilde{A}^{1:1}}\times (-1)^{\mathcal{P}}\times (-1)^{\# A^{-:\mathrm{odd}}}
\end{equation}

\begin{figure}[!hbt]
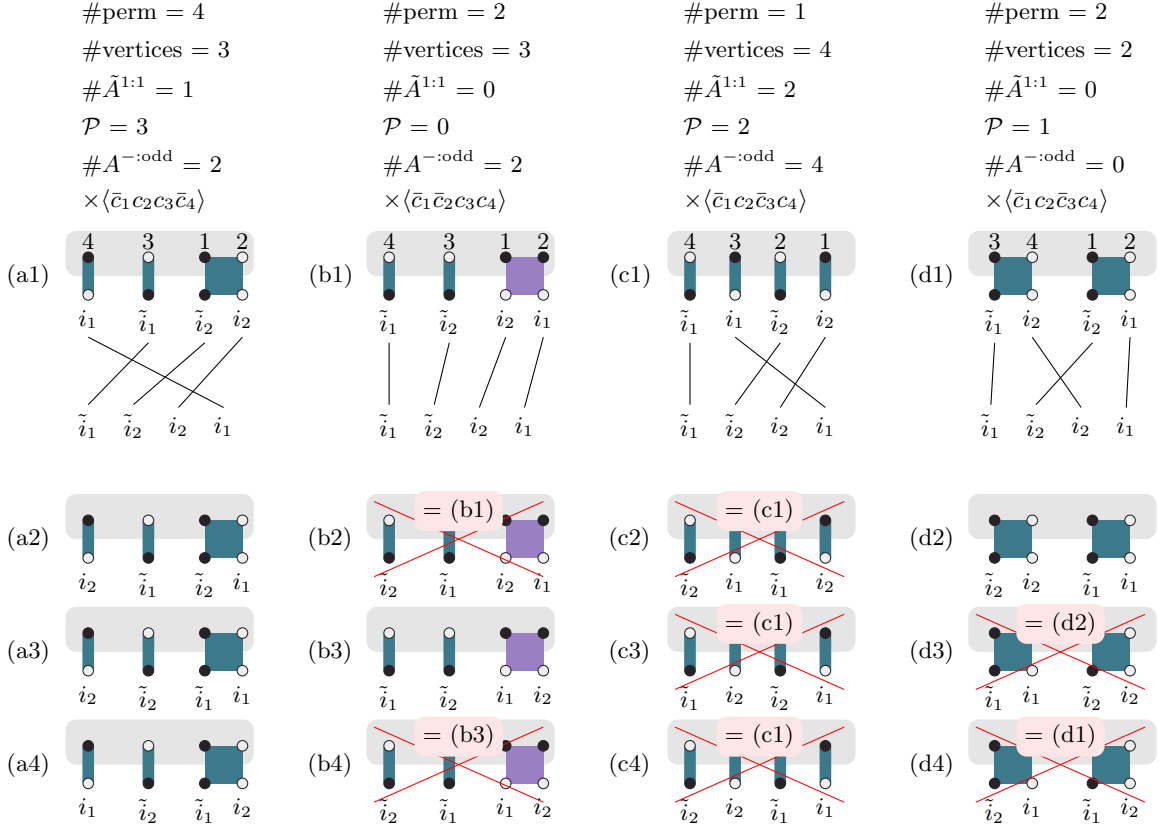

    \centering
    \constructSign
    \caption{Selection of diagrams for which we construct the prefactor and the assigned expectation value. The permutations $\mathcal{P}$ can be constructed graphically by the number of crossings between connecting the same indices in the assigned index set and the reference index set. Additionally, we draw all possible labelings and explicitly cross out those that are equivalent to another permutation under exchanging vertices.}
    \label{fig:example_rules}
\end{figure}

As examples for the prefactor, let us consider diagram (c1) in \cref{fig:diagrams}. For this, we can either label the in-(out-) going as $(i_1,i_2,\bar{i}_2,\bar{i}_1)$, $(i_2,i_1,\bar{i}_2,\bar{i}_1)$, $(i_1,i_2,\bar{i}_1,\bar{i}_2)$ or $(i_2,i_1,\bar{i}_1,\bar{i}_2)$. Each of these possibilities corresponds to a topological distinct connectivity as the number of crossings is different for each contribution and no undistinguishable vertices appear, so we have a prefactor of $4$. As another example, consider (c2). Here we replace one $B^{2:2}$ vertex by two $A^{1:1}$ vertices. These are indistinguishable and exchanging their leg labeling does not lead to a topological distinct connectivity. Therefore, the prefactor is only $2$. If we were to replace one of the $A^{1:1}$ vertices with a $A^{1:3}$ one, they are again distinguishable and we find again a prefactor of $4$. For a graphical construction, see \cref{fig:example_rules}.

While it is in principle possible to rewrite \crefrange{eq:cplact}{eq::4:5} in terms of connected $n$-particle Green's functions of the rest space, it would lead to even more terms and is therefore not instructive. Graphically, we get additional minus signs due to the permutations necessary to restructure the rest space fields. 

\section{Band space vs. orbital space}
\label{App::BvO}
We formulated the theory in orbital instead of band space for three reasons: First, the two-particle interaction is not gauge invariant and no general algorithm to comb the sign structure of the orbital-to-band transformation is known. While this is not problematic for pure density-density type interactions, in systems with delocalized orbitals there might be contributions to more general interaction matrix elements which change under gauge.
Second, bands that mix different orbitals can generate large anomalous interaction components. To see this, consider two localized orbitals that form bonding and anti-bonding states,
\begin{equation}
    \upphi^{b/a} = \alpha_{b/a} \mathsf{d}_1 + \beta_{b/a} \mathsf{d}_2 \,,
\end{equation}
with $\mathsf{d}^{(\dagger)}_{1,2}$ the two annihilation (creation) operators of the localized orbitals. When calculating Coulomb matrix elements, we immediately observe that $A^{3:1}$ in band space (the $\upphi^{b/a}$ basis) is proportional to 
\begin{equation}
    A^{3:1}_{bbba} \propto \alpha_{b}^3 \alpha_a A^{2:2}_{1111} + \beta_{b}^3 \beta_a A^{2:2}_{2222}\,,
\end{equation}
i.e., density-density interactions $A^{2:2}$ in orbital space ($\mathsf{d}_{1/2}$). Hence the unitary transformation to $\upphi^{b/a}$ can substantially alter the structure of the target space Hamiltonian. While this at first glance sounds counterintuitive we have to keep in mind that integrating out parts of the system renders the unitary transformation from orbital to band space non-invertible. In other words, the target-space model depends on the basis choice in the full space. In band space, we expect the diagrammatic expansion to be less controlled due to anomalous vertices becoming (potentially) larger. 
Lastly, the orbital to band transformation render the interaction strongly momentum dependent due to the momentum dependence of the basis change itself. 
Therefore, we mainly discuss orbital space in the main text. 
An advantage of a band space formulation on the other hand is the absence of $A^{1:1}$-type couplings.

\section{Exactly solvable toy model}
\label{App:simple}
The simplest possible example we can apply this methodology to is a $(0+0)$D field theory. While in a sense it is trivial as it is exactly solvable, it still illustrates the central ideas of the approach. As a starting point, we define
\begin{equation}
    S[f,\bar{f},c,\bar{c}] = \underbrace{{}-\mu_{f}\sum_{\sigma} \bar{f}_{\sigma}f_{\sigma} + U \bar{f}_{\uparrow}f_{\uparrow}\bar{f}_{\downarrow}f_{\downarrow}}_{S_f}
    \underbrace{{}-\mu_{c}\sum_{\sigma} \bar{c}_{\sigma}c_{\sigma} + \tilde{U} \bar{c}_{\uparrow}c_{\uparrow}\bar{c}_{\downarrow}c_{\downarrow}}_{S_c}
    + \underbrace{C \sum_{\sigma}(\bar{f}_{\sigma}f_{\sigma}\bar{c}_{\bar{\sigma}}c_{\bar{\sigma}})}_{\mathcal A}
\end{equation}
as the total action. We will in the following integrate out the $c$ fields following the formalism above. 
The full partition function is given by
\begin{equation}
    \mathcal Z = \int \mathcal Df\mathcal{D}\bar{f} \int \mathcal Dc\mathcal D\bar{c}\, e^{-S[f,\bar{f},c,\bar{c}]} = (\mu_{f}^2-U)(\mu_{c}^2-\tilde{U}) - 2\mu_{f}\mu_{c}C + C^2 \,,
\end{equation}
and the target space Green's function is given by
\begin{equation}
    G_f = \int \mathcal Df\mathcal{D}\bar{f} \int \mathcal Dc\mathcal D\bar{c}\,\bar{f}_{\sigma}f_{\sigma} e^{-S[f,\bar{f},c,\bar{c}]} = \frac{(\mu_{c}^2-\tilde{U})\mu - \mu_{c}C}{\mathcal Z} \,.
\end{equation}

We can also obtain this target space Green's function by first integrating out the rest space fields. The non-vanishing cumulants of the rest space ($c$) theory read
\begin{align}
    G_c^{(2)} &{}= \frac{1}{\mathcal Z_c}\int \mathcal Dc\mathcal D\bar{c}\, \bar{c}_\sigma c_\sigma e^{-S_c[c,\bar{c}]} = \frac{\mu_{c}}{\mu_{c}^2 -\tilde{U}} \,, \\
    G_c^{(4)} &{}= \frac{1}{\mu_{c}^2 -\tilde{U}} -  \big( G^{(1)} \big)^2 \,, \\
    G_c^{(n>4)} &{}= 0 \,.
\end{align}
Therefore, the effective field theory is given as
\begin{equation}
\begin{aligned}
    S_{\mathrm{eff}}[f,\bar{f}] &{}= S_f - G_c^{(2)} \sum_{\sigma} C \bar{f}_{\sigma}f_{\sigma} + C^2 \big(G_c^{(4)}-(G_c^{(2)})^2\big) \bar{f}_{\uparrow}f_{\uparrow}\bar{f}_{\downarrow}f_{\downarrow} \\
    &{}= \big(\mu_{f}-G_c^{(2)}C\big) \sum_{\sigma} \bar{f}_{\sigma}f_{\sigma} - \big(U-C^2 G_c^{(4)} + (G_c^{(2)})^2\big) \bar{f}_{\uparrow}f_{\uparrow}\bar{f}_{\downarrow}f_{\downarrow} \,.
\end{aligned}
\end{equation}
Thus we directly find 
\begin{equation}
    \mathcal Z_{\mathrm{eff}} = \frac{1}{\mathcal Z_c} \mathcal Z\,.
\end{equation}
The prefactor drops out of all observables and thereby we exactly recover the Green's function in the target space with the effective theory.

\section{Calculating interaction matrix elements}
\label{App:coulombmat}
To calculate the interaction matrix elements, we first perform a Wannierization and store the real-space Wannier functions on a common origin. Here, we have to ensure that the real space box is big enough such that all Wannier functions decay to zero within it. Whenever we multiply two Wannier functions with shifted centers $\bvec{R}$ we have to consider only those belonging to the same lattice distance box, as out-of-box values do not generate overlap. 
Since this box typically is determined by lattice vectors, we additionally have to include the base change Jacobian.
Next we calculate 
\begin{equation}
    U_{1234} = \braket{\psi_3 \psi_4 |\hat{H}_{pot}|\psi_2 \psi_1} = \int \dd\bvec{r} \dd\bvec{r'} \;V(|\bvec{r}-\bvec{r'}|) \psi_3(\bvec{r})^* \psi_4(\bvec{r}')^*\psi_2(\bvec{r}') \psi_1(\bvec{r}) \;,
\end{equation}
by the Fourier transformation method~\cite{schnell2002ab,PhysRevB.94.045440}. For this, we first use translational invariance to set one of the Wannier-centers to zero. Next, we insert the Fourier transformation of the Coulomb interaction in 3D resulting in
\begin{equation}
\begin{aligned}
    U_{1234} &= \frac{e^2}{2\pi^2}\int\dd\bvec{r}\dd\bvec{r}' \int\dd\bvec{q} \; \frac{1}{q^2} e^{i\bvec{q}(\bvec{r}-\bvec{r}')} \psi_3(\bvec{r})^* \psi_4(\bvec{r}')^*\psi_2(\bvec{r}') \psi_1(\bvec{r}) \\
             &= \frac{e^2}{2\pi^2}\int\dd\bvec{q} \; \frac{1}{q^2} \underbrace{\left(\int\dd\bvec{r}e^{i\bvec{q}\bvec{r}} \psi_3(\bvec{r})^* \psi_1(\bvec{r}) \right)}_{f_{13}(\bvec{q})} 
             \underbrace{\left(\int\dd\bvec{r}' e^{-i\bvec{q}\bvec{r}'}\psi_4(\bvec{r}')^*\psi_2(\bvec{r}')\right)}_{f_{24}(-\bvec{q})}  \\
\end{aligned}
\end{equation}
Where we introduced the Fourier transformation of a left and a right leg of the interaction as $f$. Notably, since the functions under the integral are continuous at the origin, we encounter a singular integral. To evaluate this expression for arbitrary orbitals, we first wannierize all orbitals within a cubic box chosen such that the orbitals have only negligible weight outside of the box. Each box is centered around the atom the atomic-like orbital is centered at. To evaluate the products of two Wannier functions, we linearly interpolate between the different grids assuming that the wavefunction is zero outside of its respective grid, i.e., we obtain the products on a regular grid. In the next step, we discretize the real-space integrals and evaluate them through an FFT procedure. Lastly, we evaluate the outermost integral by a summation over $q$ volumes for each of which we assume a constant function value allowing to regularize the $\bvec{q} = 0$ patch explicitly by performing the integral in spherical coordinates. We checked that we recover the correct values for selected numerical test functions~\cite{schnell2002ab}.
To ensure that the real-space cube is large enough and finely enough resolved, we check that all orbitals have a norm sufficiently close to one.

\section{Relation to Hamiltonian based downfolding}
\label{App::Hambaseddown}
A central question is how to relate the presented formalism to the Hamiltonian based approaches, e.g.~Löwdin downfolding~\cite{Lwdin1951} or canonical transformations~\cite{White2002}. We recapitulate, that in these approaches one alse starts with a separation of the Hilbert space into a target space and a rest space, here written in matrix form
\begin{equation}
    \hat{H} = \begin{pmatrix}
\hat{H}_{ff} & \hat{H}_{cf}\\
\hat{H}_{fc} & \hat{H}_{cc}
\end{pmatrix}\,,
\end{equation}
The aim of most of these approaches is to bring the Hamiltonian into a block diagonal form---decoupling the target from the rest space, either for the full spectrum or specific target states. Therefore, these approaches downfold the stationary Schrödinger equation by guaranteeing parts of the spectrum of the full operator to be exactly reproduced. Naturally---once the exact form is found---all observables on the target space are guaranteed to be exact (since time evolution does not couple the two anymore). As one works directly on the Hamiltonian, naively one works with exponentially large spaces and the states we downfold into are by construction true eigenstates of the initial Hamiltonian (up to a unitary transformation).

In contrast, the path integral (or traced based) downfolding approaches guarantee that the observables on a predefined target space are exactly reproduced and ensure this by dressing the quantum action accordingly. The states we downfold into are any basis states of the initial problem for which we find the theory to be convergent---in stark contrast to the Hamiltonian based downfolding. Furthermore, since in this formalism we construct the target space problem via the partition function, the target-space model technically depends on the temperature of our calculation. Due to these technical differences, we do not expect the downfolded models to agree between the approaches in a trivial way.

\section{Model for multi-orbital downfolding}
\label{App:model}
\begin{figure}
    \centering
    \includegraphics[width=0.5\linewidth]{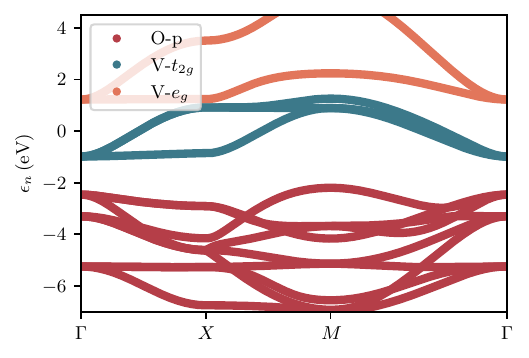}
    \caption{Bandstructure of the 14-band model for the multi-orbital downfolding example. The coloring of the bands indicate which orbital as the majority of the weight for the respective band. }
    \label{fig:sketchapp}
\end{figure}

\paragraph{Kinetics} To obtain a tight-binding model for a multi-orbital system, we first run a density-functional theory calculation of SrVO$_3$ using VASP~\cite{Kresse1996, Hafner2008} using the PBE~\cite{PhysRevLett.77.3865} parametrization of the exchange-correlation functional. We set the plane wave cutoff to $600\,$eV, use a Gaussian broadening of $0.05\,$eV and include $120$ Bands in the calculation. All calculations are performed on a $12^3$ momentum mesh. The lattice parameter is set to $a=b=c=3.84652\,$\AA. In a subsequent step, we construct a Wannier model including the nine Oxygen $2p$ and the five Vanadium $3d$ orbitals with Wannier90~\cite{Pizzi2020}. The non-interacting band-structure is visualized in \cref{fig:sketchapp}. 

\paragraph{Interactions}
On each of the atoms, we introduce a Hubbard-Kanamori type interaction, with $U=5\,$eV and $J=0.6\,$eV for the Vanadium $3$d orbitals and $U = 4\,$eV and $J=0.5\,$eV on the Oxygen $2$p orbitals of each individual oxygen atom. Additionally, we include a density-density interaction between nearest-neighbor Oxygen $2$p orbitals of $1\,$eV and a density-density interaction between the Oxygen and the Vanadium of $1.75\,$eV.

\paragraph{Downfolding}
All four downfolding variants are implemented using the divERGe library~\cite{hauck2023diverge}. For cRPA and cFRG we utilize the \texttt{gproj} feature, which projects out a defined subspace, strictly following the original proposed algorithms (for cRPA, we additionally constrain the flow equations to only include the chagre channel). To implement PCD@cRPA and PCD, we reset the Green's function by overriding the green's function generator function such that the Green's function is purely on the rest-space. Again for the RPA variant we constrain the flowing channels. The FRG code does only includes diagrams between target and rest space which end in a two-particle interaction, thus we include the missing contributions by explicitly adding them in a post-processing step. The most relevant of these is diagram (c3), see Fig.~1 in the main text. To avoid double counting, we neglect the flow of the self-energy in the rest space. We solve the flow equations on a $12^3$ momentum mesh with an additional refinement of $3^3$ for the bubble integrals. The form-factor cutoff is chosen such to include nearest neighbor atoms. We end the flow at a minimal scale of $\Lambda = 0.01\,$eV. 

\paragraph{DMFT calculation}
To test whether the different downfolding schemes are differeing in the observed physical behavior, we solve the resulting models within dynamical mean field theory. The tight-binding Hamiltonian is evaluated on a $100^3$ kmesh. We solve the DMFT self-consistency at $\beta = 20$ and use the anisimov double counting scheme~\cite{}. To calculate the $J$ imbalanced cases, we first solve the self-consistency in the averaged case and then restart from this solution keeping the double-counting fixed in the process.
\end{document}